\documentclass{aastex}
\usepackage{lscape}

\usepackage{color}
\begin{document}

\shorttitle{Ages of fifteen SMC Clusters}
\shortauthors{Parisi et al.}

\title{Age Determination of Fifteen Old to Intermediate-Age Small Magellanic Cloud Star Clusters}

\author{M.C. Parisi}
\affil{Observatorio Astron\'omico, Universidad Nacional de C\'ordoba}
\affil{Laprida 854, C\'ordoba, CP 5000, Argentina.}
\email{celeste@oac.uncor.edu}

\author{D. Geisler}
\affil{Departamento de Astronom{\'\i}a, Universidad de Concepci\'on}
\affil{Casilla 160-C, Concepci\'on, Chile.}
\email{dgeisler@astro-udec.cl}

\author{G. Carraro}
\affil{European Southern Observatory}
\affil{Casilla 19001, Santiago de Chile, Chile.}
\email{gcarraro@eso.org}

\author{J.J. Clari\'a}
\affil{Observatorio Astron\'omico, Universidad Nacional de C\'ordoba}
\affil{Laprida 854, C\'ordoba, CP 5000, Argentina.}
\email{claria@oac.uncor.edu}

\author{E. Costa}
\affil{Universidad de Chile}
\affil{Casilla 36-D, Santiago de Chile, Chile.}
\email{costa@das.uchile.cl}

\author{A.J. Grocholski}
\affil{Department of Physics and Astronomy, Louisiana State University}
\affil{202 Nicholson Hall, Tower Drive, Baton Rouge, LA 70803-4001, USA.}
\email{grocholski@phys.lsu.edu}

\author{A. Sarajedini}
\affil{Department of Astronomy, University of Florida}
\affil{PO Box 112055, Gainesville, FL 32611, USA.}
\email{ata@astro.ufl.edu}

\author{R. Leiton}
\affil{Departamento de Astronom{\'\i}a, Universidad de Concepci\'on}
\affil{Casilla 160-C, Concepci\'on, Chile.}
\email{roger.leiton@astro-udec.cl}

\and

\author{A.E. Piatti}
\affil{Observatorio Astron\'omico, Universidad Nacional de C\'ordoba}
\affil{Laprida 854, C\'ordoba, CP 5000, Argentina.}
\email{andres@oac.uncor.edu}

\begin{abstract}
We present CMDs in the $V$ and $I$ bands for fifteen star clusters in the Small Magellanic Cloud (SMC) based on data taken with the Very Large Telescope (VLT, Chile). We selected these clusters from our previous work \citep{par09}, 
wherein we derived cluster radial velocities and metallicities from Calcium II infrared triplet (CaT) spectra also taken with the VLT. 
We discovered that the ages of six of our clusters have been appreciably underestimated by previous studies, which used comparatively small telescopes, 
graphically illustrating the need for large apertures to obtain reliable ages of old and intermediate-age SMC star clusters. 
In particular, three of these clusters, L\,4, L\,6 and L\,110, turn out to be amongst the oldest SMC clusters known, with ages of 7.9 $\pm$ 1.1, 8.7 $\pm$ 1.2 and 
7.6 $\pm$ 1.0 Gyr, respectively, helping to fill a possible ``SMC cluster age gap'' \citep{gla08a}. Using the present ages and metallicities from \citep{par09}, we analyze the 
age distribution, age gradient and age metallicity relation (AMR) of a sample of SMC clusters measured homogeneously. There is a suggestion of bimodality in the age 
distribution but it does not show a constant slope for the first 4 Gyr \citep{pia11b}, and we find no evidence for an age gradient. Due to the improved ages of 
our cluster sample, we find that our AMR is now better represented in the intermediate/old period than that we derived in \citealt{par09}, where we simply took ages 
available in the literature. Additionally, clusters younger than $\sim$ 4 Gyr now show better agreement  with the bursting model of \citet{pag98}, but we confirm that this model is not a good representation of the AMR during the intermediate-age/old period. A more complicated model is needed to explain the SMC chemical evolution in that period.

\end{abstract}

\keywords{galaxies: star clusters - Magellanic Clouds}

\section{Introduction}

As it is widely known, the Small Magellanic Cloud (SMC) is the closest dwarf irregular galaxy to the Milky Way. For this reason, 
it is the ideal local analog to this type of primeval distant galaxies. Due to its low metallicity, the SMC constitutes an excellent 
field of study to explore the role played by metallicity in these galaxies. Star clusters of the Magellanic Clouds (MCs), 
particularly those of the SMC, are important to trace their chemical evolution and star formation history. Because of the 
richness and variety of the SMC star clusters, they are also important testbeds for theoretical models of stellar evolution 
at intermediate age and moderately low metallicity (e.g., \citealt{fer95}). Besides, the SMC clusters have been 
used as empirical templates to interpret the unresolved spectra and colors of distant galaxies (e.g., \citealt{betal02}). 

In order to study the early star formation history of a galaxy, it is crucial to characterize well the oldest
stellar populations. Globular Clusters (GCs) are ideal objects as tracers of  the oldest
populations. In addition, although the GC system of a galaxy exhibits an age range, the oldest GCs clearly
belong to the oldest stellar systems that were formed in the early Universe and still survive. In those
galaxies where the age of the oldest field populations can be determined and compared with the age of the
Galactic GCs, it was found that, in general, the age difference between these  two populations is not
significant within the errors and that the oldest GCs in different galaxies in the Local
Group are coeval. This indicates a common  epoch of initial
star formation in the Milky Way (MW) and its companions in the Local Group. However, the SMC is the only Local Group
galaxy with a significant cluster population which  seems not to share this  common epoch of early cluster formation.  In fact, although this
galaxy has a large number of young and intermediate-age clusters, it contains only one
cluster, NGC\,121, confirmed to be older than 10 Gyr and considered the only known SMC GC, in the
sense that its age is comparable to those of the bulk of the Galactic GCs.  This has  given rise to a
series of speculations and controversies regarding the chemical evolution of the SMC and that of its companion,
the Large Magellanic Cloud (LMC).  NGC\,121, however, is clearly not as old as the oldest GCs of the LMC or
of the MW \citep{gla08a}. In fact, NGC\,121 has an estimated age of only 10.5 Gyr \citep{gla08a} instead of
12-12.5 Gyr found for the oldest clusters of the LMC and  MW (e.g \citealt{dot10}).  Even though recent efforts have uncovered
at least one other SMC cluster approaching the age of NGC 121 (HW42, with an age of 9.3 Gyr - \citealt{pia11a}),
 it is now clear that the SMC lacks the very oldest GCs that are found in other massive Local Group galaxies.
The reason for this is presently unknown.\\

The next oldest cluster in the SMC currently known after NGC\,121 and HW\,42, L\,1,  has an age of only 7.5 Gyr, based on deep HST
data \citep{gla08a}.
Interestingly, the LMC has a well-known large cluster age gap between 3 and 12 Gyr within which only one cluster exists.
We also know, however, that the LMC contains 15 or 16 clusters older
than 10 Gyr (see, e.g., \citealt{dut99}). The SMC indeed also has a potential cluster age gap (as suggested by \citealt{gla08a}), 
with only two clusters presently known between 7.5 and 13.7 Gyr, a gap covering almost half the age of the galaxy since the Big Bang.\\

At present,  there is no  clear explanation for the lack of clusters in the so-called age gap in the LMC.
Likewise, no scenario has been proposed for why there are no truly ancient clusters in the SMC or for its possible cluster
age gap, although this latter has not been well studied and the sample size is quite small. The fact that
NGC\,121 is the only known relatively old SMC cluster means that  all conclusions related to the early
evolution of this galaxy's cluster system are based only on this single object. It is evident, in this context, that the
discovery of other old cluster in the SMC would have important implications related to its early
formation and chemical evolution, even clusters inhabiting the possible age gap between NGC 121 and L1. \\

Our group began investigating MC clusters using the powerful technique of CaII triplet spectroscopy (CaT)
with FORS2 on the VLT several years ago. As shown by \citet{col04}, CaT is a very efficient and accurate
metallicity indicator, with minimal age effects. Our first study \citep{gro06} yielded excellent data for
28 LMC clusters. We found a very tight metallicity distribution for intermediate age clusters, no metallicity
gradient and confirmed that the clusters rotate with the disk. We followed this up with an initial study
of SMC clusters \citep {par09} (hereafter Paper 1). We used FORS2 + VLT to obtain CaT spectra of more than
100 stars in 15 populous SMC clusters, spanning a wide range of ages and metallicities. We derived velocities
to a few km s$^{-1}$  per star and mean cluster metallicity (from $>$7 high probability members) to 0.05 dex. We
examined the metallicity distribution, metallicity gradient, the age-metallicity relation (AMR) and kinematics
for the cluster sample. We found a suggestion of bimodality in the metallicity distribution, no
evidence for a gradient and a hint of rotation in our cluster sample, but the kinematics are dominated by
the velocity dispersion.\\

While the cluster metallicities constituted a reliable as well as homogeneous source of information, since  
they were determined by applying the same well-proven technique, based on excellent data, cluster ages unfortunately  are on a
heterogeneous and relatively poorly-determined scale. This results from the fact that we took the ages
available from the literature, which were derived by applying several different methods. These clusters
were selected mainly from CMDs in the Washington system published by \citet{pia05a,pia07b,pia07c} which
were based on data obtained on the Cerro Tololo Inter-American Observatory 0.9m telescope. Unfortunately, such data are not optimum
for deriving accurate ages for old and intermediate-age SMC clusters, given the small aperture and large
pixel scale. The lack of accuracy and homogeneity in the cluster ages limits our confidence in the study of the chemical evolution of the SMC, despite the excellent metallicities in hand. \\

In order to improve that situation, we decided to redetermine the ages of the 15 SMC clusters studied in
Paper I on a homogeneous scale. As explained there, before obtaining the CaT spectroscopic observations, the VLT staff obtained
images of the selected clusters in the $V$ and $I$ bands. These images, usually called {\it pre-images},
were necessary to build the instrumental color-magnitude diagrams (CMDs) which allowed us to select the cluster red giants that were observed spectroscopically and to perform the requisite astrometry for these targets in order to ensure good slit positioning. {\it Pre-images} are not generally used for purposes  other than this and thus are usually taken
with  the minimum exposure times necessary, in this case to obtain reasonable photometry for the upper red
giant branch of each cluster. This unfortunately prohibits using them for something as demanding as determining accurate ages,
which requires good data to well below the MS. However, we realized that for a minimum additional cost in overhead
we could lengthen the preimages from the usual few second exposures to several minutes. Given the 8-m
aperture and generally good to excellent seeing available, this ensured that these data would go much deeper
than any available and would yield excellent material from which to derive ages. Thus, we put these preimages
to very good astrophysical use! However, in Paper 1 we only used the literature ages as time did not permit
us to obtain the required photometry. Also, Paranal Observatory does not provide standard star images in order to determine the
transformation of magnitudes and colors to the standard system. Thus, standardizing the preimages was
problematic and we waited until we could find an appropriate solution. We have subsequently solved 
this either by obtaining independent calibration photometry and/or obtaining the ages differentially, 
without the need for standardization. Hence, the time was ripe for investigating the cluster ages. \\

In this Paper, we report on the ages we derive for our cluster sample from the preimages and related results. In Section 2, we
detail the procedure used to obtain the uncalibrated photometry. Section 3 describes our initial age
determination procedure. Next we present the CMDs of each cluster and investigate two existing age calibrations.
Here we find that several of our clusters are substantially older than indicated by previous studies 
and indeed turn out to be amongst the oldest star clusters known in the SMC. Section 5
describes the importance of such clusters for examining the early chemical and dynamical evolution of the SMC.
In addition, we derive a new age calibration appropriate for old to intermediate-age SMC clusters, which yields
ages for our oldest clusters supporting their extreme age. In order to further strengthen these results, we also
present isochrone ages based on calibrated photometry we derive for the three most interesting old cluster candidates.
In Section 6 we discuss our results, while Section 7 summarizes our findings.

\section{Observations and Reductions}

Paper I gives all relevant details about the cluster sample.
Basic information is given in Table \ref{t:sample},  wherein we list the different
cluster designations followed by their right ascension and declination, and the metallicity and semi-major
axis $a$ reported in Paper I (see section 6.2 for details about calculation of $a$). The metallicities of
the 15 clusters presented in Paper I were derived from the equivalent  widths of the CaT
lines, measured on the spectra of a number of cluster member red giants. \\

As we will see in the next section, there exist age calibrations that permit one to derive the age of a cluster
from the difference between the  magnitude of the red clump (RC) and the main sequence turn-off (MSTO). Since
this involves a difference in magnitudes, we do not need calibrated photometry in order to
derive cluster ages through this procedure. When we programmed the spectroscopic observations, we deliberately
had in mind the need to rederive cluster ages in a homogeneous way, so we requested VLT staff to extend the
{\it pre-image} exposure times. We used  exposure times of 200 sec and 100 sec in the $I$ and $V$ bands,
respectively, in addition to the usual short exposures required to not saturate the brightest giants.\\

FORS2 has two 2k $\times$ 4k CCDs. Pixels were binned 2$\times$2, providing a plate scale of 0.25'' pixel$^{-1}$. 
Target clusters were centered on the master CCD (upper) while the secondary CCD (lower) 
was used to observe field stars. The observations were performed with a typical seeing less than 1''
The {\it pre-images} were corrected for {\it bias} and {\it flat-field} following the usual procedure using
IRAF \footnote{Image Reduction and Analysis Facility, distributed by the National Optical Astronomy
Observatories, which is operated by the Association of Universities for Research in Astronomy, Inc., under
contract with the National Science Foundation.}, while the Point Spread Function (PSF) photometry was performed
using the DAOPHOT/ALLSTAR packages, independent from IRAF \citep{ste87}.

Figure 1 show, as example, the typical behavior of color and magnitude errors (upper and lower panels, respectively) as a function of instrumental magnitude. 
For magnitudes less than 15, most stars have $\sigma_v <$ 0.03 and $\sigma_{v-i} <$ 0.05, for the bulk of cluster. For some clusters,
for example, L\,17, L\,27, L\,111, thes error are larger probably due to the greater degree of crowding in the core regions. 

\section{Morphological Age Index }

One way to derive cluster ages uses a morphological index that quantifies observational parameters
in the CMD that are sensitive to age \citep{att85,jp94,ros99}. A differential method, which utilizes {\it
differences} between well-observed parameters to derive age, is particularly powerful as it is insensitive to
photometric problems, reddening, distance, etc. This technique allows the derivation of cluster ages avoiding
the well known difficulties present in the isochrone fitting method \citep{van90,sd90,sw97} but requires
accurate calibration with fiducial clusters.\\

\citet{pjm94} and \citet[hereafter JP94]{jp94}, respectively, defined the $\delta V$ parameter as the difference
between the visual magnitude $V$ of the MSTO ($V_{TO}$) and that of the RC ($V_C$) (see Figure 1 of
\citealt{pjm94}). It is well known that the MSTO luminosity depends on the cluster age but that the RC luminosity 
is almost age independent \citep{can70}, therefore  $\delta V$ turns out to be a good and reliable age indicator. 
JP94 showed that there is a good correlation between the parameter $\delta V$ and the
cluster age, in the sense that the younger a cluster, the smaller $\delta V$.  Due to the heterogeneity of 
the calibration material, JP94 emphasize the fact that their calibration can only be used to make a ranking of 
clusters in terms of age and is not robust for absolute age determinations. For this reason, \citet{cc94} (hereafter CC94),
and, more recently, \citet[hereafter S04]{swp04} recalibrated the relation between age, metallicity and $\delta V$,
based on homogeneous Galactic cluster samples. \\

 Both calibrations include a metallicity dependence but, because this dependence is small, in cases where there is no information regarding
the cluster metallicity, the metallicity-dependent term can be neglected (CC94). In spite of
having information about the metallicity of our clusters, we decided to neglect the metallicity term because
those clusters used for the derivation of all available $t-\delta V-[Fe/H]$ calibrations (for example S04 and CC94)
are galactic clusters that cover a metallicity range very different from the one covered by our cluster sample, which 
are considerably more metal-poor. Since we are trying to derive ages of SMC clusters which have a very different
chemical evolution from those of our Galaxy, the metallicity issue is relevant. It is for this reason plus the fact that metallicity terms are
small that we will not consider the metallicity effect on age. We note that the systematic effects from 
uncertainty in the basis of the CC94 (and S04) relations and the neglect of any metallicity compensation means that the uncertainties 
in the derived ages are likely larger in many cases than the errors stemming from determining $\delta V$. A revised  $\delta V$ age calibration explicitly taking into account the chemical
composition of both SMC and LMC star clusters is needed.

One of the most reliable methods to derive cluster ages is isochrone fitting. However, the CMDs necessary
to apply such a method must be based on calibrated photometry. As noted above, our PSF photometry was performed on VLT
{\it pre-images} and we did not have the calibration images necessary to transform magnitudes and colors to the
standard system. We then decided to use the $\delta V$ method as our primary age determinant since it does not rely on calibrated photometry.

\section{Cluster Ages}

\subsection{Color Magnitude Diagrams  and Age Determination}

Before starting the analysis of the observed cluster CMDs, it is necessary to minimize their contamination by
field stars. Such contamination clearly constitutes an important factor which limits precision in the
determination of all cluster parameters, particularly $V_{TO}$. To decrease the field star contamination in
the CMDs, we followed the procedure described by \citet{pia05a}, that is, to build CMDs including stars located
in different concentric circular areas centered on the cluster  center. Then we qualitatively analyzed the
ratio of field to cluster stars within each ring by studying their variation as a function of radius.\\

Figures \ref{f:bs121M.bin} - \ref{f:l111M.bin} show the instrumental CMDs built from the DAOPHOT output,
corresponding to circular regions centered on each
cluster. For each cluster, we built four different CMDs which include stars belonging to four radial ranges
of 100 pixels width each. The first diagram (left top panels) includes all of the stars located within a circle of 100
pixel radius. The following two diagrams (right top and left bottom panels), include all of the stars between 100 and 200 pixels and from 200 to 300 pixels from the cluster center, respectively. The
fourth diagram (right bottom panels) includes all of the stars with distances from the cluster center greater than
300 pixels.\\

It can be clearly seen that the first diagram (inner interval) is, as expected, the least contaminated by field
stars. The second annular CMD is still dominated by cluster stars, although the field contamination has increased 
considerably and is beginning to dominate the cluster stars in the third annular CMD. Therefore,
to identify the CMD features of interest  ($V_C$ and $V_{TO}$), we decided to use the CMDs constructed with stars
belonging to the two innermost regions, that is, those located within 200 pixels from the
center. Figures \ref{f:bin12.1} - \ref{f:bin12.4} show the CMDs of
our cluster sample constructed using only the aforementioned  annuli. These CMDs were used to determine $V_C$ and
$V_{TO}$. As can be seen in Figures \ref{f:bin12.1} - \ref{f:bin12.4}, all clusters exhibit reasonably well defined  RCs and  MSTOs,
except clusters BS\,121, L\,17 and L\,27, in which the dispersion near the MSTO is comparatively higher.
In these cases, of course, the determination of the MSTO is clearly more uncertain.
To quantify contamination by field stars in the CMD of Figures \ref{f:bin12.1} - \ref{f:bin12.4}, we calculated 
the ``contamination index'': since we know the field star surface density 
surrounding the cluster, i.e., in the r $>$ 300 pixel region, we calculated the number of field stars within a r = 200 pixel circle. 
The ratio between this number and the real number of (cluster + field) stars inside the said circle is a measure of the field star contamination. 
This parameter is listed in the last column of Table \ref{t:ages}. It is seen that contamination is very high 
for BS\,121, strong for HW\,47, L\,17, L\,19 and L\,27, relatively small for HW\,84, HW\,86, L\,5, L\,6 and L\,7, and very low for the remaining clusters. Clearly, ages derived for clusters in the two most contaminated categories will have larger errors.
  
Because the determination
of $\delta V$ includes some degree of subjectivity, three members of our team determined this parameter
independently. Since these three independent determinations generally showed good agreement
(total spread of 0.1-0.5 in $\delta V$),
we then adopted the value of $\delta V$ corresponding to the average of such determinations. We estimated that
a typical error in the determination of both $V_C$ and $V_{TO}$
is 0.2 magnitudes, which implies an uncertainty of $\sim$ 0.3 magnitudes for $\delta V$. However, the range in such factors as contamination among our sample means
 that this adopted typical error is likely too small in
 some cases and too large in others.
Then we estimated ages using both the S04 and CC94 calibrations. From here on, we
will refer to the ages derived from the calibrations of S04 and CC94 $t_S$ and $t_C$, respectively.\\

In Table \ref{t:ages} we list the cluster designations, $\delta V$, $t_S$ and $t_C$, and their respective errors. 
The errors were calculated using error propagation on the corresponding equation.
We also include in the table the age values reported in the literature, $t_L$, as well as the reference of the
corresponding work where $t_L$ was determined. Finally, we list the ``contamination index'' described above.\\

As can be seen from Table \ref{t:ages}, $t_S$ is systematically larger than $t_C$ for all clusters. This tendency
becomes considerably larger for older clusters. We  compared ages derived for both the CC94 and S04 prescriptions 
to those found by \citet{gla08b} for seven old to intermediate-age SMC clusters using deep Hubble Space Telescope (HST) 
data and state-of-the-art isochrones. These ages are
among the best constrained of any SMC clusters. This comparison is presented in Figure \ref{f:cc94Mvss04_glatt}.
It is clear that CC94 ages are in much better agreement with those of \citet{gla08b} for all clusters. Based on
these arguments, for the subsequent analysis, we use ages estimated from the CC94 calibration. However, it is important 
to note that, although CC94 ages are in better agreement with Glatt's ages than those of S04, there is a significant 
offset even  between ages from \citet{gla08b} and those derived by
using the CC94 calibration, with the CC94 ages being larger by about a Gyr on average, independent of age. More precisely,  
the difference between both values goes from 0.29 to 2.02 Gyr with a mean value of 1.22 (standard deviation of 0.61). This 
difference may arise from the fact that,
as already mentioned, the SMC and Galactic clusters used by CC94 as calibrators have very different chemical
compositions. Again, a new age calibration for the chemical composition of the Magellanic Clouds is sorely needed.\\

\subsection{Comparison with previous age determinations}

To compare our age determinations with those from other authors, we show $t_L$ in Figure \ref{f:age_lit_cc94M}
as a function of the age values found in the current study ($t_C$). The solid line represents the 1:1 relation
between both quantities. From this diagram, we classified the clusters in three different groups: the ages of
the 9 clusters identified with circles are in reasonably good agreement with those previously derived by other
authors, while the remaining clusters of the sample exhibit age values reported in the literature lower than
the ones derived in the present work, which is more remarkable for squares than for triangles. Triangles and
squares represent clusters where $t_L$ is moderately and considerably smaller than $t_C$, respectively. To make
this discrimination, we adopted the following criteria: we first calculated the difference between
$t_L$ and $t_C$ and then we compared that difference with the mean error of $t_C$ ($\bar{\sigma}_{t_C}$),
which is 0.54 Gyr. If the difference $t_L - t_C$ is lower than 3$\bar{\sigma}_{t_C}$, we consider that the
value of $t_L$ is in agreement with the value of $t_C$ (circles). If the difference is larger than
3$\bar{\sigma}_{t_C}$ but smaller than 6$\bar{\sigma}_{t_C}$, we consider that the disagreement between $t_L$
and $t_C$ is moderate (triangles). Finally, if the difference $t_L - t_C$ is larger than 6$\bar{\sigma}_{t_C}$,
then both age determinations exhibit a considerable disagreement (squares). The mean
differences between $t_L$ and $t_C$ and their corresponding standard deviations are  0.7  $\pm$ 0.4 , 2.7 $\pm$
0.5  and 5.1  $\pm$ 0.4  Gyr for {\it consistent clusters}, {\it moderately inconsistent clusters}
 and {\it highly inconsistent clusters}, respectively. Note that all of the inconsistent clusters have 
ages derived here that are larger than those found in the literature. The two highly inconsistent clusters
are L\,4 and L\,6, which clearly stand out from the other clusters as being much older than previous indications. Much
more will be said about these intriguing clusters in the next Section. \\

The values of $t_L$ for clusters L\,4, L\,6, L\,27 and L\,19 were derived by \citet{pia05a} following a
procedure similar to ours. Using the ($T_1$,$C-T_1$) CMDs, they derived the parameter $\delta T_1$ (difference
between the $T_1$ magnitude of the RC and the MSTO). This parameter allows one to determine the age of a cluster
from the calibration of \citet{gei97}.  Note that this calibration included a wide range of metallicity, 
covering that of the majority of our cluster sample here, so that metallicity effects should be minimal. 
The most likely explanation for the differences found between $t_L$ and
$t_C$ is that the ($T_1$,$C-T_1$) CMDs of these clusters are not deep enough to reach the corresponding MSTO
with sufficient photometric accuracy to define this feature well. This fact may have led \citet{pia05a}
to underestimate $\delta T_1$, which, in turn, implies a significant underestimation of the $t_L$ value. The
photometric observations used in the current study were obtained with an  8 m telescope, while those of
\citet{pia05a} used an 0.9 m telescope. At a given magnitude, the photometric errors of the 8 m telescope
observations are roughly a factor of 3 less than those from the 0.9 m telescope. As such, not only do the 8 m 
data extend to deeper magnitudes, but they do so with greater precision. Given the much smaller telescope, 
the difficulties to reach the fainter
portions of the cluster CMDs are more problematic, especially for older clusters in which the MSTO occurs at
fainter magnitudes. It is then to be expected that the difference between $t_L$ and $t_C$ is more noticeable
for the older clusters of our sample (L\,4 and L\,6), as shown in Figure \ref{f:age_lit_cc94M}. In addition,
both the {\it seeing} as well as the telescope scale represent important factors that tend to decrease the
quality of  Piatti et al.'s data compared to ours. In fact, the photometric data used by \citet{pia05a}
were obtained with a {\it seeing} of 1.5'' and a scale of 0.4''px$^{-1}$, while our {\it pre-images} were
obtained with a {\it seeing} typically smaller than 1'' and a scale of 0.25''px$^{-1}$.\\

Regarding cluster L\,17, the value of $t_L$ was estimated by \citet{rz05} from the comparison of the cluster
integrated colors ($U-B$, $B-V$ and $V-I$) with models of simple stellar populations.  As these authors point
out, however, this work lacks accuracy in the calculated ages, mainly because the integrated colors have low
precision and the models used do not predict consistent colors for a given age. As \citet{rz05} emphasize,
their cluster ages should only be taken as a guide for future investigations. As for the cluster L\,108, its $t_L$
value was derived by \citet{pia07c} by fitting theoretical isochrones to the Washington system CMD. Recently,
\citet{pia11b} derived ages for L\,108 of 1.10 $\pm$ 0.25 and 1.60 $\pm$ 0.30 Gyr, by fitting theoretical
isochrones and by using the $\delta T_1$ parameter, respectively. According to the above described criteria, the 
L\,108 isochrone age is still moderately inconsistent but its morphological age is consistent with the value
derived in the present work. Again, the most likely reason for the discrepancy is that the Piatti et al. data 
are from a much smaller telescope (in this case a 1.5m) with inferior seeing and scale.\\

\section{Old SMC clusters}

As mentioned in the previous section, two of our clusters (L\,4 and L\,6) appear much older than indicated 
by previous studies. Indeed, our data strongly suggest they are indeed very old SMC clusters, helping to 
fill the  possible age gap between L\,1 and NGC\,121. 

To corroborate the very old nature of L\,4 and L\,6, we decided to perform independent age determinations. In 
addition, although the age of L\,110  here derived is consistent with the value reported in the literature 
(within the errors), our age of 7.6 Gyr based on the CC94 calibration officially makes L\,110 
slightly older than L1, providing another ``age-gap'' cluster and a third very interesting object to be more 
thoroughly investigated, which we now do.\\

It is worth mentioning that \citet{pia12b} recently found that the SMC cluster ESO\,51-SC09 appears to be 
quite old, with an age (7.0 Gyr) comparable to that of L\,1. He calculated the parameter $\delta V$ and used the 
JP94 calibration to derive the cluster age. He then transformed $\delta V$ into $\delta T_1$ according 
to equation (3) of \citet{gei97} and calculated ESO 51-SC09's age from equation (4) of the same work. 
Finally, he fitted theoretical isochrones of \citet{mar08}. He derived age values of 8.3 $\pm$ 1.6, 6.5 $\pm$ 1.0 and 6.3 $\pm$ 1.0 Gyr from 
$\delta V$, $\delta T_1$ and isochrone procedures, respectively. \citet{pia12b} averaged these three values, 
obtaining an age of 7.0 $\pm$ 1.3 Gyr. However, as we have already mentioned, the JP94 calibration  suffers 
from serious problems that render it less reliable for deriving accurate ages, and in particular, it overestimates ages, 
especially for older clusters. Using the $\delta V$ value reported by 
\citet{pia12b} (2.5 $\pm$ 0.15 Gyr) and the CC94 calibration, we derive an age of 5.2 $\pm$ 0.4 Gyr for 
ESO\,51-SC09, which is 3 Gyr less than the value reported by \citet{pia12b} using the JP94 calibration.
This value turns out to be in better agreement with those derived from $\delta T_1$ and isochrone fitting. If we 
calculate the average of the three values from $\delta V$, $\delta T_1$ and isochrone fitting, using the CC94 instead 
of the JP94 calibration, ESO\,51-SC09 has an age of 6.0 $\pm$ 1.4 
Gyr. Though the average age derived in the present work is in agreement with the average age derived by 
\citet{pia12b}, within the errors, ESO\,51-SC09 is now younger than L\,1, which supports the view that the JP94 calibration 
needs to be used with caution. \\

NGC\,361 is another cluster \citet{pia12b} suggested may be older than L\,1. He reports for this cluster 
an age of 8.1 $\pm$ 1.2 Gyr derived by \citet{mig98}, but this value corresponds to the age NGC\,361 
would have if we assumed that L\,1 is 9 Gyr. This is obviously not the case because L\,1 now has a very 
well determined age of 7.5 Gyr \citep{gla08b}. \citet{mig98} actually derived  an age of 6.8 
$\pm$ 0.5 Gyr for NGC\,361 and then added 1.3 $\pm$ 1.1 Gyr to this value in order to place its age on the scale of \citet{ols96}. 
This allows us to conclude that the only currently known cluster candidates to lie in the possible ``SMC age gap'' between 
L\,1 and NGC\,121 are L\,4, L\,6, L\,110 and HW\,42.\\

Based on samples of catalogued and well studied SMC clusters, \citet{pia11a} performed a statistical study of clusters 
older than 2.5 Gyr. He found that only about seven old or relatively old clusters have yet to be found. We note, however, that a considerable number of clusters (used by \citealt{pia11a}) whose ages were determined 
from the MSTO were 
observed with small telescopes and relatively poor seeing conditions and pixel scale. Therefore, it is likely that the photometry of some clusters 
studied by these authors is not deep enough to delineate the MSTO clearly. 
In consequence, some
of these clusters could have ages that are underestimated, as shown clearly in Figure \ref{f:age_lit_cc94M}, and which is the case for both L4 and L6.

\subsection{The SMC $\delta$ calibration}
 
In order to obtain an independent age determination for our new candidate old SMC clusters L\,4, L\,6 and L\,110 based on the best observed SMC clusters,
we derived a new calibration (the ``SMC $\delta 
$ calibration''), using the SMC clusters studied by \citet{gla08a,gla08b}. These authors observed  the clusters 
Lindsay\,1, Kron\,3, NGC\,339, NGC\,416, Lindsay\,38, 
NGC\,419 and NGC\,121 with HST. From a photometric analysis of their $m_{555}$ vs. 
$m_{555} - m_{814}$ CMDs, they reported the  $m_{555}$ magnitude of the MSTO and of the RC (see Table 2 of 
\citealt{gla08b}). The difference between these values should be identical to $\delta$V. They also derived cluster 
ages using Padova \citep{gir08,gir00}, Teramo \citep{pie04} and Dartmouth isochrones \citep{dot07}. We adopted their 
ages given by the  Dartmouth isochrones because, as \citet{gla08a,gla08b} note, this set of isochrones are 
the best approximation to the CMDs, except for NGC\,419,  in which Padova isochrones appear to be the best 
representation of its CMD. The Dartmouth age for NGC\,121 is 10.5 $\pm$ 0.5 Gyr \citep{gla08a}, while the Dartmouth ages 
for the other six observed clusters are given in Table 6 of \citet{gla08b}. The age of NGC\,419  is rather 
uncertain because its CMD presents more than one MSTO. Hence, we decided not to include this cluster in the SMC 
$\delta $ calibration. \\

We present in Figure \ref{f:glatt} the behavior of age as a function  of $\delta m_{555}$ by using the Glatt et 
al. values. The best linear fit (solid line) is:

\begin{equation}
Age = 8.1 (\pm 1.9) \times \delta (m_{555}) - 16.6 (\pm 5.7),                                                 
\label{eq:cal_smc}
\end{equation}

\noindent where age is given in Gyr. The fit has an rms of 0.83 Gyr. This ``SMC $\delta$ calibration'' is valid 
only for clusters older than $\sim$ 6 Gyr  with similar metallicities ([FeH] $\sim$ -1) and can be applied to 
any V-band-like photometric system because it only involves a magnitude difference. \\ 

We then applied this calibration to our three old clusters, which indeed have 
similar metallicities to the Glatt sample \citep{par09}. We found ages for L\,110, L\,4 and L\,6  of 6.6, 6.9 
and 7.7 Gyr, respectively. These values are in very good agreement with those found using the CC94 calibration 
(albeit showing the same offset found in Figure \ref{f:cc94Mvss04_glatt}) 
but are more than 4 Gyr younger than the age calculated from the S04 calibration. This  reaffirms our earlier suggestion that the S04 calibration 
overestimates age, which prompted us to choose the CC94 calibration. The error in age due to the error  in 
$\delta V$ is 2.42 Gyrs. We added in quadrature this value and the rms dispersion of the linear fit which renders 
an age error of 2.5 Gyr. Note that, according to this procedure, all 3 clusters are about the same age as L\,1 at 
7.5 Gyr, with L\,6 slightly older, and L\,4 and L\,110 a bit older than the next oldest clusters in the Glatt sample at 6.5 Gyr.\\

\subsection{Fitting isochrones}

For  another independent age determination for L\,4, L\,6 and L\,110, we also fit isochrones to the observed CMDs. 
As previously mentioned, our FORS2 photometry is uncalibrated because we do not have images of standard stars
taken during the same night as the {\it pre-images}. We emphasize, however,  that the three clusters were observed with FORS2
during the same night of August 8, 2005, in a time interval shorter than one hour. The night, according to VLT logs, was declared
clear. For these reasons, the three clusters have been observed under the same, good sky conditions and at virtually the same airmass. \\
\noindent

To tie the FORS2 images to the standard system, we obtained independent images of these 
three clusters, as well as images of standard stars to perform the corresponding calibration of our PSF photometry 
using local standard stars. From now on, for the sake of clarity, we will call CPIs the cluster {\it pre-images} 
and CCIs (cluster calibration images) the images obtained to calibrate the PSF photometry of the CPIs. \

CCIs have been obtained at  Cerro Tololo InterAmerican observatory (CTIO), using the Yale 1.0m telescope, operated by 
the SMARTS consortium\footnote{\tt http://http://www.astro.yale.edu/smarts} during an observation run in Dec 8-10, 
2010. The telescope works only in imaging mode with the Y4KCAM. This camera is equipped with an STA 4064$\times$4064 
CCD\footnote{\texttt{http://www.astronomy.ohio-state.edu/Y4KCam/ detector.html}} with 15-$\mu$m pixels, yielding a scale 
of 0.289$^{\prime\prime}$/pixel and a field-of-view (FOV) of $20^{\prime} \times 20^{\prime}$ at the Cassegrain focus of the CTIO 1-m telescope. 
 The CCD was operated without binning, at a 
nominal gain of 1.44 e$^-$/ADU, implying a readout noise of 7~e$^-$ per quadrant (this detector is read
by means of four different amplifiers).\\

The clusters were observed with exposures  of 10, 60, and 600 secs in both the V and I filters. To tie photometry 
to the standard system we observed the \citet{l92}  fields SA98 and T Phe 3 times during the night, spanning an 
airmass range from 1.03 to 2.10 . SA98 is a particularly useful field since it contains more than 50 standard 
stars with a wide range in color.\\

After basic pre-processing, instrumental magnitudes were extracted following the point-spread function (PSF) 
method \citep{ste87}. Aperture corrections were then determined large aperture photometry of 
a suitable number (typically 10 to 20) of bright, isolated, stars in the field. These corrections were found to 
vary from 0.190 in V to 0.200 in I. The PSF photometry was then aperture-corrected, filter by filter. Finally, 
calibrated magnitudes were obtained using the calibration solution inferred from the Landolt standards.

Once  the PSF photometry of the CCIs has been transformed into the standard system, we used it to calibrate the 
CPIs to the standard system. We proceeded as follows: 1) Firstly, we selected 
as local standard stars (LSSs) 97 stars from the CCIs standard system photometry. (2) Secondly, we identified 
the LSSs in our CPI PSF photometry. (3) Next, we used the standard magnitudes and colors ($V$ and $V-I$) of the 
LSSs (obtained from the CCIs) and the instrumental magnitudes and colors ($v$ and $v-i$) of the LSSs (obtained 
from the CPIs) to find the transformation coefficients to the standard system. (4) Finally, we applied the 
obtained transformation equations to all stars of the CPIs.   \\

To select LSSs, we carefully chose bright stars as isolated as possible, avoiding stars located near the 
cluster centers, which fell within our color range of interest (from the upper RGB to MSTO). 
LSSs have $V$ magnitudes from $\sim$ 
17.5 to $\sim$ 21 and colors $V-I$ from $\sim$ -0.5 to $\sim$ 1.2.  

We used IRAF FITPARAM task to derive the transformation coefficients. Because we are using LSSs, we fixed 
the airmass coefficients at zero. The equations and coefficients are: \\

\begin{equation}
v = (-2.413 \pm 0.011) + V + (-0.037 \pm 0.012)  \times (V-I) 
\label{eq:transf.v}
\end{equation}

\begin{equation}
i = (-2.137 \pm 0.016) + I + (0.023 \pm 0.019) \times (V-I) 
\label{eq:transf.i}
\end{equation}

Capital and lowercase letters represent standard and instrumental magnitudes, respectively. The rms are 0.036 and 0.096, for $v$ and $i$ fits, respectively.
The standard magnitudes 
and colors of our observed stars were calculated inverting equations \ref{eq:transf.v} and  \ref{eq:transf.i}.
The resulting CMDs for L\,4, L\,6 and L\,110 are presented in Figures \ref{f:l4_fiducial}, \ref{f:l6_fiducial} and 
\ref{f:l110_fiducial}, respectively. 

To derive ages from isochrone fits, we followed the prescriptions given by \citet{gla08b}. Using the 
$m_{555}$ vs $m_{555} - m_{814}$ CMDs, they fit three different models of isochrones: Padova,
Teramo and Dartmouth. We also fit, when possible, these three sets of isochrones to our 
own CMDs. Isochrone fits depend of course on many parameters, in particular cluster metallicity, reddening
and distance modulus. We adopted cluster metallicities from Paper I, which were very accurately determined. Therefore, 
we have only two free parameters to adjust, reddening and distance modulus. Although the SMC distance modulus is 
assumed to be 18.88 \citep{sto04}, small adjustments to this parameter could be made in order to take into account the 
depth of the SMC in the line of sight (LOS). 

According to \citet{cro01}, the LOS depth of the SMC ranges between 6 and 12 kpc. Consequently, it would 
seem reasonable to adopt 9 kpc as the mean value of the LOS depth. If we adopt 60 kpc for the mean SMC 
heliocentric distance and consider that star clusters can be located in front or behind the SMC, then 
their apparent distance moduli may present differences as large as 0.2 mag.\\

We fit Teramo and Dartmouth isochrones (upper and lower panels of Figures \ref{f:l4.iso} and \ref{f:l110.iso}, respectively) to the CMD of L\,4 and L\,110, and Teramo and Padova 
 isochrones  (upper and lower panels of Figure \ref{f:l6.iso}, respectively) to the CMD of L\,6. 
We adopted a chemical
composition $Z=0.002$ ([Fe/H]=-1.02, \citealt{ber94}) for L\,4 and L\,110, and $Z=0.001$ ([Fe/H]=-1.33, \citealt{ber94}) for L\,6, which are the closest to the observed values of these clusters. 

 In all cases, we adopted a set 
of isochrones and superimposed them on the CMDs, once they were 
shifted by the corresponding E(V-I) colour excess and by the apparent distance modulus (DM). 
Then we adopted, as the cluster age, the one corresponding to the
isochrone which best reproduced the cluster fiducial ridgeline. The fitting parameters are shown on the corresponding plot (Figures \ref{f:l4.iso} and \ref{f:l6.iso}).  

We had to face the following issue when we fitted isochrones to L\,4 and L\,6. 
The CMD of L\,4 is shifted to the blue by 0.035 magnitudes (in $V-I$) with respect to the CMD of L\,6, as judged by their RCs and RGBs. These two clusters have virtually the 
same reddening  \citep{pia05a}, and indeed lie very close together in the outer parts of the SMC, but L\,4 is more metal-rich than L\,6. Therefore, the  L\,4 fiducial  should be redder than that of L\,6.
For this reason, we had to adopt E(B-V) values
larger than the ones reported in the literature (0.03 and 0.04 for L\,4 and L\,6, respectively,  \citealt{pia05a}). This problem is likely due to the fact that L\,4 and
L\,6 were observed on the same CTIO 1m frame, which was centered on L\,6, leaving L\,4 located near  the image border. It is well known that objects observed in the border of such images may 
have a difference of up 0.05 magnitudes with respect to the centered objects.

From this procedure we find than the two clusters have ages of $\sim$ 7 Gyrs. While our photometric calibration  is not precise enough to consider these ages robustly determined, 
it is important to note that they constitute an independent confirmation of the truely old nature of these clusters.

\section{Age Analysis}

In the previous section, it was demonstrated that the combination of the $\delta$V parameter and CC94 calibration 
is a very good indicator of cluster age. For the subsequent analysis, we then adopted values of cluster ages 
derived in the present work using the CC94 calibration and metallicities from Paper I. As a consequence, the ages and 
metallicities of our sample of 15 SMC clusters were determined homogeneously and on the same scale. 
We believe our errors in both age and metallicity to be significantly smaller than in other works (e.g. \citealt{pia11a,pia11b,pia11c}), 
as well as being homogeneously derived. However, even though our sample is homogeneous, it is 
limited to only 15 clusters. In order to study global effects, a larger sample is necessary. We now proceed to enlarge the cluster sample as carefully as possible.

\subsection{Age Distribution}

The age distribution (AD) of a sample of star clusters is a very valuable tool that allows us to understand cluster formation processes. Until 
about a decade ago, it was widely believed that the SMC formed its clusters continuously over the last 12 Gyr 
(\citealt{ols96,vdb91,dch98}-hereafter DH98), while the AD of star clusters in the LMC was known to have the infamous age gap \citep{dac91}. \citet{ri00} presented the first compelling evidence for bursts in cluster formation in SMC clusters. They reported HST photometry and MSTO ages for seven SMC 
clusters older than 1 Gyr and they analyzed the AD of their cluster sample together with the sample previously 
studied by DH98. They found two peaks in the AD at 2 $\pm$ 0.5 and 8 $\pm$ 2 Gyr. More recently, however, 
\citet{gla08b}, using superior HST photometry for 7 clusters found no evidence of two cluster formation bursts, although their AD reveals a 
slight enrichment in the cluster formation process at 6 Gyr. Although \citet{gla08b} 
ages were accurately determined, their sample is statistically very small. On the other hand, \citet{pia11b} 
obtained the most complete SMC AD using 43 star clusters with age estimates (see their Table 
19) derived in their paper and supplemented by ages  taken from the literature. The AD derived by these authors presents bimodality, showing two peaks of 
enhanced star formation processes at $t  \sim$ 2 and 5 Gyr. Their AD also shows that the slope in the cluster formation/destruction rate is constant from 12 to $\sim$ 
8 Gyr. According to these authors, this may reflect the possibility that in this period there existed a constant cluster formation rate, in agreement with the simple closed box model presented by DH98. \\

The crosshatched histogram in Figure \ref{f:AD} shows the AD obtained for the present fifteen cluster sample. 
Unfortunately, our sample is too small to draw definitive conclusions. For this reason, we decided to add other previously studied clusters to our sample, trying to 
maintain age homogeneity as far as possible. Piatti et al. systematically studied SMC star clusters from Washington 
photometry. In a series of papers they calculated the $\delta  T_1$ parameter as the difference between the 
$T_1$ magnitude of the RC and the MSTO. We selected 27 clusters from \citet{pia01,pia07b} and 
\citet{pia11a,pia11c}. We transformed their $\delta T_1$ values into $\delta V$ by inverting equation (3) of 
\citet{gei97}. We also included the age value of  ESO\,51-SC09 (5.2 $\pm$ 0.4  Gyr) derived in
section 5 by using the CC94 calibration. Next, we calculated cluster ages using the CC94 calibration to place Piatti et al.'s cluster ages 
on the same scale as ours. \citet{pia11b} derived the age of 14 SMC clusters (8 of them not included
in our sample) from the $\delta T_1$ parameter.  These authors did not include the $\delta T_1$                        
values but we estimated  them from the published CMDs. We did not include the cluster IC\,1708 because its
CMD does not have an evident clump. In Table \ref{t:piatti}, we list cluster designation, $\delta T_1$ with
its error, the reference of the work from where $\delta T_1$ was taken, $\delta V$ and the corresponding
age with their respective errors.  

Therefore, our full sample now consists of a total of 50 clusters with ages on 
the same scale, comparable to the sample used by \citet{pia11b}. 
In relation to \cite{pia11b} work, it is worth pointing out that they use an age scale where L\,1 is 10 
Gyr \citep{mig98}, although the best age determination for L\,1 is 7.5 Gyr  \citep{gla08b}. Similarly, Piatti gives an age for NGC 121 of 12 Gyr 
and K3 of 9 Gyr, while the corresponding Glatt et al values are 10.5 and 6.5, respectively.
Thus, it appears that the Piatti et al. scale is indeed somewhat different from the one adopted here, especially for the oldest clusters.\\

The unhatched histogram of Figure \ref{f:AD} represents the AD of the extended sample. 
Figure \ref{f:AD}  reproduces the two peaks found in \citet{pia11b} at $\sim$ 2 and $\sim$ 5 Gyr. 
They suggest the two peaks correspond to bursting cluster formation episodes. Very recently, \citet{wei13} have derived large 
scale field star formation histories of both Clouds from deep HST imaging of a number of regions. They find two notable peaks 
in the SMC, at  $\sim$ 4.5 and $\sim$ 9 Gyr ago,
which are not present in the LMC. They also find sharp increases in the star formation history of both
galaxies starting about 3.5 Gyr ago. One of course expects that the field and cluster formation histories
should be very similar. We do find good correspondence with the strong increase starting 3.5 Gyr ago,
and reasonable agreement with the peak of about 4.5 -5 Gyr ago. The field stars also appear to show a
peak around 2 Gyr ago, as we find in the clusters. However, the cluster sample is too small to discern 
any structure earlier than 6.5 Gyr ago, in  particular to corroborate or not the field star peak at 9 Gyr.

Our sample does not appear to 
show a constant slope for the first  $\sim$ 4 Gyr of the SMC, as  posed by \cite{pia11b}. 
In any case, the numbers are very small and statistics therefore very uncertain. 
Note that the ages of L\,4 and L\,6 included in the sample 
of \cite{pia11b} are the ones derived by \citet{pia05a}, which have been considerably underestimated, as we have 
shown in section 4. Consequently, the conclusions reached by \citet{pia11b} 
could substantially change if these authors used our ages for L4 and L6, which should be much more accurate. Similarly, some of the cluster ages included in our extended 
sample could also have been underestimated because the clusters were observed with small telescopes (especially 
those in \citealt{pia01} and \citealt{pia07b}). In these cases, the real MSTO $T_1$ magnitude may not have been clearly  
detected. Although our sample of 15 star clusters is not large enough to either confirm or deny the existence 
of the first burst of star formation at 5 Gyr and/or the constant slope for the first 
$\sim$ 4 Gyr, the photometry of our 15 clusters is deep enough to measure the MSTO with small errors. This fact, together 
with the homogeneous nature of our 15 cluster ages, supports the existence of the second peak of star formation 
at 2 Gyr suggested by \citet{pia11b} (crosshatched AD). It is clearly advisable to continue studying clusters in a homogeneous way, applying the same techniques. 
With this in mind, we have obtained $V$ and $I$ images and spectra in the 
CaT region of an additional sample of 15 clusters. Currently, we are in the process of deriving ages and 
metallicities following the same procedures used here and in Paper I. We are confident that such a homogeneous 
sample of 30 clusters will enable us to better understand the processes of cluster formation  
in the SMC.  \\

\subsection{Age Gradient}

To take into account the orientation of the SMC and the probable projection effects necessary to analyze 
the possible existence of an SMC gradient, we adopted the elliptic system defined in \citet{pia07a}. 
In this system, the semimajor axis $a$ is used instead of the cluster true galactocentric distance 
(see {\citealt{pia07a} and P09 for details on the calculation of $a$).

In Figure \ref{f:AG} we show how cluster ages vary as a function of $a$. Filled circles represent our 15 clusters 
which show no evidence of an age gradient. We also included in Figure \ref{f:AG} the additional 35 clusters studied by 
Piatti et al. (see the above description and figure caption for details). The extended sample also supports the nonexistence 
of an age gradient as has been previously suggested by other authors (see, e.g., \citealt{pia11c,gla08b}). Note that in Paper I we find no metallicity gradient, either. 
It is necessary to note, however, that the spatial distribution of age and metallicity are built based on the current positions of the clusters. 
To find out where they were born  it would be necessary to know their orbits, which at present is not possible.\\
 
The \citet{wei13} study of the large scale field star formation history of the Clouds finds that after the onset
of the recent strong increase in star formation some 3.5 Gyr ago in the SMC, subsequent star formation
was concentrated to the central regions of the galaxy and little or no field star formation occurred outside
this. Their sampling is small,  however, as they only have fields at a galactocentric radius of about 0.5 kpc
or 2.3 kpc, which corresponds to a $\sim$ 1 and 4.6$^{\circ}$. However, we clearly have a number of 
young clusters located beyond this radius in our diagram. Either the Weisz et al. sample is too small and
does not reveal real variations in the galaxy or the field and cluster star formation is uncoupled, which
is less likely. We believe the clusters are telling us more of the true star formation history and that recent
star formation is not solely concentrated to the central regions.\\

\section{Age Metallicity Relation}

The relationship between the age of a stellar population and overall metallicity, and its time variation,
is one of the best tools to study the chemical evolutionary history of a galaxy. Both the ages and metallicities 
of the tracer population need to be as accurately determined as possible. It is of vital importance for those 
parameters not only to be on the same scale but also to be determined in a homogeneous way.\\

Several investigations have aimed at deriving the SMC AMR using various stellar populations, such as star clusters, 
field  giants and planetary nebulae (see, e.g., \citealt{str85,dac91,ols96}). 

After these first attempts, DH98 obtained  what can be considered the first well constrained SMC AMR. 
DH98 used a sample of 6 clusters studied by them and others. Their metallicities were determined 
from CaT spectra and their ages from deep, ground-based CMDs. According to DH98, the metallicity variation with age in the SMC
is consistent with a simple closed box model. However, subsequent analyses seemed to be more 
in agreement with only a small increase
in the intermediate period followed by a faster evolution more recently. \\

A series of papers summarized in \citet{pia07c}, wherein cluster ages and abundances were derived from Washington
photometry, showed an excellent agreement with the bursting model of \citet{pag98} (hereafter PT98), even in the intermediate period ranging from 5 
to 10 Gyr. The PT98 model assumes an early burst, followed by no star formation from $\sim12 - 4$ Gyr ago, 
and then an ensuing burst of recent star formation, leading to a significantly enhanced subsequent enrichment. 
However, in Paper I, we showed that there is poor agreement between our CaT observations and the PT98 model for the 
intermediate period. \\

The recent investigation of \citet{pia11a} confirms our point of view that, although the PT98 model is a very 
good representation of the observations for ages smaller than 4 Gyr, this model fails to explain the AMR for 
larger ages. In fact, the 9 SMC clusters studied by \citet{pia11a} are older than 4 Gyr but cover a considerable 
metallicity range. More precisely, 8 of their 9 clusters are included within a very small age range ($\Delta t = 
$ 2.2 Gyr) but their metallicities vary from $-$0.7 to $-$1.3 dex, five of which have metallicity values 
considerably larger than those predicted by PT98. \\

In Paper I we also found that some of our clusters have metallicities in good agreement with the models of \citet{car05} 
and \citet{car08}, which do not assume any bursts but use instead both infall and outflow. Carrera et al. used 
the CaT to derive metallicities on the \citet{cg97} scale for a large number of field giants in regions spread across 
the SMC. In Paper I we concluded that the model of PT98 is a good representation of the AMR for ages $<$ 3-4 Gyr but the ideal 
model for larger ages would predict metallicities between PT98 and the \citet{car05} and \citet{car08} models. 

As we have already mentioned, in Paper I we analized the AMR of the SMC using our sample of 15 SMC clusters as 
tracers. The metallicities of these clusters were accurately determined following the CaT technique so that they 
are on a homogeneous scale. Unfortunately, at the time of Paper I, we could not determine 
ages for these objects from our own data, instead simply taking them from the literature. So the AMR of Paper I  is 
inhomogeneous regarding ages. The CaT method is among the most accurate procedures for the determination of 
cluster metallicities \citep{col04,gro06}, and the $\delta V$ parameter is an excellent age indicator (e.g. 
\citealt{cc94}) so the AMR derived in the current work can be considered a good indicator of the chemical 
evolution of the SMC and a significant improvement over Paper I given the now homogeneous ages. \\

In Figure \ref{f:age_met} we present our new AMR. Solid circles represent our 15 cluster sample considering ages derived in the
 present work and metallicities from Paper I. 

As in Paper I, we have also included in Figure \ref{f:age_met} six clusters having metallicities derived by other authors 
on the same scale than ours, namely three from DH98 (open circles) and other three from \citet{gla08b}, having their metallicities 
been derived by CaT method. We only included clusters which we do not have in common with other studies, using in these 
cases the metallicity values on the \citet{cg97} scale to match ours. We have also included NGC\,330, whose metallicity (-0.94 ± 0.2) 
was derived by  \citet{gon99} from high resolution spectroscopy. Squares in Figure \ref{f:age_met} represent the mean CaT metallicities 
calculated by \citet{car08} in different age bins for their sample of field stars. Different lines represent: (i) the model of closed 
box continuous star formation computed by DH98 (short-dashed line). (ii) the bursting model of PT98 (solid line), (iii)  
the best-fit model derived by \citet{car05}, long-dashed line), and (iv) the AMR obtained by \citet{har04}.

If we first consider only our 15 cluster sample, the main differences that we can find between the present AMR and the one derived in Paper I, are: 
1) ages of clusters younger than 4 Gyrs are now in better agreement with the model of PT98, 2) we have now 6 star clusters with ages larger than 4 Gyr 
so we have improved the representation of the intermediate/old SMC period. Four of these 6 clusters follow reasonably well both
DH98 and \citet{car05} models, but the other 2 intermediate/old clusters are better represented by the model of PT98, 3) we have found 3 new very old SMC clusters
and two of them, L\,4 and L\,6,  are right next to each other in the very outer reaches of the SMC, and 4) these three old clusters are candidates for filling the possible  ``SMC age gap''. Accurate ages and abundances for them now allows improved understanding of the AMR for the old period, particularly between L\,1 and NGC\,121. \\

Considering the full cluster sample, we corroborate that the model of PT98 is a good indicator of the chemical enrichment of the SMC for ages less that 4 Gyrs (except for the anomalous
cluster NGC\,330) but a more complicated model is needed to  account for the intermediate/old SMC period. The PT98 model underestimates metallicities during this period. It is clear that the chemical evolution was slow during this period and then experienced an increase during the last few Gyrs, as discussed extensively in Paper I. However, sample size is still small, especially for older clusters. Continued
searches for such new old clusters will be important to try and pin down the chemical evolution and
cluster formation history during this time, and see how real the possible SMC age gap really is. Accurate ages and metallicities for more clusters, on a homogeneous scale, are also of paramount importance to 
derive a definitive understanding of the chemical evolution of this important galaxy. 

\section{Summary and Conclusions}

We measured the parameter $\delta V$ in the $V$ vs. $V-I$ CMD of 15 SMC star clusters and we used the calibrations 
of \citet[CC94]{cc94} to derive  their ages. We found that three of them, L\,110, L\,4 and L\,6,
are newly discovered very old SMC clusters ( $t =$  7.6 $\pm$ 1.0, 7.9 $\pm$ 1.1 and 8.7 $\pm$ 1.2, respecively).  Consequently,  we have enlarged considerably the number 
of known old clusters in this galaxy. We corroborate the old nature of these new old clusters, from two independent methods. First, we derived a new calibration 
(SMC $\delta$ calibration), using the SMC clusters studied by \citet{gla08a,gla08b} with deep HST photometry. Secondly, we calibrated our VLT photometry with independent observations
and we fitted different sets of isochrones to the calibrated CMD of these old clusters. 

Finally, using ages derived from the combination of $\delta V$ and CC94's calibration, and metallicities from \citet{par09} we analyzed the age distribution (AD), 
age gradient and the age metallicity relation of the SMC. Our main results are: 

\begin{itemize}

\item We combined our 15 SMC cluster with 35 clusters studied by other authors. The AD of this
combined sample of 50 clusters reproduces the two peaks found in \citet{pia11b} at $\sim$ 2 and $\sim$ 5 Gyr (corresponding to possible episodes of cluster
formation) but it does not show a constant slope in the AD  for the first  $\sim$ 4 Gyr. The peak at 5 Gyr is
in reasonable agreement with the peak in the field star formation rate at about 4.5 Gyr found recently by the 
HST study of \citet{wei13} but their second peak at 9 Gyr is not found in our data.

\item  The extended sample supports the idea of the nonexistence of an age gradient as has been previously suggested by other authors (see, e.g., \citealt{pia11c,gla08b}). In particular, we find clusters
younger than 3.5 Gyrs at a large range of galactocentric distances and not just near the center, as suggested by the field star data of \citet{wei13} 

\item Combining our cluster sample with those observed by DH98 and Glatt et al. (also with CaT) and the one cluster with high resolution metallicity, we compiled a 
sample of 25 clusters on a homogeneous metallicity scale with relatively small errors and study the cluster AMR. We corroborate our Paper I result that for ages smaller than $\sim$ 4 Gyr, 
the chemical enrichment of the SMC is well represented by the PT98 model but for explaining the intermediate/old SMC period a more complicated model is needed. 
We now have seven clusters older than 6 Gyr  homogeneously studied so we have doubled the representation
of that period.

\item According to the calibration of CC94, our three new old clusters lie between L\,1 and NGC\,121, which suggest that the suggested SMC cluster age gap claimed by \citet{gla08a} is probably 
not real.

\end{itemize}
  
\acknowledgments

This work is based on observations collected at the European Southern Observatory,
Chile, under program number 076.B-0533. We would like to thank the
Paranal Science Operations Staff.  Specially we would like to thank the referee for the valuable comments made on the manuscript.
M.C.P. acknowledges Dr. Kenneth Janes for his orientation about age determination. 
M.C.P. and J.J.C. gratefully acknowledge financial support from the Argentinian
institutions CONICET and SECYT (Universidad Nacional de C\'ordoba). D.G. gratefully acknowledges support from the Chilean 
BASAL   Centro de Excelencia en Astrof{\'\i}sica
y Tecnolog{\'\i}as Afines (CATA) grant PFB-06/2007.

{\it Facilities:} \facility{VLT: Antu (FORS2)}.

\begin{deluxetable}{lcccc}
\tablewidth{0pt}
\tablecaption{Cluster sample}
\tablehead{
\colhead{Cluster}                & \colhead{RA (J2000.0)}  &
\colhead{Dec (J2000.0)}          & \colhead{[Fe/H]} &
\colhead{$a$}        \\
                                 & \colhead{($h$ $m$ $s$)} &
\colhead{($^{\circ}$ $'$ $''$)}  & \colhead{dex} &
\colhead{($^{\circ}$)}           }
\startdata
BS\,121 = SMC\,OGLE\,237                             & 01 04 22 & -72 50 52 & -0.66 $\pm$ 0.07 & 1.496\\
HW\,47                                               & 01 04 04 & -74 37 09 & -0.92 $\pm$ 0.04 & 3.502\\
HW\,84                                               & 01 41 28 & -71 09 58 & -0.91 $\pm$ 0.05 & 5.513\\
HW\,86                                               & 01 42 22 & -74 10 24 & -0.61 $\pm$ 0.06 & 7.345\\
L\,4 = K\,1, ESO\,28-15                              & 00 21 27 & -73 44 55 & -1.08 $\pm$ 0.04 & 3.265\\
L\,5 = ESO\,28-16                                    & 00 22 40 & -75 04 29 & -1.25 $\pm$ 0.05 & 3.092\\
L\,6 = K\,4, ESO\,28-17                              & 00 23 04 & -73 40 11 & -1.24 $\pm$ 0.03 & 3.124\\
L\,7 = K\,5, ESO\,28-18                              & 00 24 43 & -73 45 18 & -0.76 $\pm$ 0.06 & 2.888\\
L\,17 = K\,13, ESO\,29-1                             & 00 35 42 & -73 35 51 & -0.84 $\pm$ 0.03 & 1.718\\
L\,19 = SMC\,OGLE 3                                  & 00 37 42 & -73 54 30 & -0.87 $\pm$ 0.03 & 1.564\\
L\,27 = K\,21, SMC\,OGLE 12                          & 00 41 24 & -72 53 27 & -1.14 $\pm$ 0.06 & 1.392\\
L\,106 = ESO\,29-44                                  & 01 30 38 & -76 03 16 & -0.88 $\pm$ 0.06 & 7.877\\
L\,108                                               & 01 31 32 & -71 57 10 & -1.05 $\pm$ 0.05 & 4.460\\
L\,110 = ESO\,29-48                                  & 01 34 26 & -72 52 28 & -1.03 $\pm$ 0.05 & 5.323\\
L\,111 = NGC\,643, ESO\,29-50                        & 01 35 00 & -75 33 24 & -0.82 $\pm$ 0.03 & 7.830\\
\enddata
\label{t:sample}
\end{deluxetable}

\begin{deluxetable}{lcccccc}
\tablewidth{0pt}
\tablecaption{Cluster ages}
\tablehead{
\colhead{Cluster}                & \colhead{$\delta V$} &
\colhead{$t_S$}                  & \colhead{$t_C$}	& 
\colhead{$t_L$}   		 & \colhead{$t_L$}      &
\colhead{Contamination}                                 \\
                                 & \colhead{mag}        &
\colhead{Gyrs}                   & \colhead{Gyrs}       & 
\colhead{Gyrs}                   & \colhead{Reference}  & 
\colhead{index} }
\startdata
BS\,121  & 1.89 $\pm$ 0.30 & 3.5   $\pm$ 0.5   & 2.8  $\pm$ 0.5 & 2.30  & 1 & 0.86 \\
HW\,47   & 2.07 $\pm$ 0.30 & 4.3   $\pm$ 0.7   & 3.3  $\pm$ 0.5  & 2.80  & 1 & 0.59 \\
HW\,84   & 1.39 $\pm$ 0.30 & 2.1   $\pm$ 0.3   & 1.6  $\pm$ 0.2  & 2.40  & 1 & 0.35 \\
HW\,86   & 1.23 $\pm$ 0.30 & 1.7   $\pm$ 0.2   & 1.4  $\pm$ 0.2  & 1.60  & 1 & 0.40 \\
L\,4     & 2.91 $\pm$ 0.30 & 12.3  $\pm$ 2.1   & 7.9  $\pm$ 1.1  & 3.10  & 1 & 0.21 \\ 
L\,5     & 2.17 $\pm$ 0.30 & 4.9   $\pm$ 0.8   & 3.7  $\pm$ 0.5  & 4.10  & 1 & 0.30 \\ 
L\,6     & 3.00 $\pm$ 0.30 & 13.8  $\pm$ 2.4   & 8.7  $\pm$ 1.2  & 3.30  & 1 & 0.32 \\ 
L\,7     & 1.45 $\pm$ 0.30 & 2.2   $\pm$ 0.3   & 1.6  $\pm$ 0.2  & 2.00  & 1 & 0.29 \\ 
L\,17    & 2.34 $\pm$ 0.30 & 5.9   $\pm$ 0.9   & 4.4  $\pm$ 0.6  & 1.26  & 2 & 0.54 \\ 
L\,19    & 2.43 $\pm$ 0.30 & 6.6   $\pm$ 1.1   & 4.8  $\pm$ 0.7  & 2.10  & 1 & 0.50 \\ 
L\,27    & 2.39 $\pm$ 0.30 & 6.3   $\pm$ 1.0   & 4.6  $\pm$ 0.6  & 2.10  & 1 & 0.52 \\ 
L\,106   & 1.58 $\pm$ 0.30 & 2.5   $\pm$ 0.4   & 2.0  $\pm$ 0.3  & 0.89  & 3 & 0.13 \\ 
L\,108   & 1.93 $\pm$ 0.30 & 3.7   $\pm$ 0.6   & 2. 9 $\pm$ 0.4  & 0.89  & 3 & 0.17 \\ 
L\,110   & 2.87 $\pm$ 0.30 & 11.6  $\pm$ 1.9   & 7.6  $\pm$ 1.0  & 6.40  & 4 & 0.09 \\ 
L\,111   & 1.57 $\pm$ 0.30 & 2.5   $\pm$ 0.4   & 2.0  $\pm$ 0.3  & 1.00  & 3 & 0.06 \\ 
\enddata
\tablerefs{
(1) \citealt{pia05a}; (2) \citealt{rz05}\\ 
(3) \citealt{pia07c} (4) \citealt{pia07b}}.
\label{t:ages}
\end{deluxetable}

\begin{deluxetable}{lcccc}
\tablewidth{0pt}
\tablecaption{Washington Cluster sample}
\tablehead{
\colhead{Cluster}                & \colhead{$\delta T_1$}  &
\colhead{Ref.}                   & \colhead{$\delta V$}    &
\colhead{$t_C$}                                             }
\startdata
K\,28    & 1.70 $\pm$  0.30 &   1 &  1.57 $\pm$  0.30 &  2.0  $\pm$ 0.3 \\
K\,44    & 2.10 $\pm$  0.20 &   1 &  1.97 $\pm$  0.20 &  3.0  $\pm$ 0.3 \\
L\,116   & 2.00 $\pm$  0.40 &   1 &  1.87 $\pm$  0.40 &  2.7  $\pm$ 0.5 \\
L\,32    & 2.50 $\pm$  0.10 &   1 &  2.37 $\pm$  0.10 &  4.6  $\pm$ 0.2 \\
L\,38    & 2.70 $\pm$  0.10 &   1 &  2.58 $\pm$  0.10 &  5.6  $\pm$ 0.3 \\
L\,112   & 2.60 $\pm$  0.15 &   2 &  2.47 $\pm$  0.15 &  5.1  $\pm$ 0.3 \\
L\,113   & 2.35 $\pm$  0.15 &   2 &  2.22 $\pm$  0.15 &  3.9  $\pm$ 0.3 \\
AM\,3    & 2.70 $\pm$  0.15 &   3 &  2.58 $\pm$  0.15 &  5.6  $\pm$ 0.4 \\
HW\,31   & 2.50 $\pm$  0.15 &   3 &  2.37 $\pm$  0.15 &  4.6  $\pm$ 0.3 \\
HW\,40   & 2.60 $\pm$  0.15 &   3 &  2.47 $\pm$  0.15 &  5.1  $\pm$ 0.3 \\
HW\,41   & 2.70 $\pm$  0.15 &   3 &  2.58 $\pm$  0.15 &  5.6  $\pm$ 0.4 \\
HW\,42   & 3.10 $\pm$  0.15 &   3 &  2.98 $\pm$  0.15 &  8.5  $\pm$ 0.6 \\
HW\,59   & 2.80 $\pm$  0.15 &   3 &  2.68 $\pm$  0.15 &  6.2  $\pm$ 0.4 \\
HW\,63   & 2.60 $\pm$  0.15 &   3 &  2.47 $\pm$  0.15 &  5.1  $\pm$ 0.3 \\
L\,91    & 2.40 $\pm$  0.15 &   3 &  2.27 $\pm$  0.15 &  4.1  $\pm$ 0.3 \\
NGC\,339 & 2.60 $\pm$  0.15 &   3 &  2.47 $\pm$  0.15 &  5.1  $\pm$ 0.3 \\
L\,3     & 0.80 $\pm$  0.25 &   4 &  0.66 $\pm$  0.25 &  0.8  $\pm$ 0.1 \\
L\,28    & 0.75 $\pm$  0.25 &   4 &  0.61 $\pm$  0.25 &  0.7  $\pm$ 0.1 \\
HW\,66   & 2.20 $\pm$  0.25 &   4 &  2.08 $\pm$  0.25 &  3.4  $\pm$ 0.4 \\
L\,100   & 1.80 $\pm$  0.25 &   4 &  1.67 $\pm$  0.25 &  2.2  $\pm$ 0.3 \\
HW\,79   & 2.40 $\pm$  0.25 &   4 &  2.27 $\pm$  0.25 &  4.1  $\pm$ 0.5 \\
L\,109   & 2.20 $\pm$  0.25 &   4 &  2.07 $\pm$  0.25 &  3.3  $\pm$ 0.4 \\
HW\,85   & 1.75 $\pm$  0.25 &   4 &  1.62 $\pm$  0.25 &  2.1  $\pm$ 0.2 \\
B\,34    & 1.20 $\pm$  0.15 &   5 &  1.06 $\pm$  0.15 &  1.2  $\pm$ 0.1 \\
B\,39    & 1.20 $\pm$  0.20 &   5 &  1.06 $\pm$  0.20 &  1.2  $\pm$ 0.1 \\
B\,47    & 1.10 $\pm$  0.30 &   5 &  0.96 $\pm$  0.30 &  1.1  $\pm$ 0.1 \\
B\,112   & 1.30 $\pm$  0.15 &   5 &  1.16 $\pm$  0.15 &  1.3  $\pm$ 0.1 \\
BS\,88   & 1.60 $\pm$  0.15 &   5 &  1.46 $\pm$  0.15 &  1.8  $\pm$ 0.1 \\
HW\,22   & 2.70 $\pm$  0.20 &   5 &  2.58 $\pm$  0.20 &  5.6  $\pm$ 0.5 \\
HW\,55   & 1.90 $\pm$  0.30 &   5 &  1.77 $\pm$  0.30 &  2.4  $\pm$ 0.3 \\
HW\,67   & 2.00 $\pm$  0.20 &   5 &  1.87 $\pm$  0.20 &  2.7  $\pm$ 0.3 \\
K\,38    & 2.10 $\pm$  0.35 &   5 &  1.97 $\pm$  0.35 &  3.0  $\pm$ 0.4 \\
L\,58    & 1.60 $\pm$  0.20 &   5 &  1.46 $\pm$  0.20 &  1.8  $\pm$ 0.2 \\
NGC\,419 & 1.10 $\pm$  0.25 &   5 &  0.96 $\pm$  0.25 &  1.1  $\pm$ 0.1 \\
\enddata
\tablerefs{
(1) \citealt{pia01}; (2) \citealt{pia07b};
(3) \citealt{pia11a}; (4) \citet{pia11b}, (5) \citealt{pia11c}.}

\label{t:piatti}
\end{deluxetable}

\newpage

\begin{figure}
\begin{centering}
\includegraphics[width=15.cm]{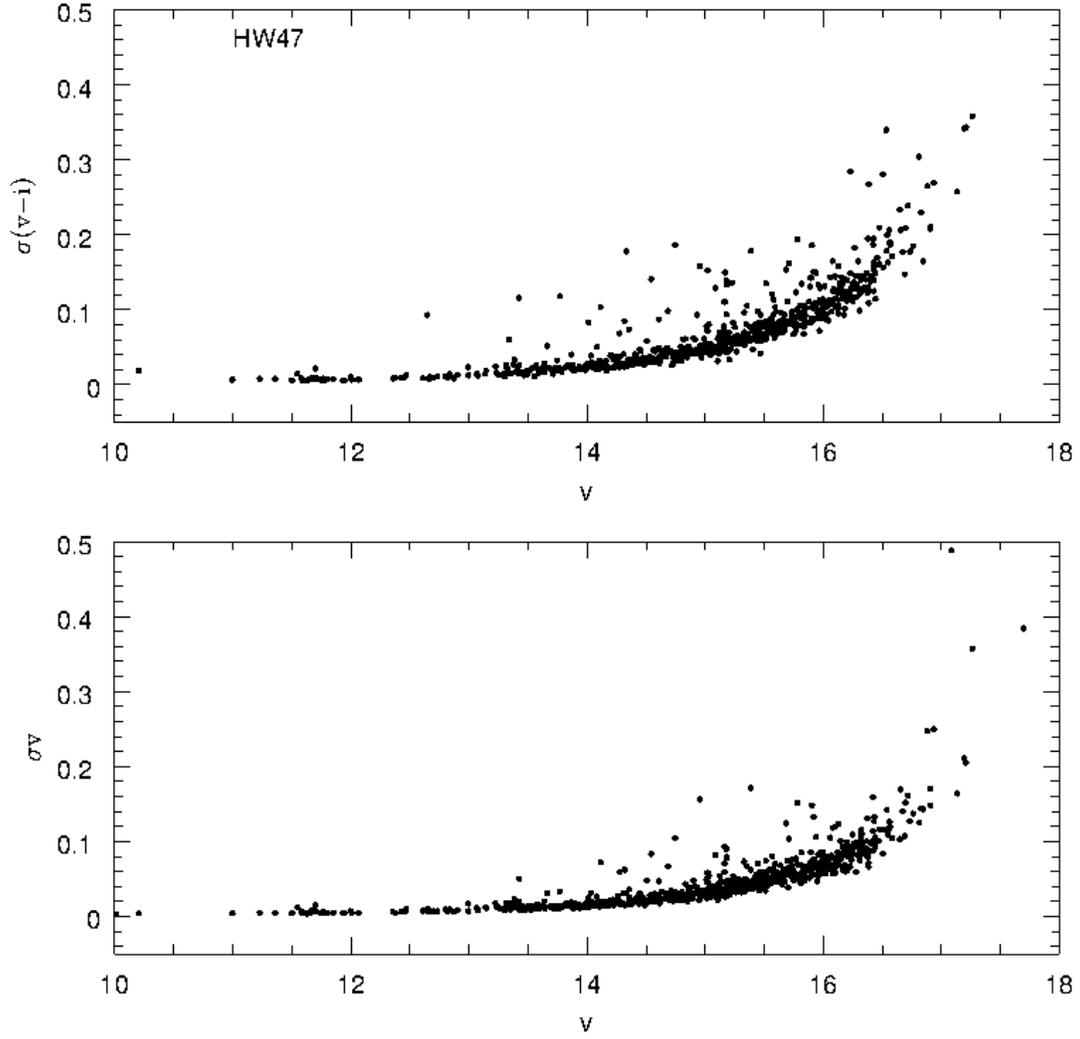}
\caption{\small{ Photometric errors in $v$ and $v-i$ as a function of instrumental $v$ magnitude for cluster HW\,47.
}}
\label{f:err}
\end{centering}
\end{figure}

\begin{figure}
\begin{centering}
\includegraphics[width=15.cm]{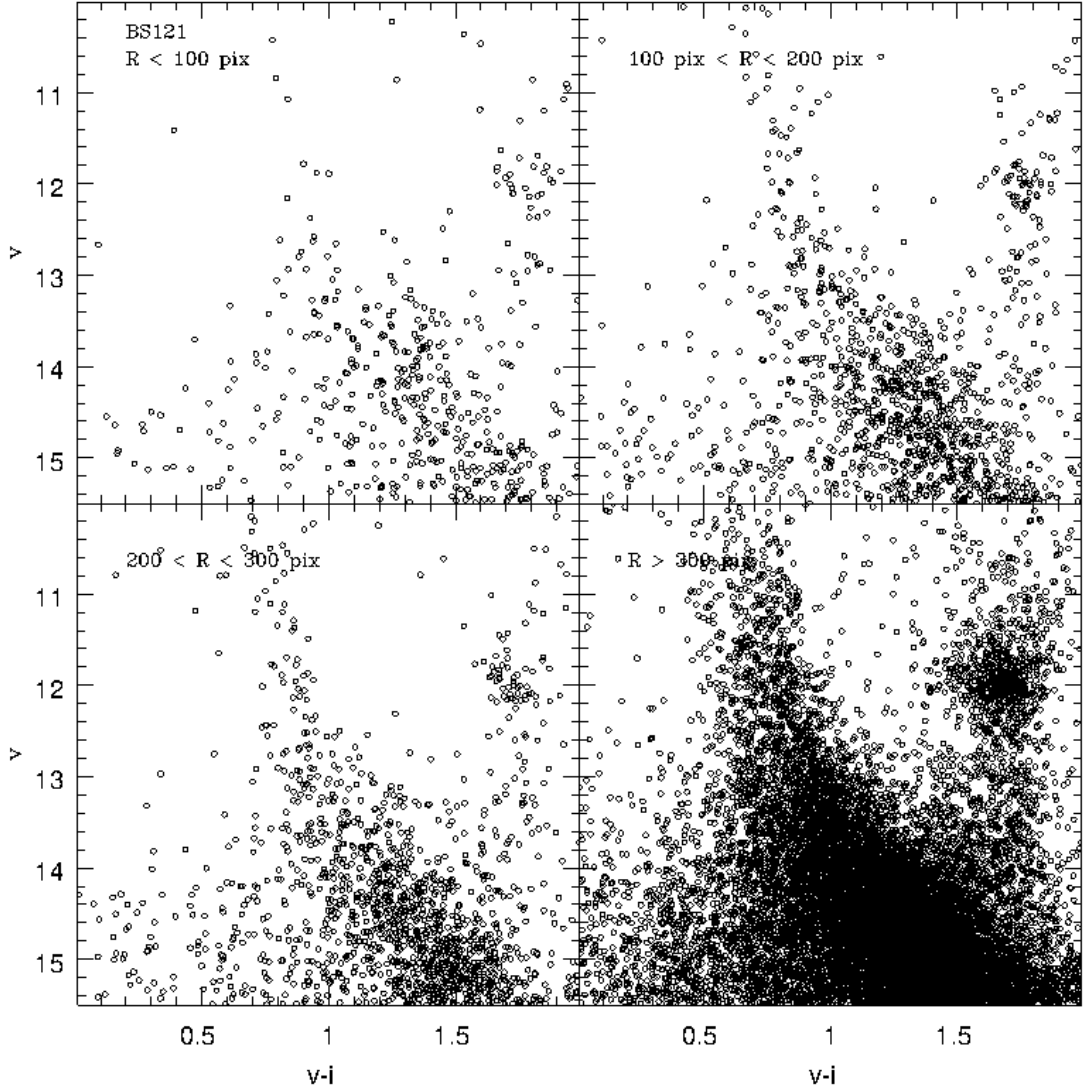}
\caption{\small{Instrumental color magnitude diagram for cluster BS\,121, built from the DAOPHOT output. Each panel corresponds to a radial extraction.
The selected radial intervals for each extraction are shown on the plot.
}}
\label{f:bs121M.bin}
\end{centering}
\end{figure}

\begin{figure}
\begin{centering}
\includegraphics[width=15.cm]{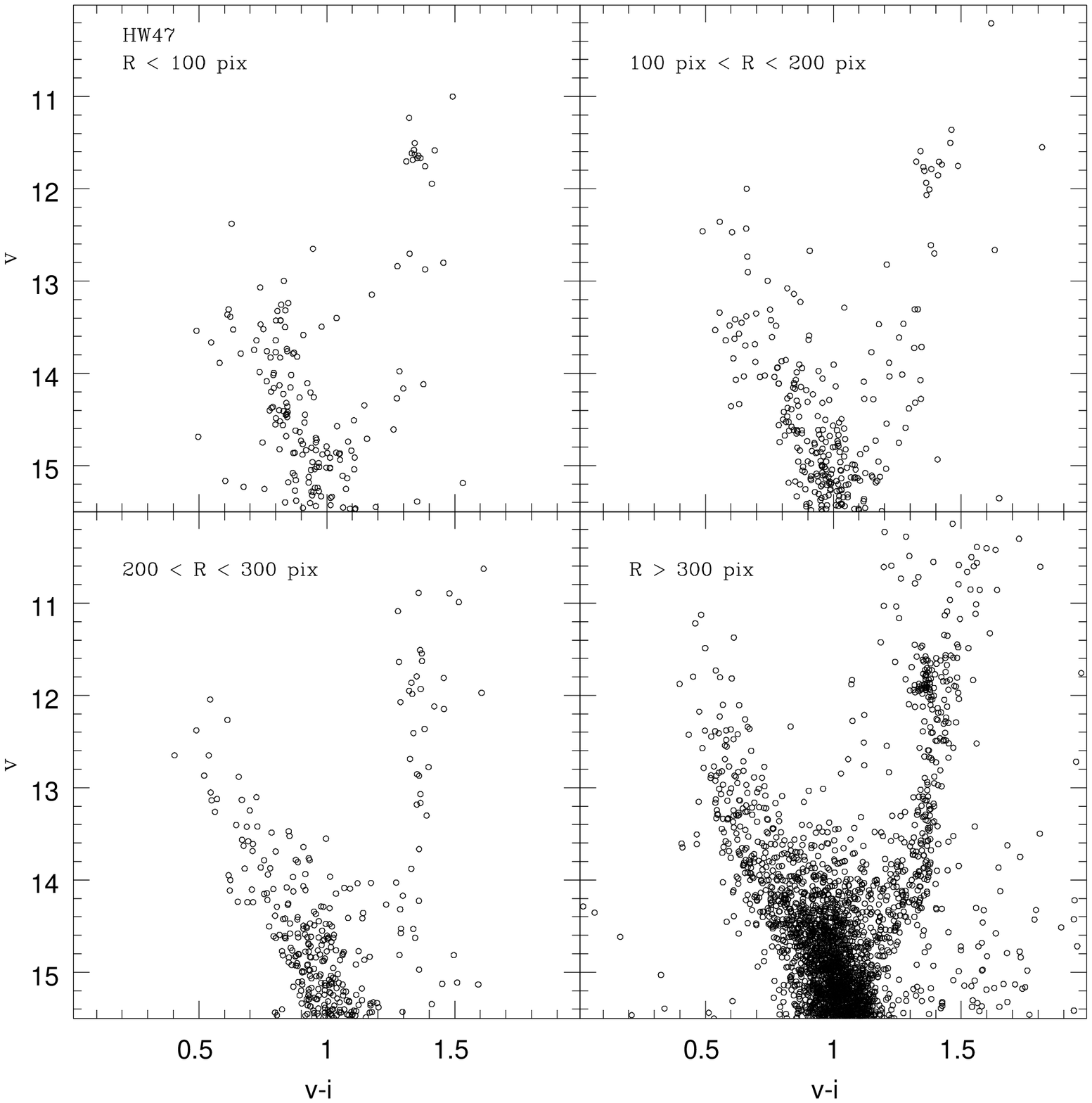}
\caption{\small{Idem Figure \ref{f:bs121M.bin} but for cluster HW\,47.
}}
\label{f:hw47M.bin}
\end{centering}
\end{figure}

\begin{figure}
\begin{centering}
\includegraphics[width=15.cm]{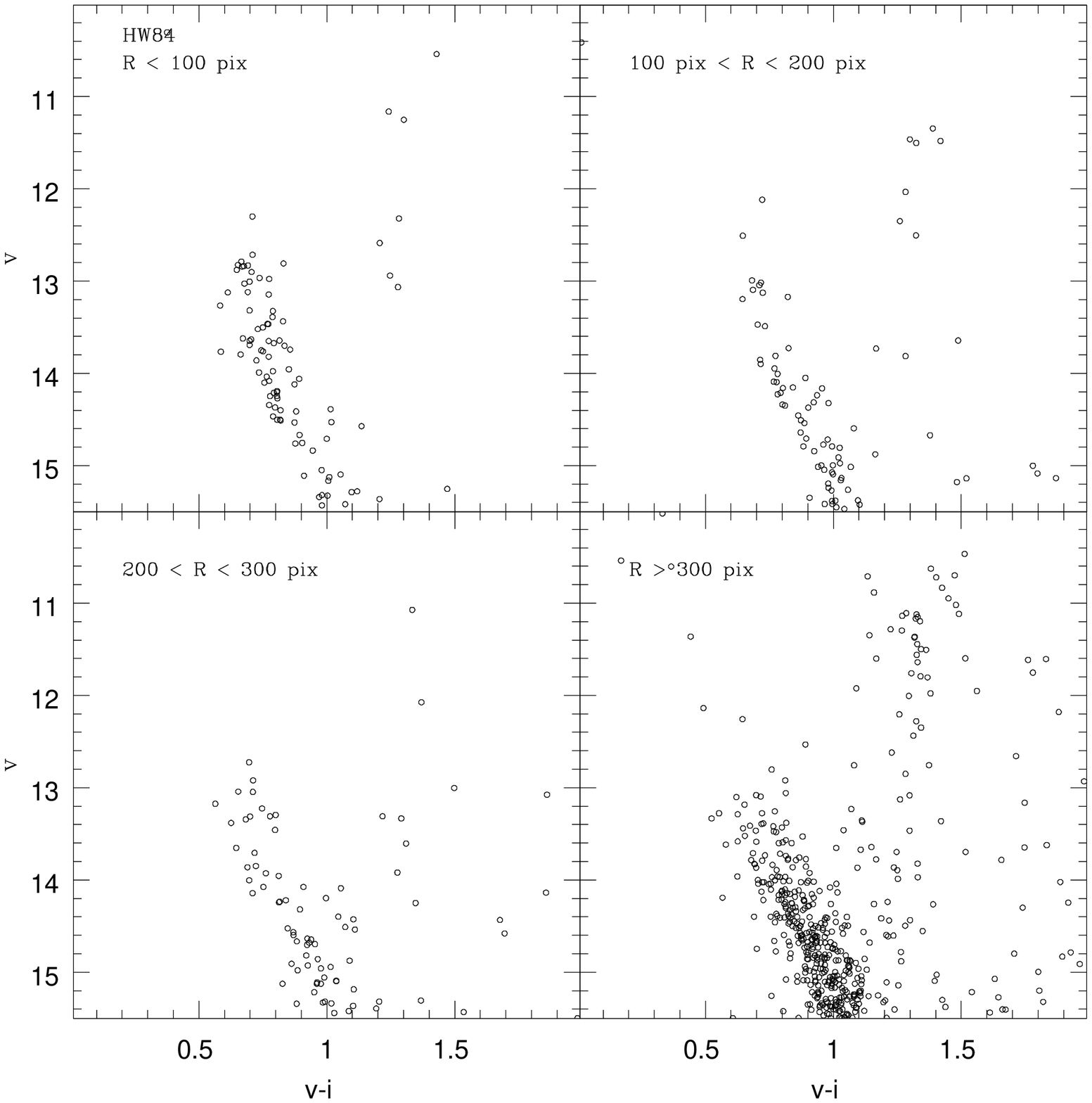}
\caption{\small{Idem Figure \ref{f:bs121M.bin} but for cluster HW\,84.
}}
\label{f:hw84M.bin}
\end{centering}
\end{figure}

\begin{figure}
\begin{centering}
\includegraphics[width=15.cm]{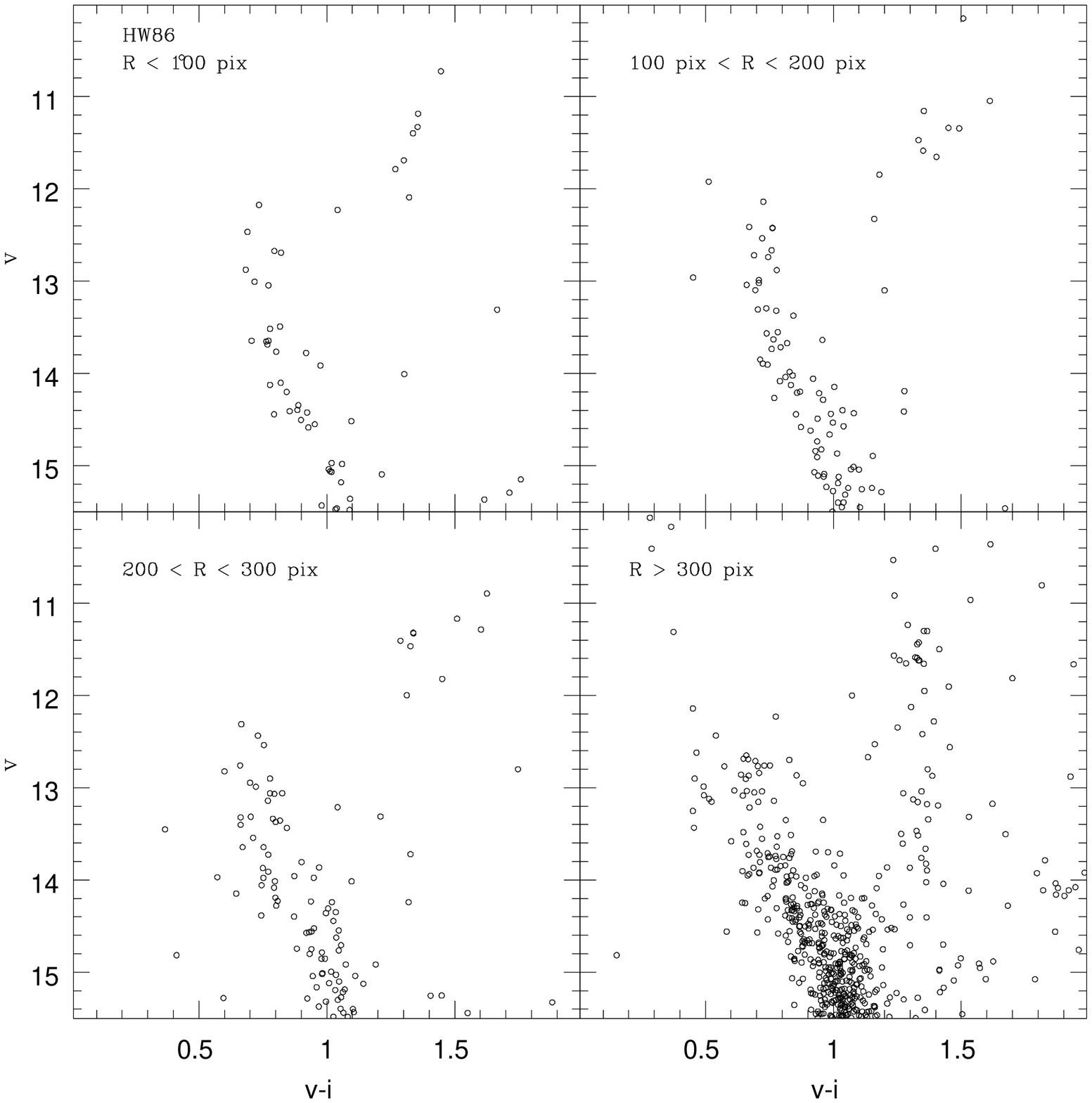}
\caption{\small{Idem Figure \ref{f:bs121M.bin} but for  cluster HW\,86.
}}
\label{f:hw86M.bin}
\end{centering}
\end{figure}

\begin{figure}
\begin{centering}
\includegraphics[width=15.cm]{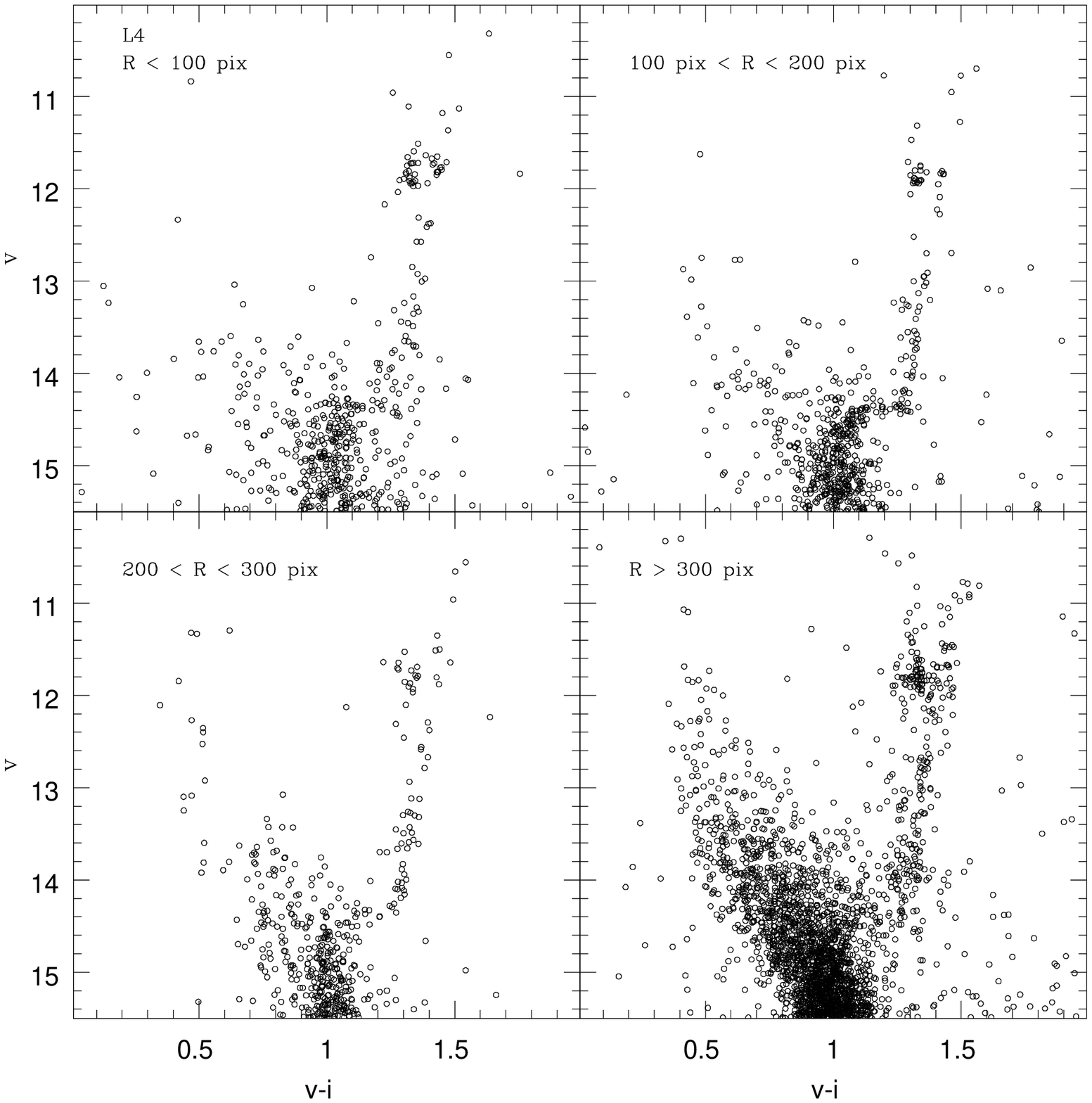}
\caption{\small{Idem Figure \ref{f:bs121M.bin} but for cluster L\,4.
}}
\label{f:l4M.bin}
\end{centering}
\end{figure}

\begin{figure}
\begin{centering}
\includegraphics[width=15.cm]{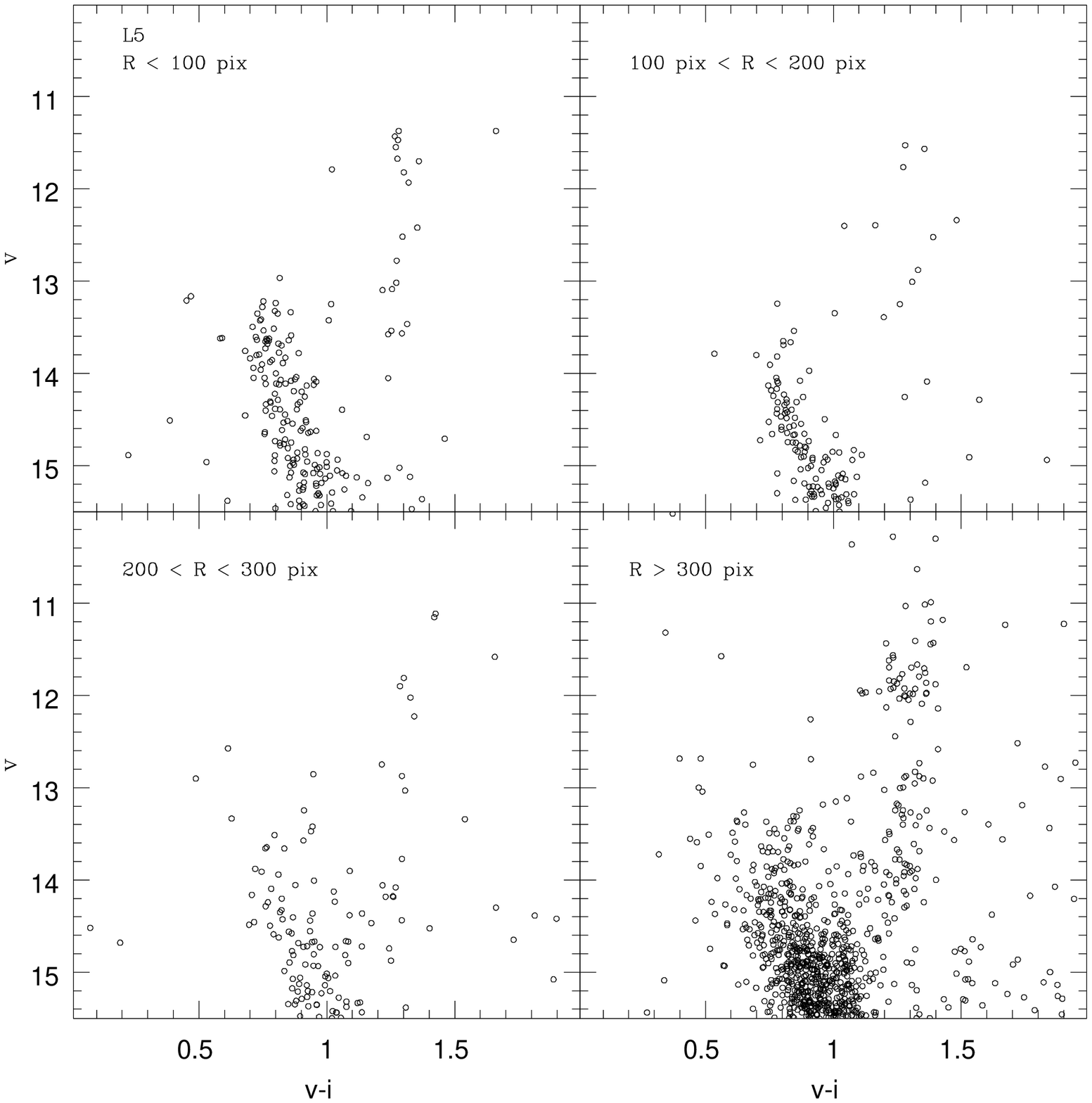}
\caption{\small{Idem Figure \ref{f:bs121M.bin} but for cluster L\,5.
}}
\label{f:l5M.bin}
\end{centering}
\end{figure}

\begin{figure}
\begin{centering}
\includegraphics[width=15.cm]{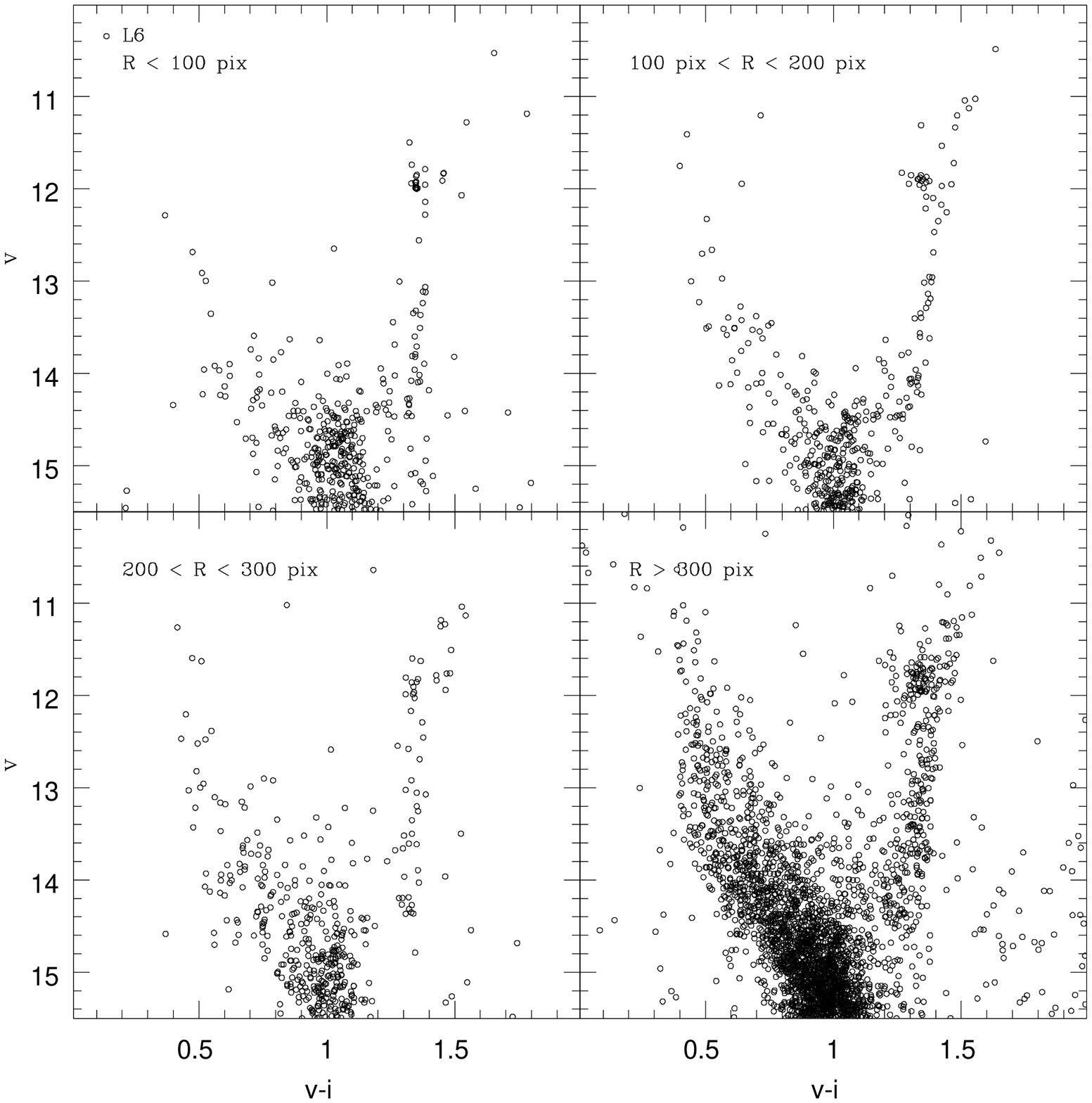}
\caption{\small{Idem Figure \ref{f:bs121M.bin} but for cluster L\,6.
}}
\label{f:l6M.bin}
\end{centering}
\end{figure}

\begin{figure}
\begin{centering}
\includegraphics[width=15.cm]{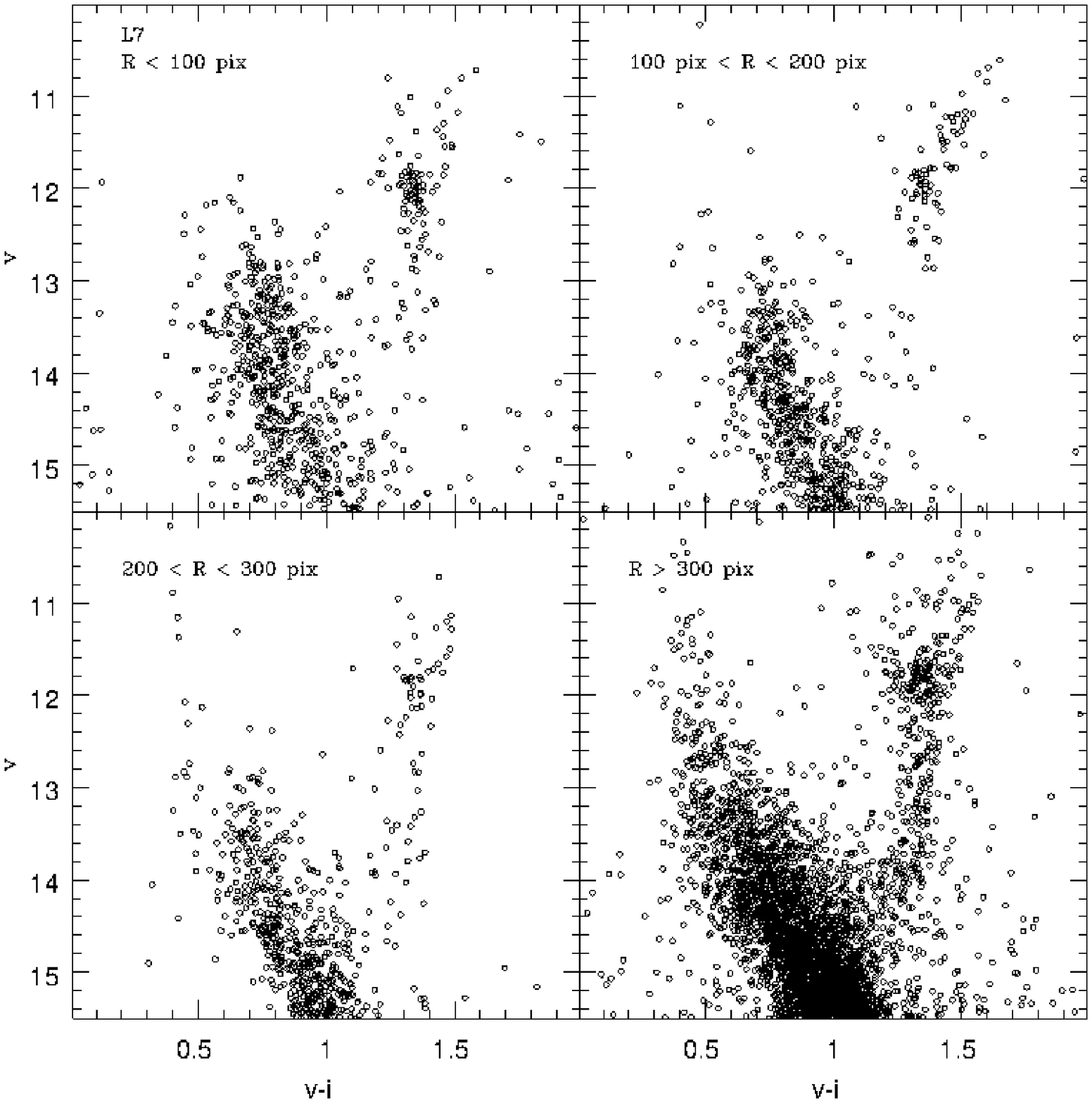}
\caption{\small{Idem Figure \ref{f:bs121M.bin} but for cluster L\,7.
}}
\label{f:l7M.bin}
\end{centering}
\end{figure}

\begin{figure}
\begin{centering}
\includegraphics[width=15.cm]{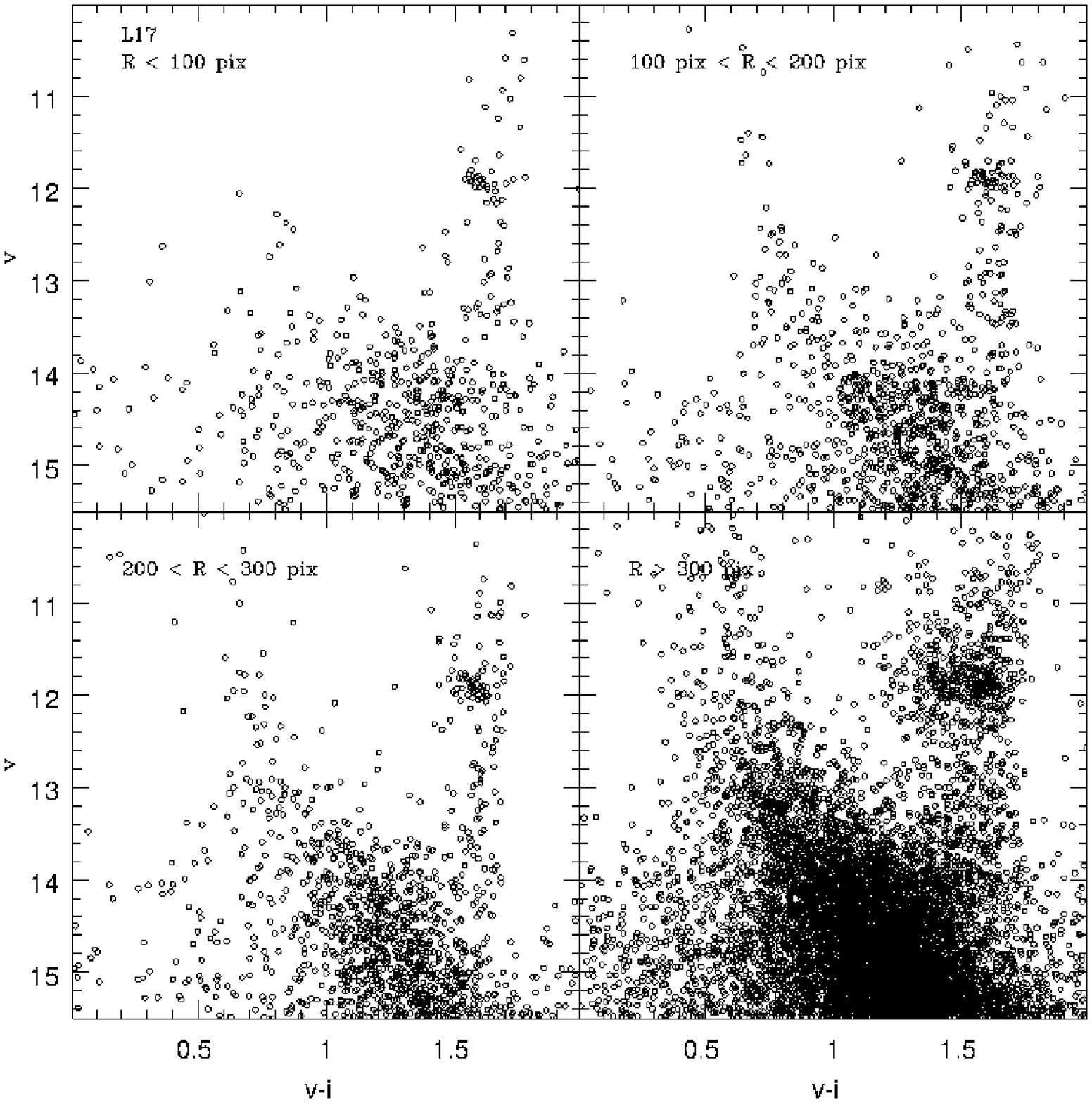}
\caption{\small{Idem Figure \ref{f:bs121M.bin} but for cluster L\,17.
}}
\label{f:l17M.bin}
\end{centering}
\end{figure}

\begin{figure}
\begin{centering}
\includegraphics[width=15.cm]{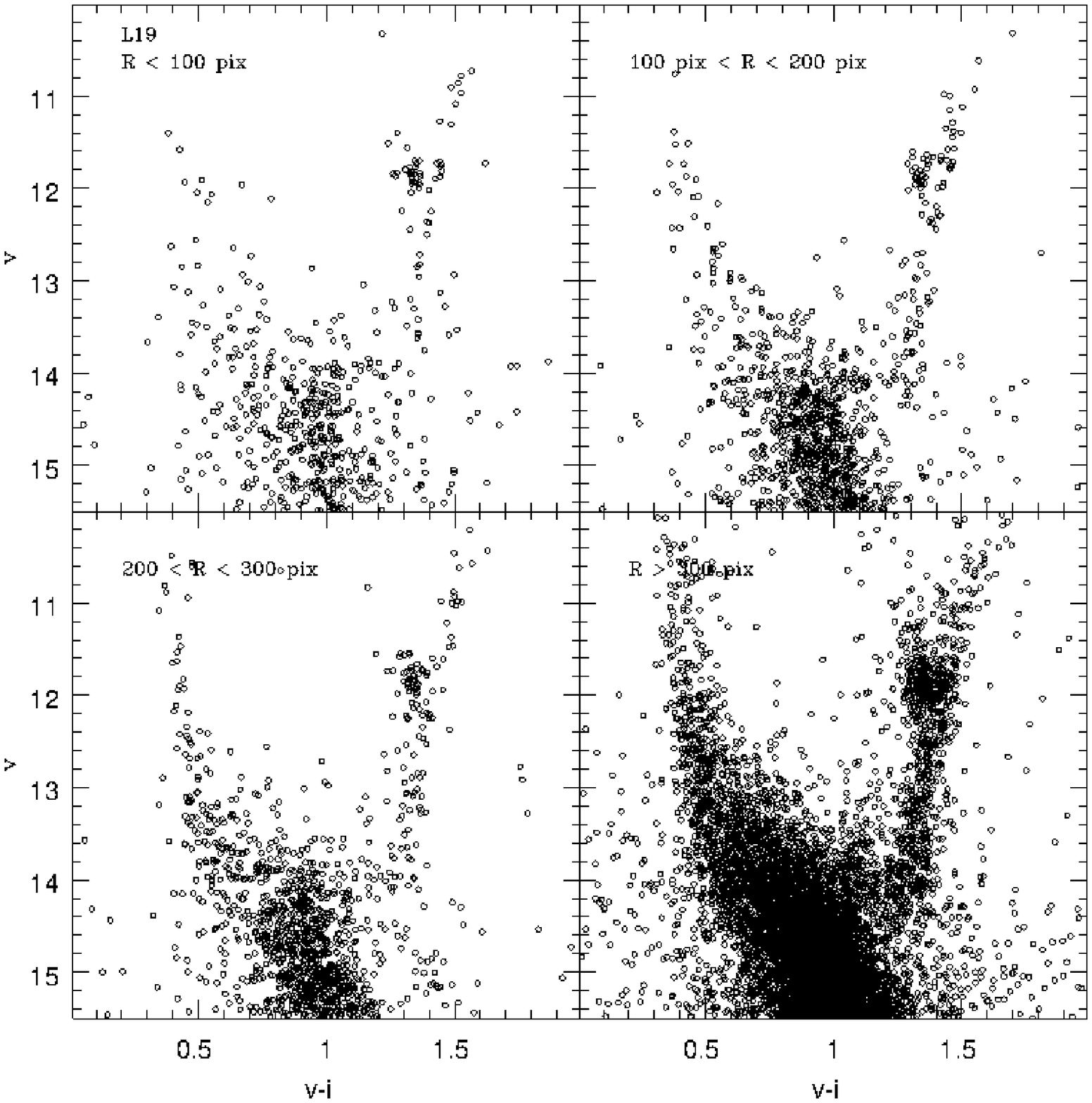}
\caption{\small{Idem Figure \ref{f:bs121M.bin} but for cluster L\,19.
}}
\label{f:l19M.bin}
\end{centering}
\end{figure}

\begin{figure}
\begin{centering}
\includegraphics[width=15.cm]{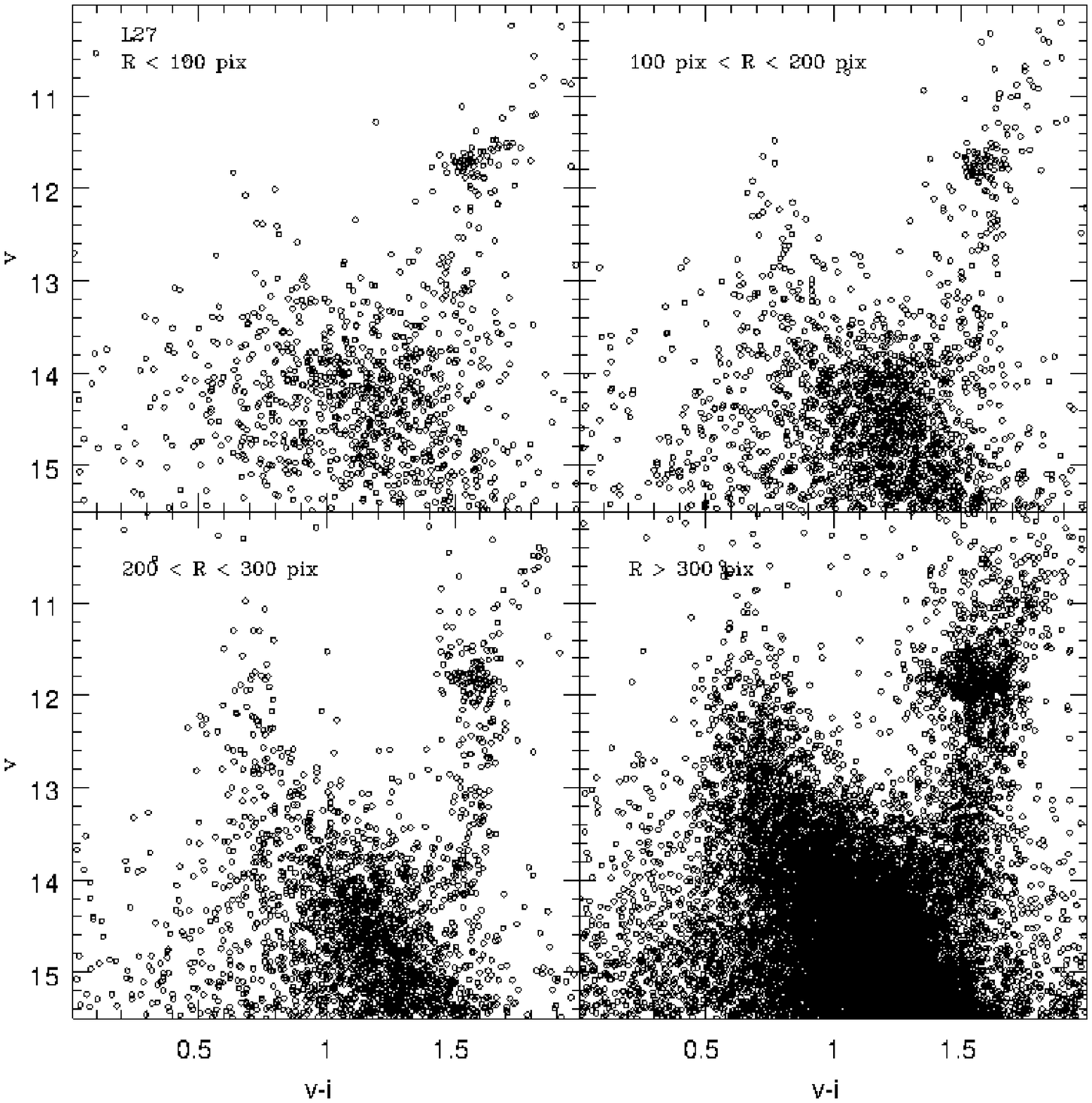}
\caption{\small{Idem Figure \ref{f:bs121M.bin} but for cluster L\,27.
}}
\label{f:l27M.bin}
\end{centering}
\end{figure}

\begin{figure}
\begin{centering}
\includegraphics[width=15.cm]{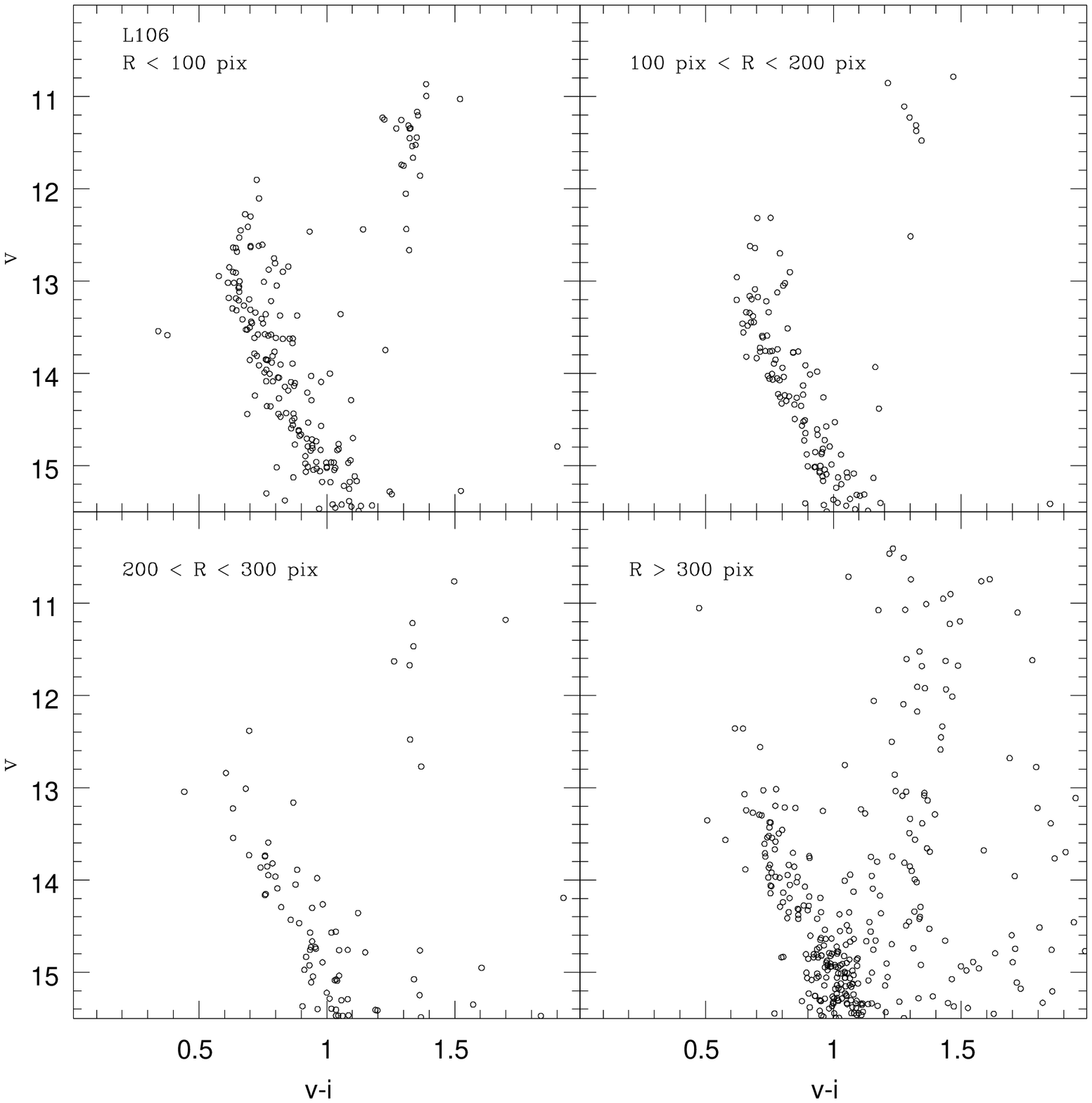}
\caption{\small{Idem Figure \ref{f:bs121M.bin} but for cluster L\,106.
}}
\label{f:l106M.bin}
\end{centering}
\end{figure}

\begin{figure}
\begin{centering}
\includegraphics[width=15.cm]{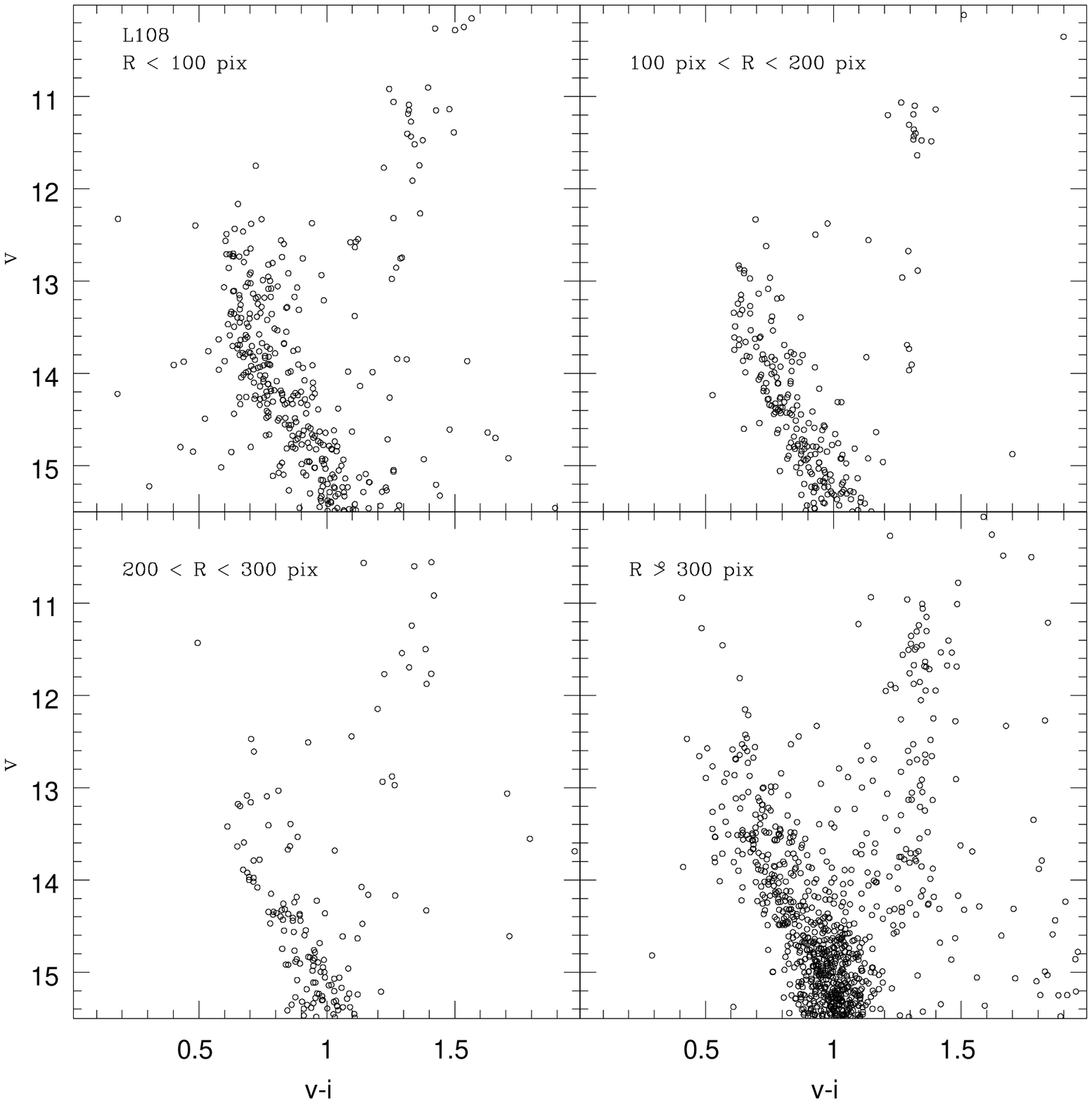}
\caption{\small{Idem Figure \ref{f:bs121M.bin} but for cluster L\,108.
}}
\label{f:l108M.bin}
\end{centering}
\end{figure}

\begin{figure}
\begin{centering}
\includegraphics[width=15.cm]{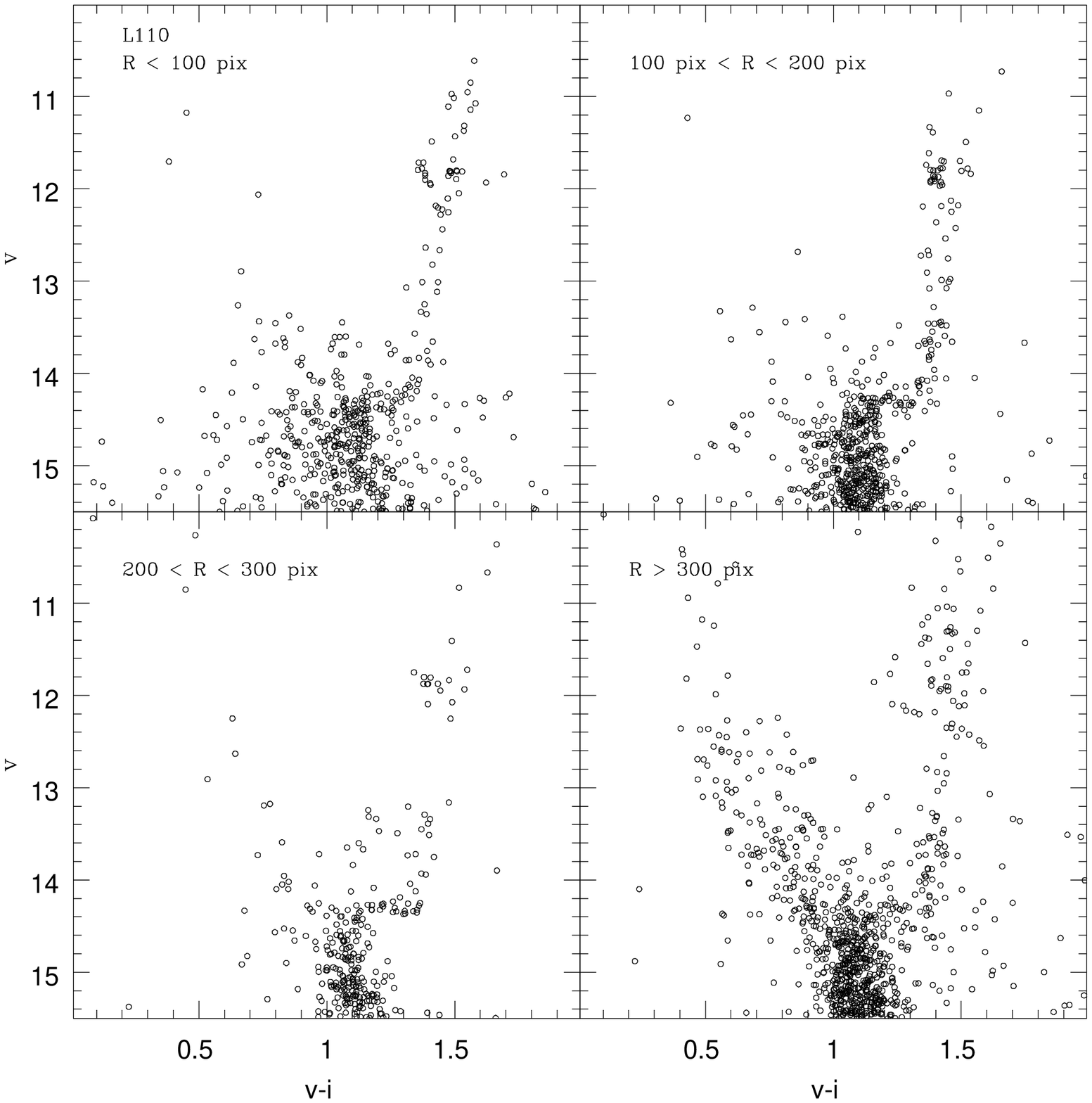}
\caption{\small{Idem Figure \ref{f:bs121M.bin} but for cluster L\,110.
}}
\label{f:l110M.bin}
\end{centering}
\end{figure}

\begin{figure}
\begin{centering}
\includegraphics[width=15.cm]{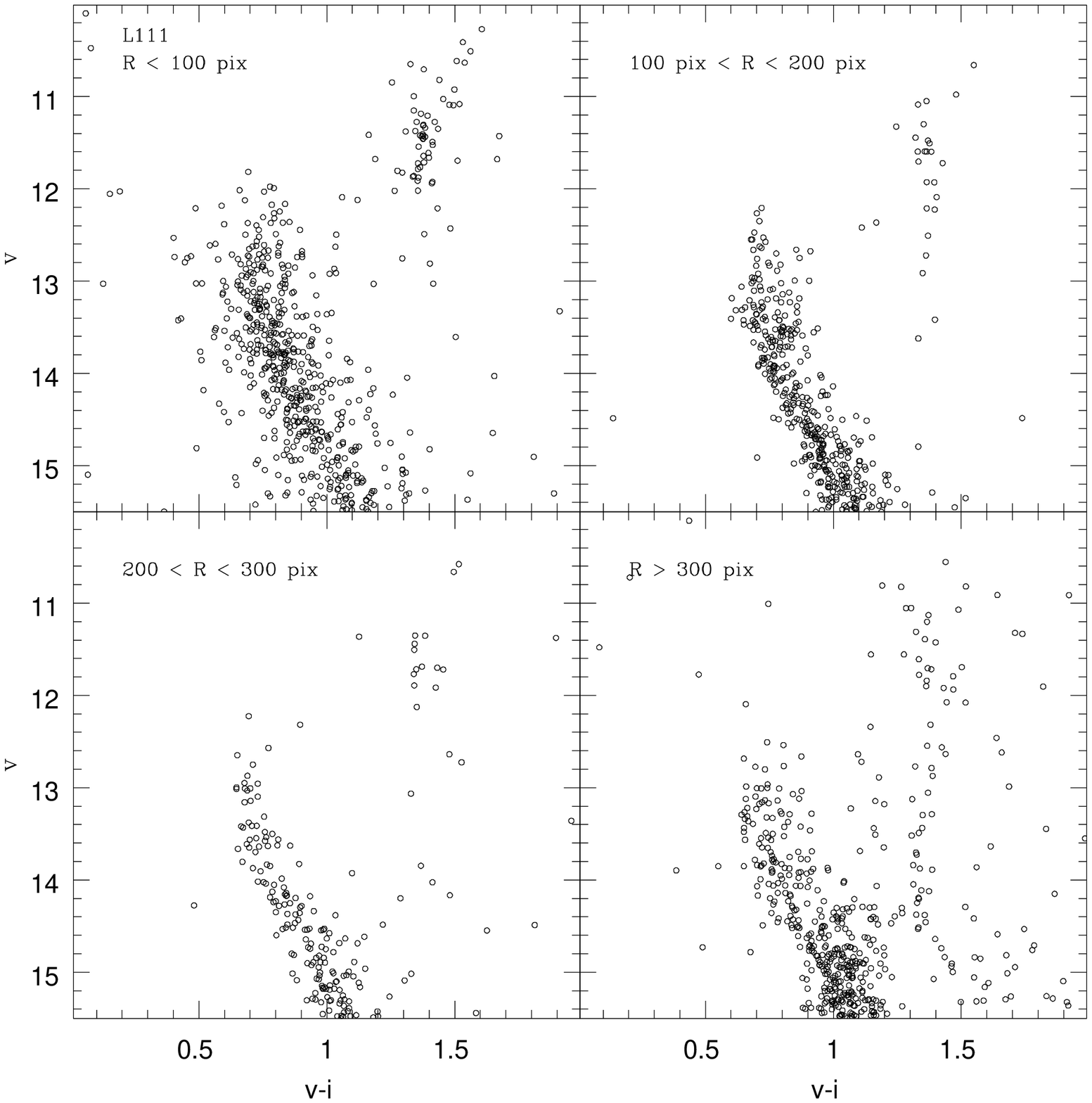}
\caption{\small{Idem Figure \ref{f:bs121M.bin} but for cluster L\,111.
}}
\label{f:l111M.bin}
\end{centering}
\end{figure}

\begin{figure}
\begin{centering}
\includegraphics[width=15.cm]{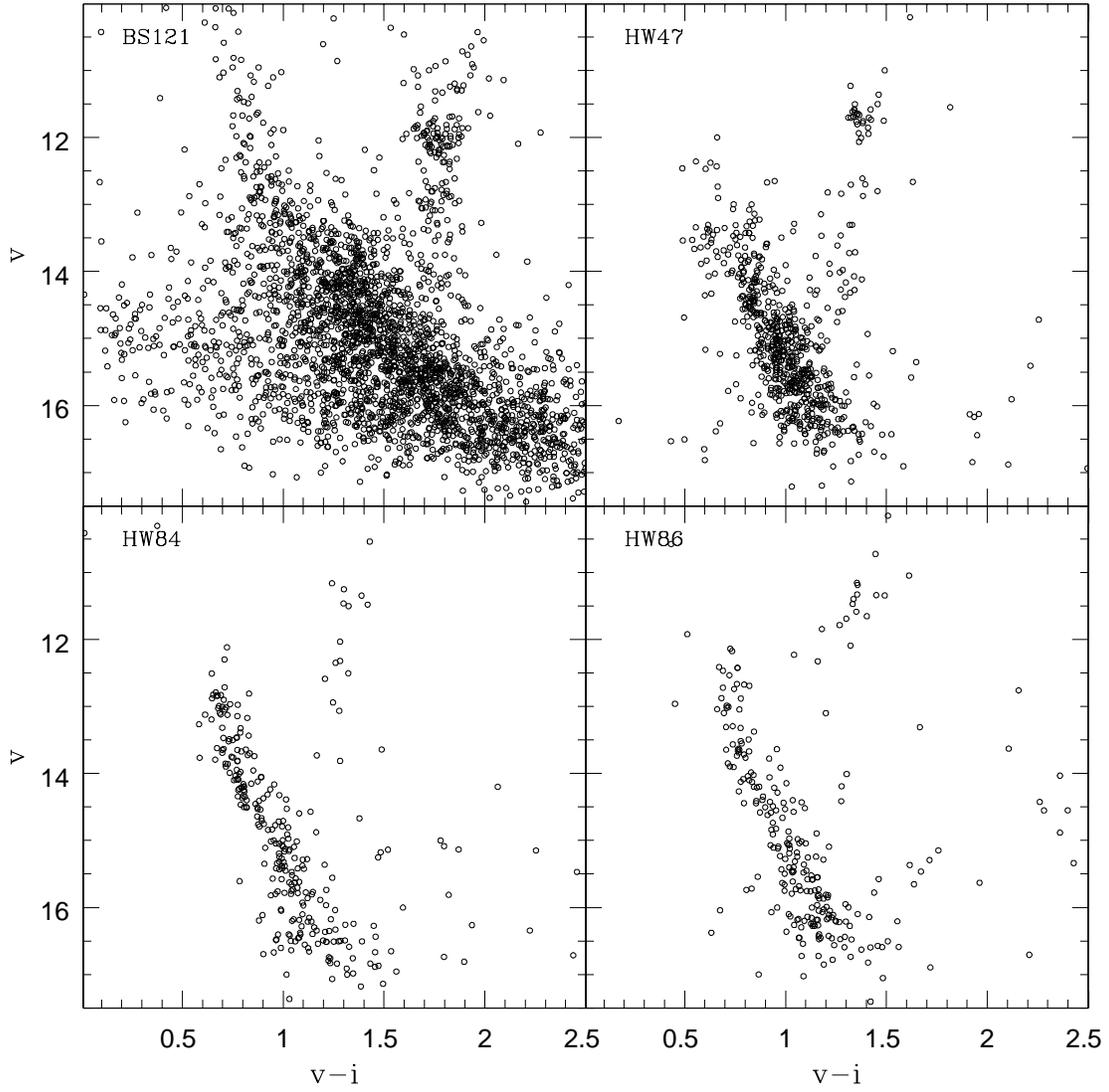}
\caption{\small{Color magnitude diagram built with stars belonging
to the two first rings (0 pixels $<$ r $<$ 200 pixels) for clusters BS\,121, HW\,47, HW\,84 and
HW\,86.  
}}
\label{f:bin12.1}
\end{centering}
\end{figure}

\clearpage

\begin{figure}
\begin{centering}
\includegraphics[width=15.cm]{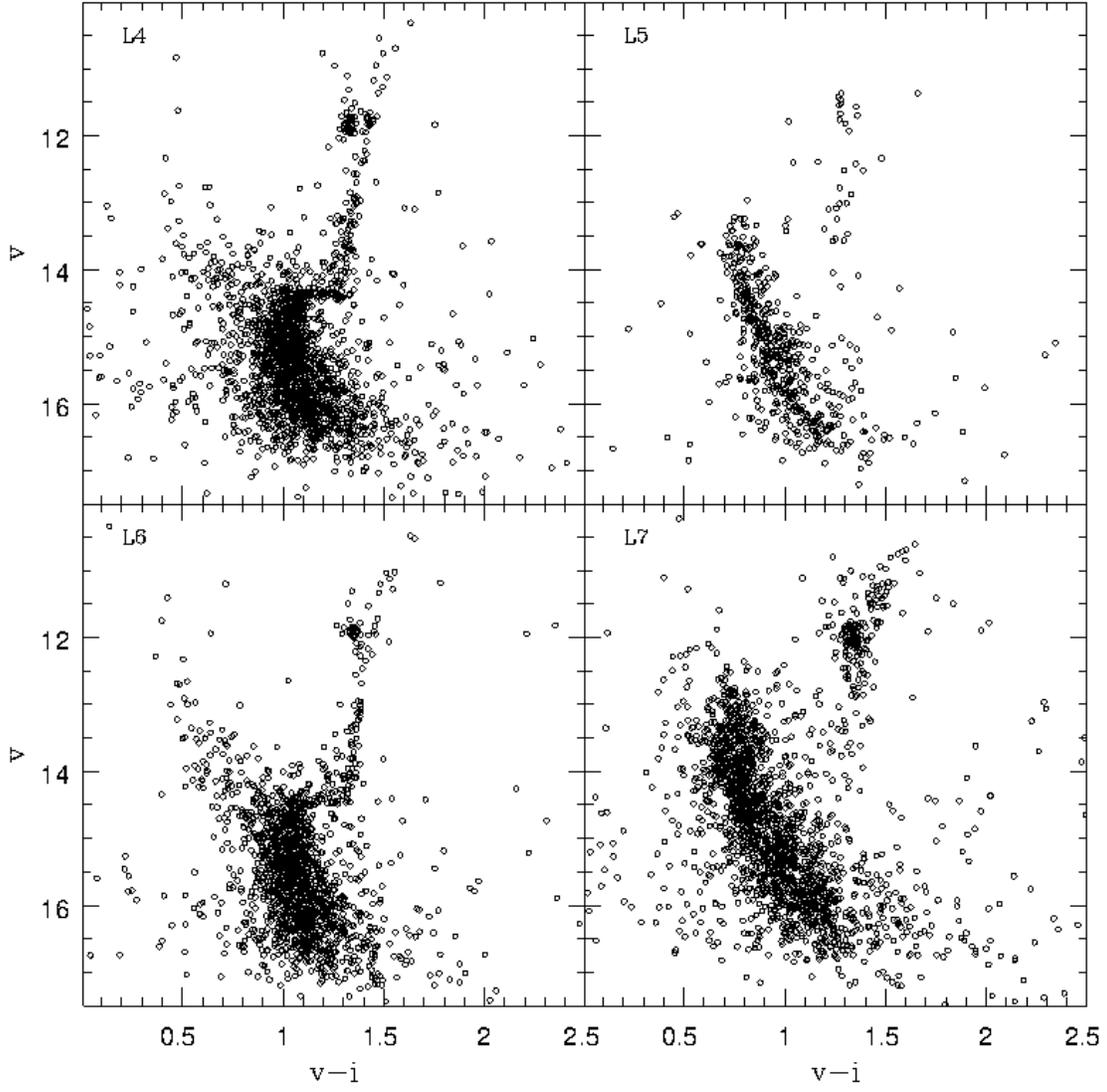}
\caption{\small{Idem Figure \ref{f:bin12.1} but for clusters L\,4, L\,5, L\,6 and L\,7.
}}
\label{f:bin12.2}
\end{centering}
\end{figure}

\clearpage 

\begin{figure}
\begin{centering}
\includegraphics[width=15.cm]{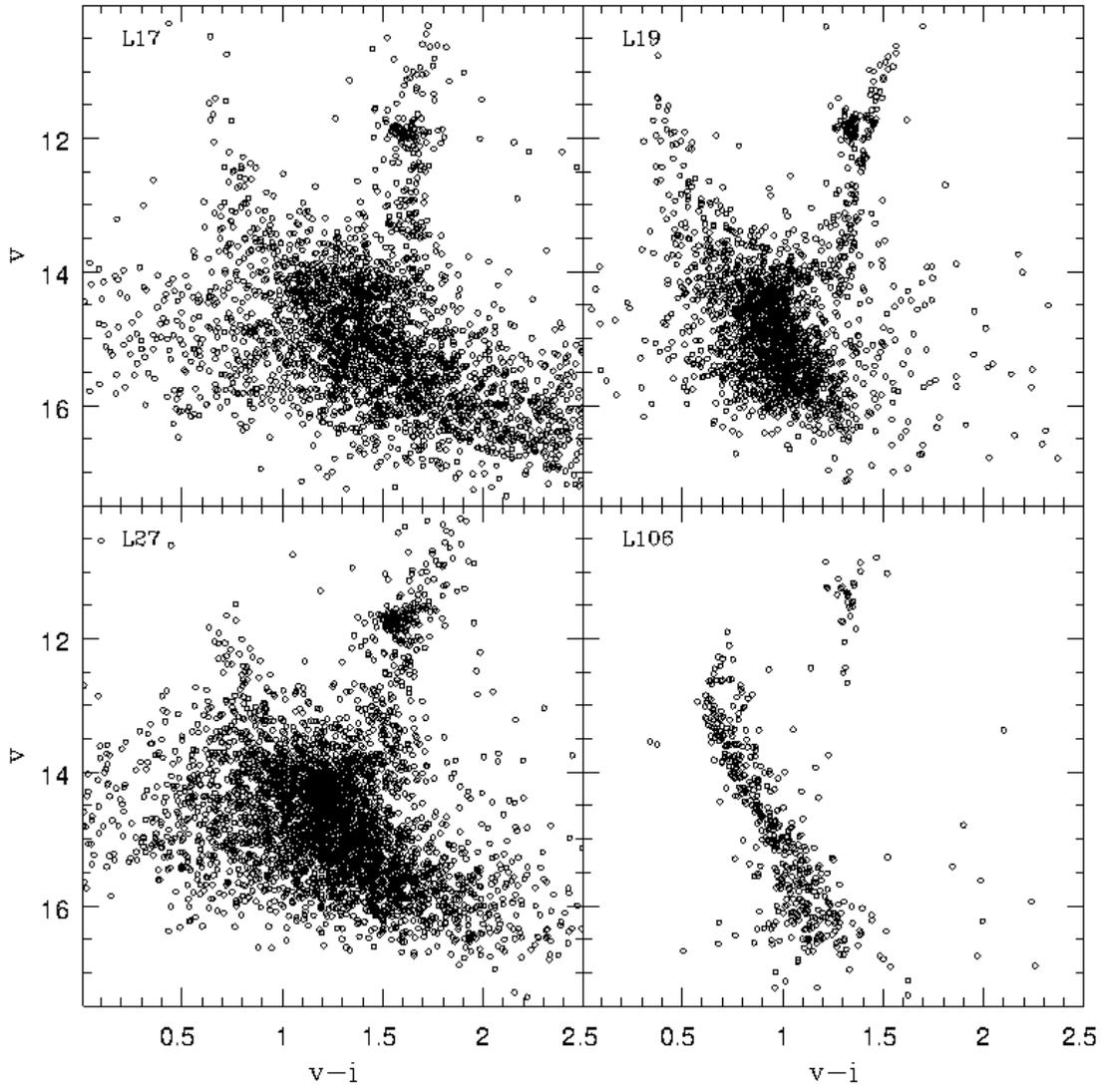}
\caption{\small{Idem Figure \ref{f:bin12.1} but for clusters L\,17, L\,19, L\,27 and L\,106.
}}
\label{f:bin12.3}
\end{centering}
\end{figure}

\begin{figure}
\begin{centering}
\includegraphics[width=15.cm]{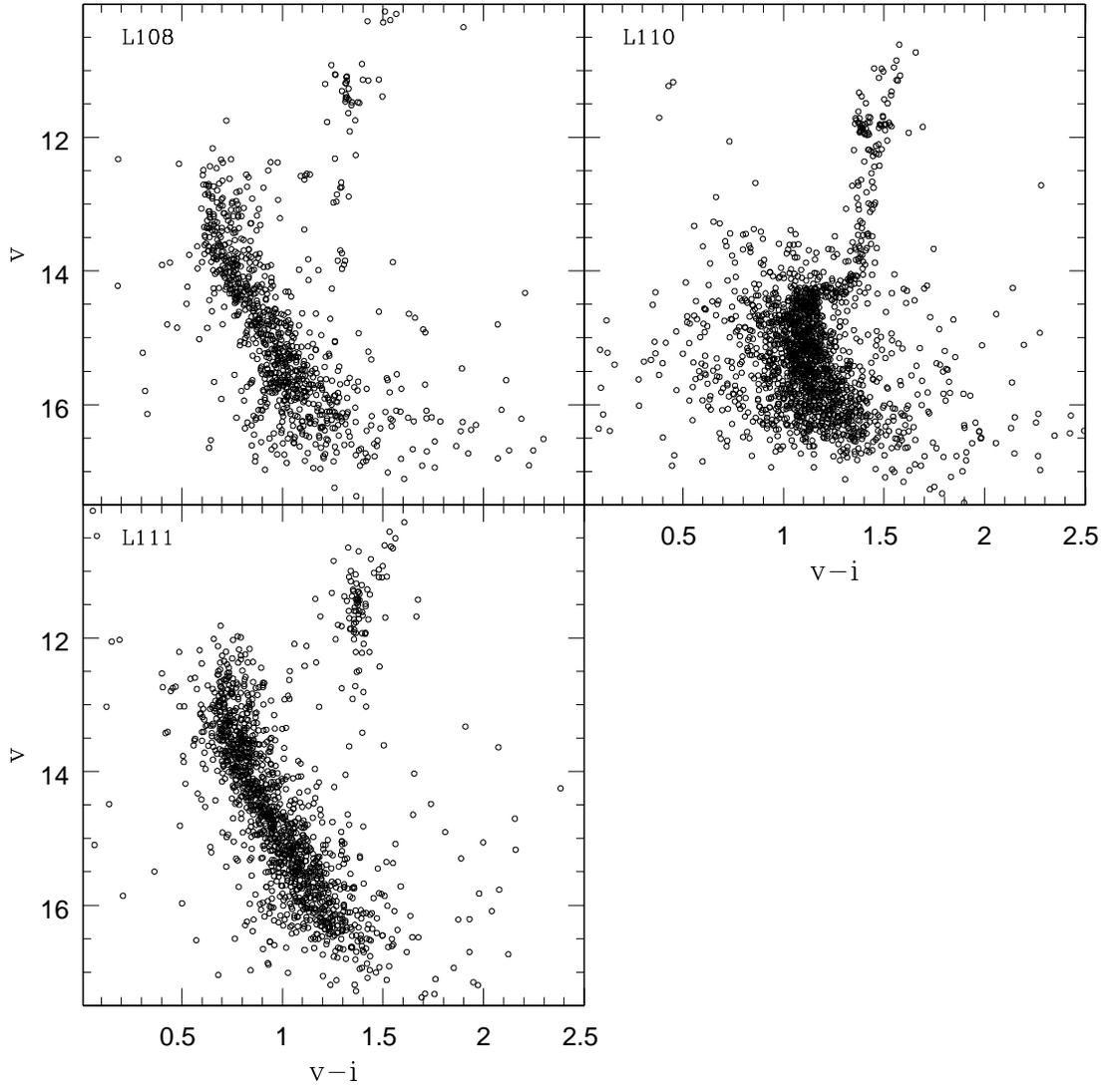}
\caption{\small{Idem Figure \ref{f:bin12.1} but for clusters L\,108, L\,110 and  L\,111.
}}
\label{f:bin12.4}
\end{centering}
\end{figure}

\begin{figure}
\begin{centering}
\includegraphics[width=15.cm]{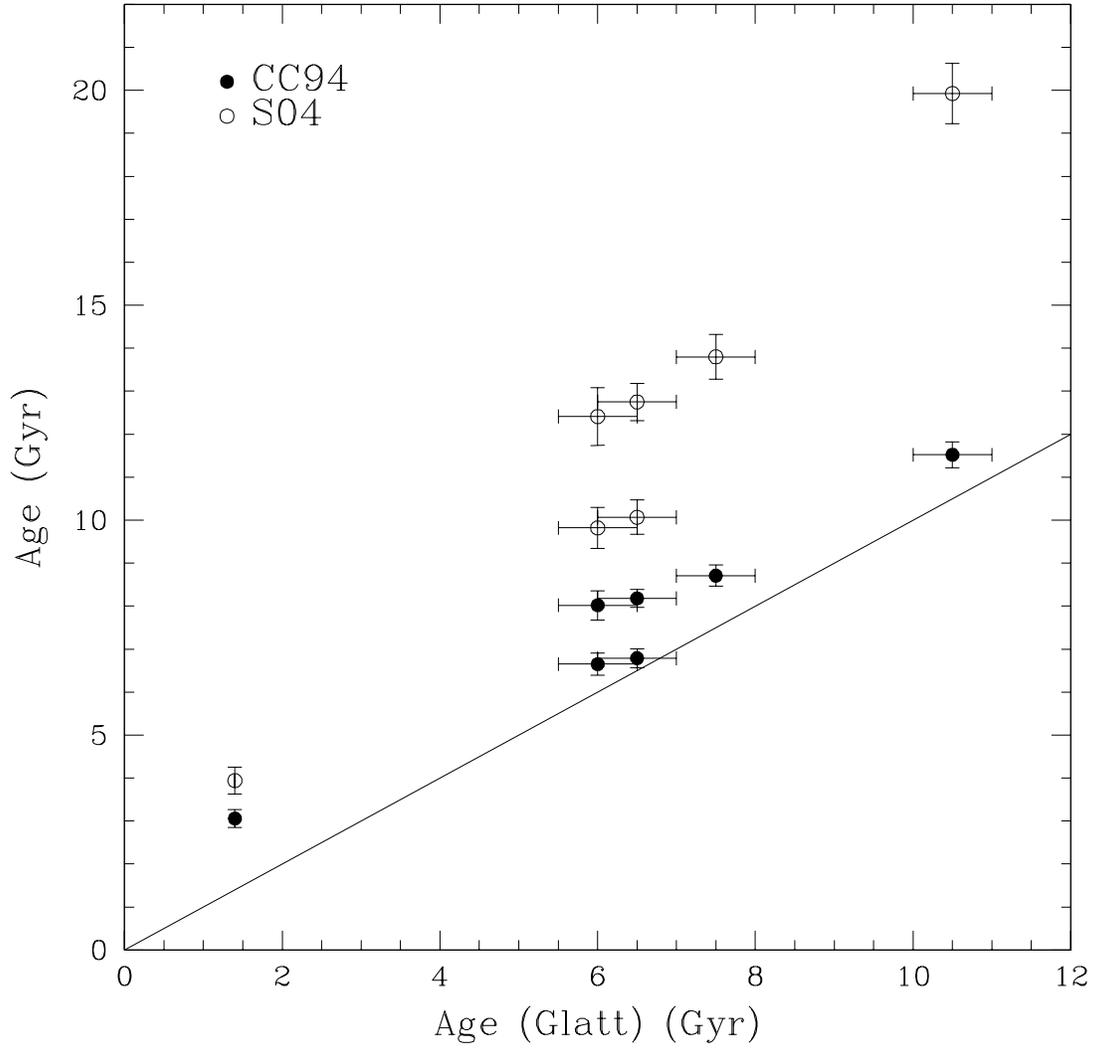}
\caption{\small{ Comparison of ages derived with (\citealt{cc94}, CC94) and (\citealt{swp04}, S04) calibrations for clusters from \citet{gla08b}.
Solid line represents the 1:1 relation.
}}
\label{f:cc94Mvss04_glatt}
\end{centering}
\end{figure}

\begin{figure}
\begin{centering}
\includegraphics[width=15.cm]{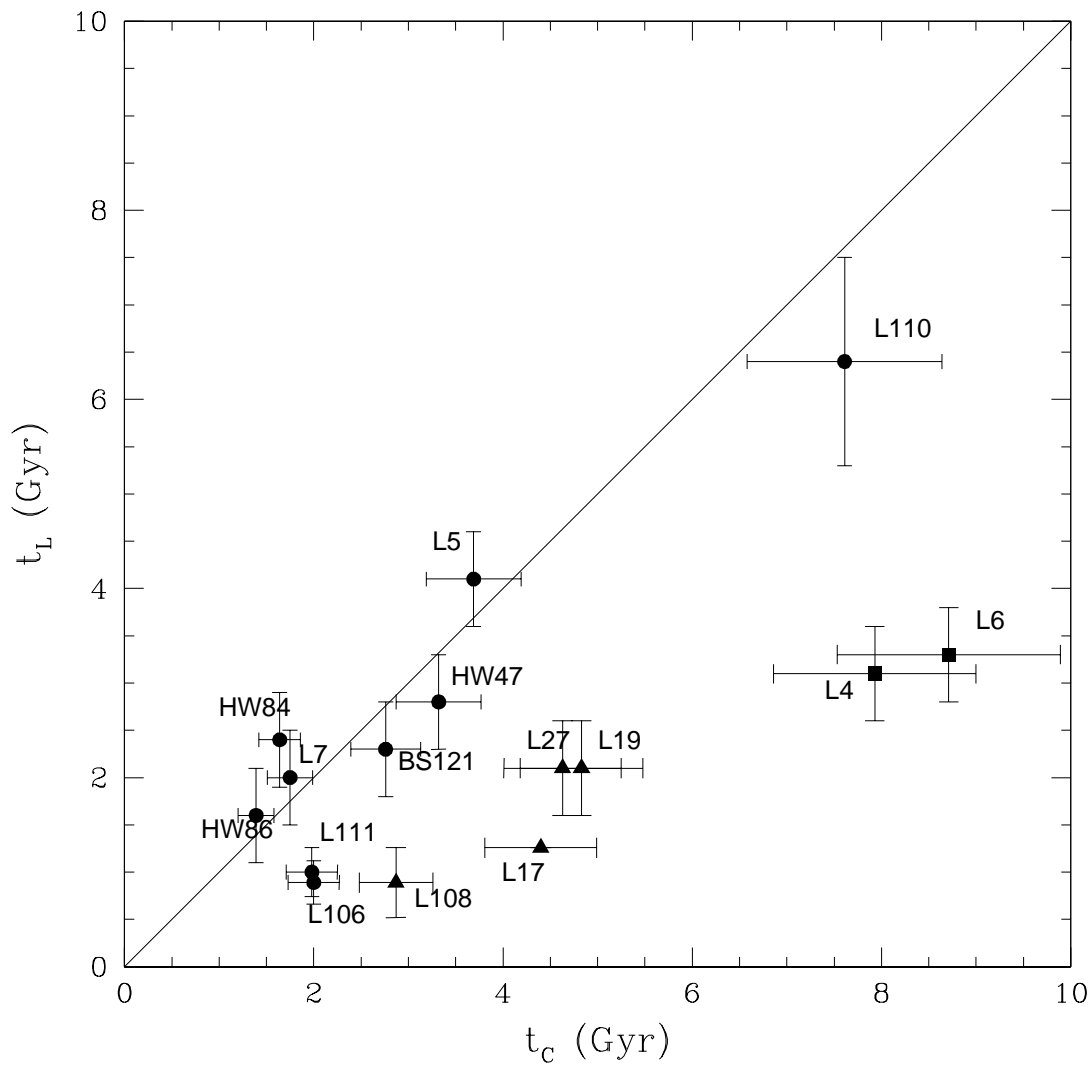}
\caption{\small{Comparison of  the ages reported in the literature ($t_L$) and the ages derived in the present 
work ($t_C$) for the clusters of our sample. Clusters presenting good agreement between both age values 
have been plotted with circles. Triangles and squares represent clusters having a moderate and a considerable
disagreement between $t_L$ and $t_C$, respectively (see text for more details).
}}
\label{f:age_lit_cc94M}
\end{centering}
\end{figure}

\begin{figure}
\begin{centering}
\includegraphics[width=15.cm]{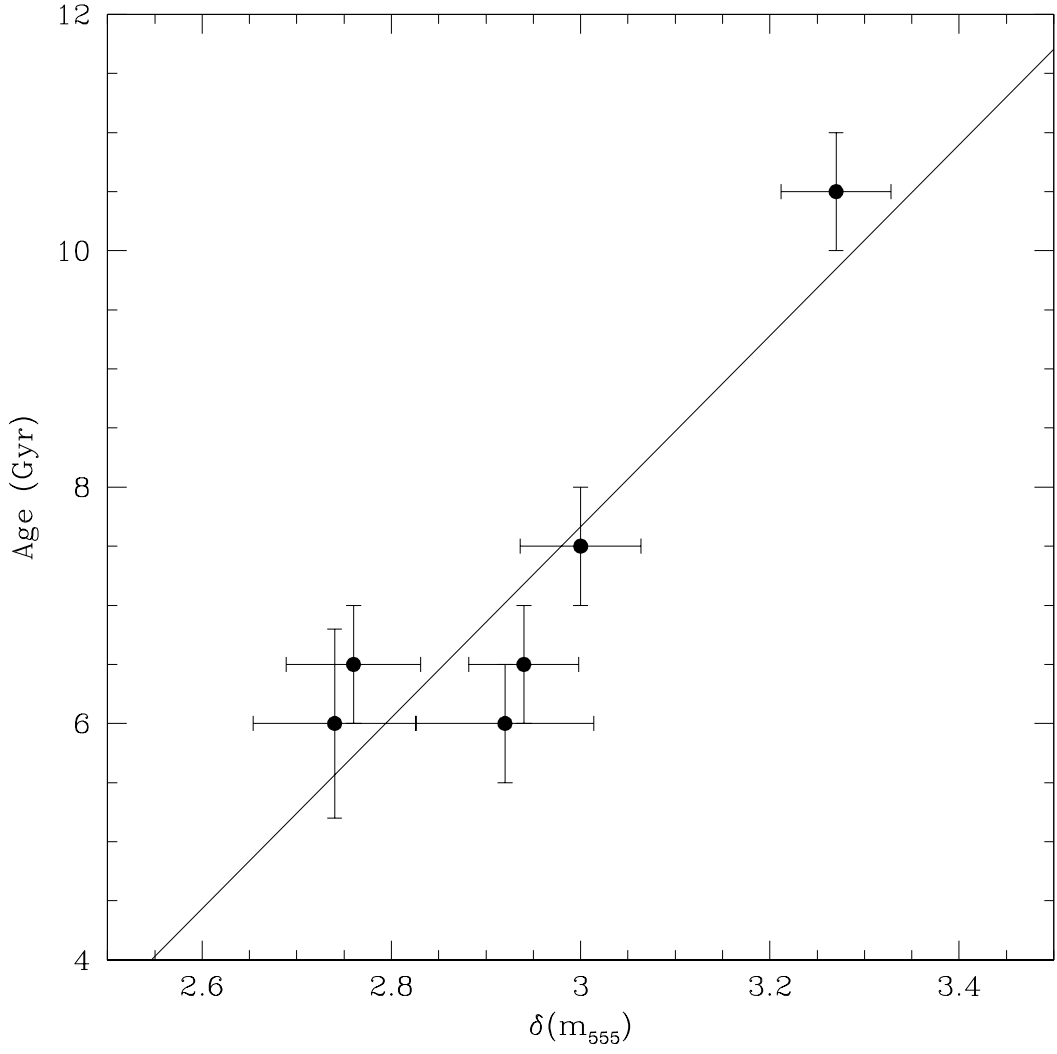}
\caption{\small{SMC $\delta $ calibration built with data taken from \citet{gla08a,gla08b}. 
}}
\label{f:glatt}
\end{centering}
\end{figure}

\begin{figure}
\begin{centering}
\includegraphics[width=15.cm]{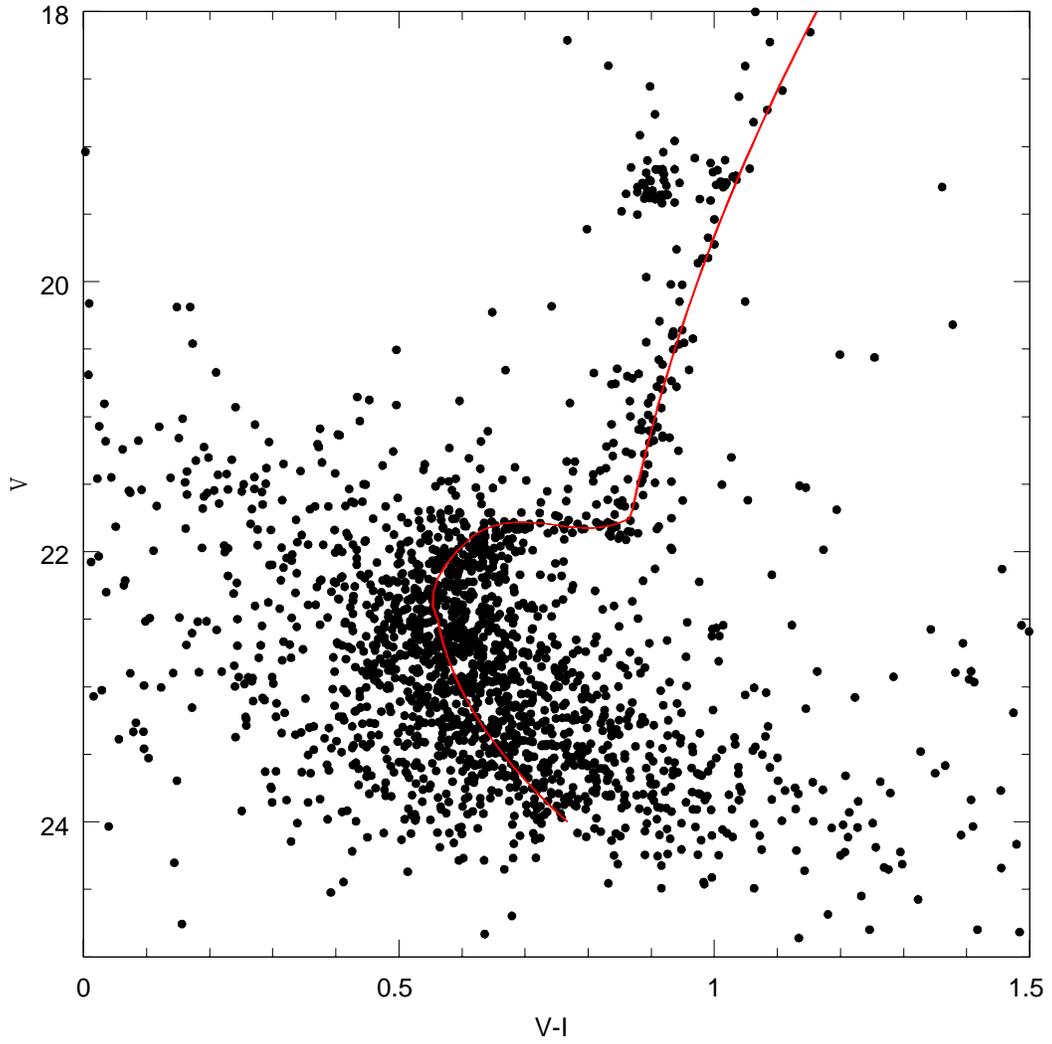}
\caption{\small{Calibrated CMD of L\,4 built with stars having a distance from the cluster center
smaller than 200 pixels. The solid line (colored red in the online version) represents the fiducial ridgeline. 
}}
\label{f:l4_fiducial}
\end{centering}
\end{figure}

\begin{figure}
\begin{centering}
\includegraphics[width=15.cm]{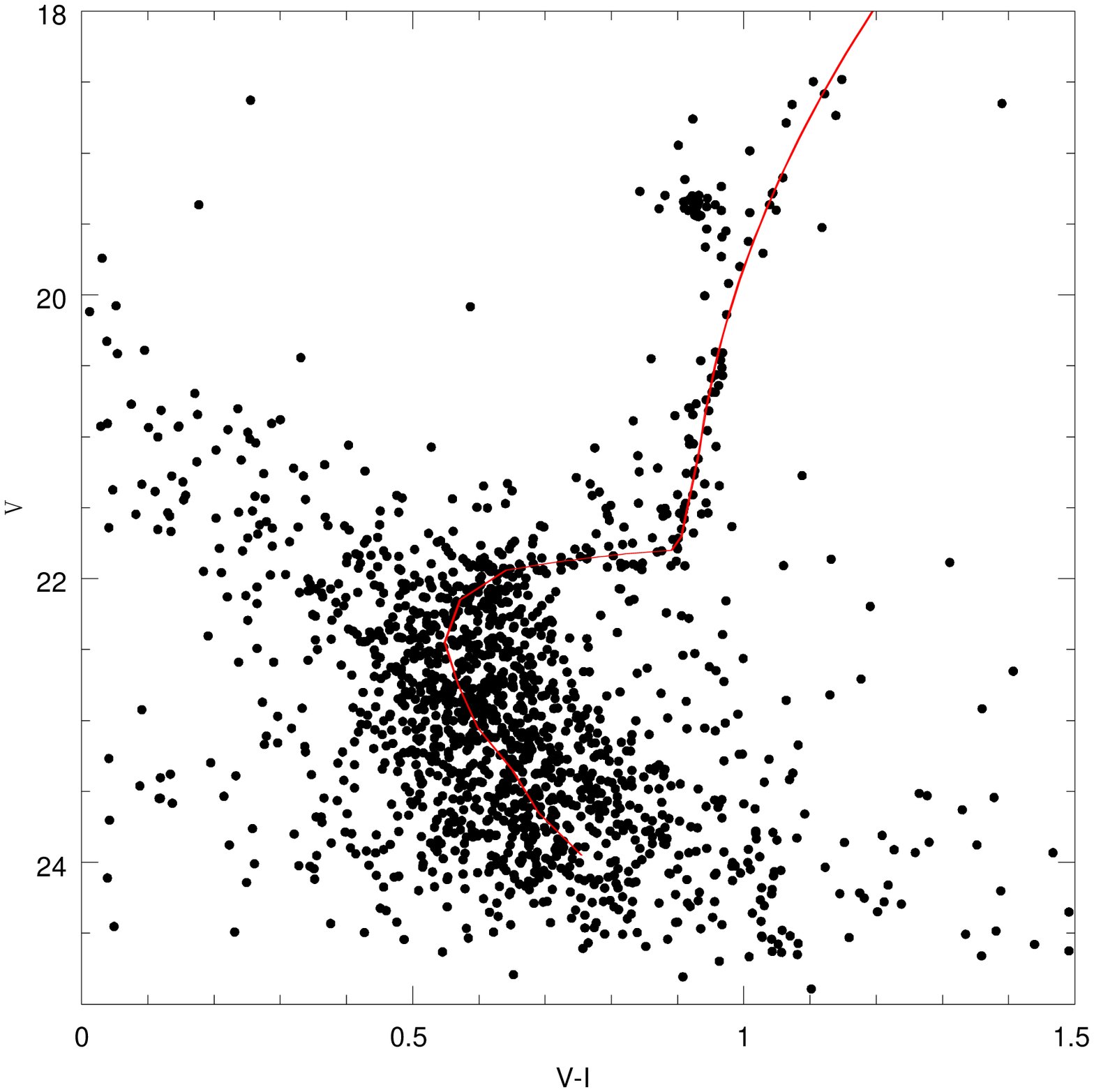}
\caption{\small{Same as Figure \ref{f:l4_fiducial} but for cluster L\,6.
}}
\label{f:l6_fiducial}
\end{centering}
\end{figure}

\begin{figure}
\begin{centering}
\includegraphics[width=15.cm]{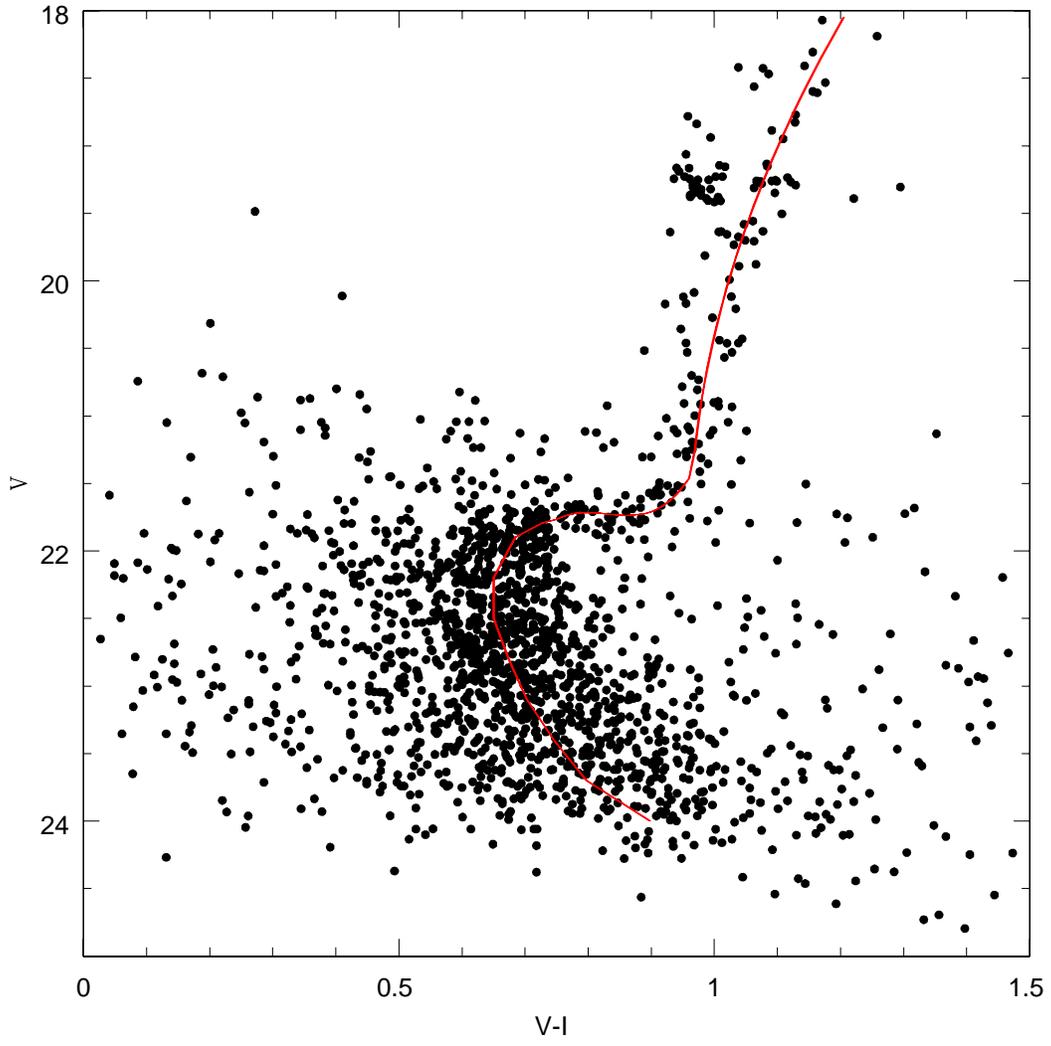}
\caption{\small{Same as Figure \ref{f:l4_fiducial} but for cluster L\,110.
}}
\label{f:l110_fiducial}
\end{centering}
\end{figure}

\begin{figure}
\begin{centering}
\includegraphics[width=10.cm]{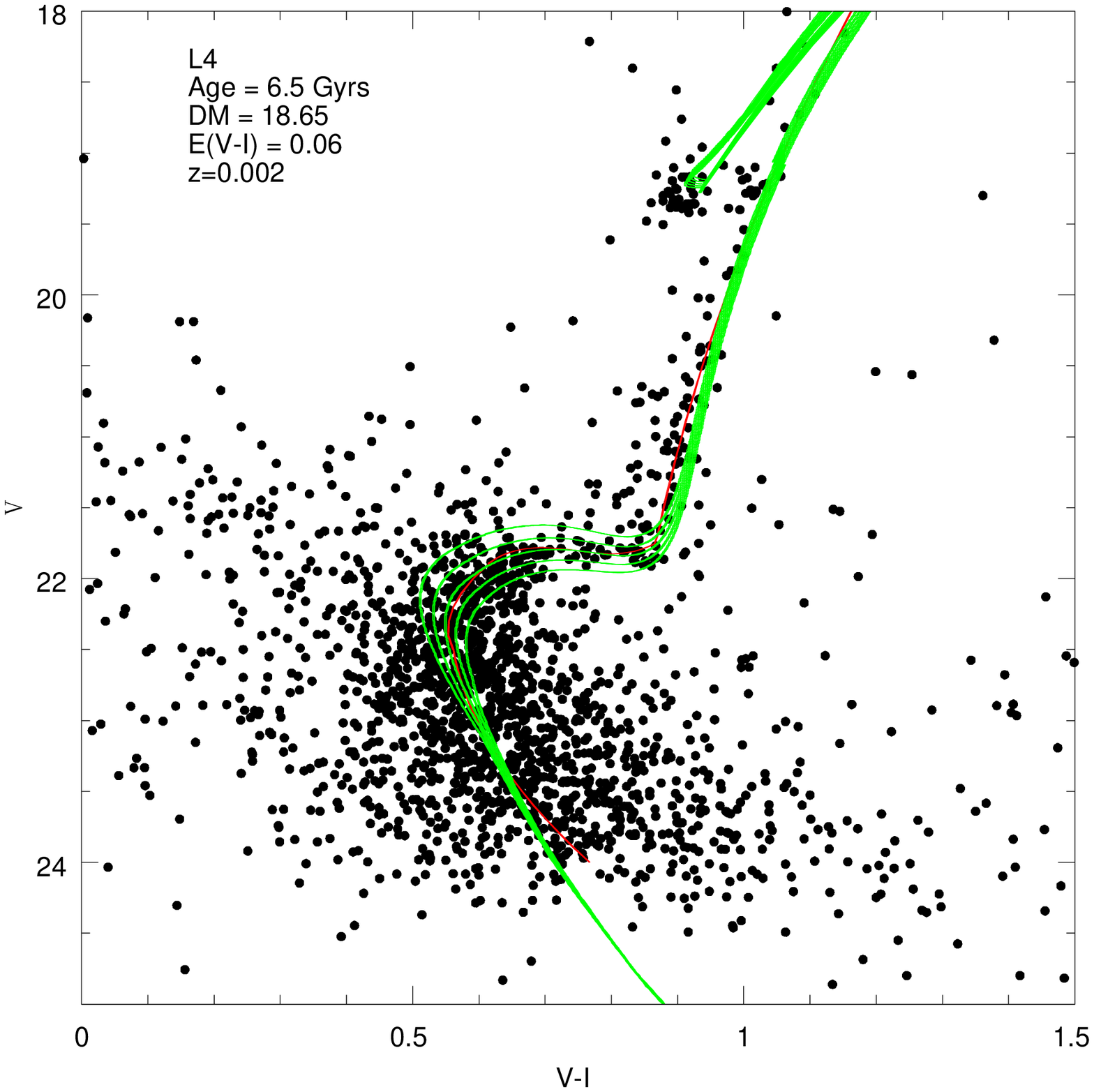}
\includegraphics[width=10.cm]{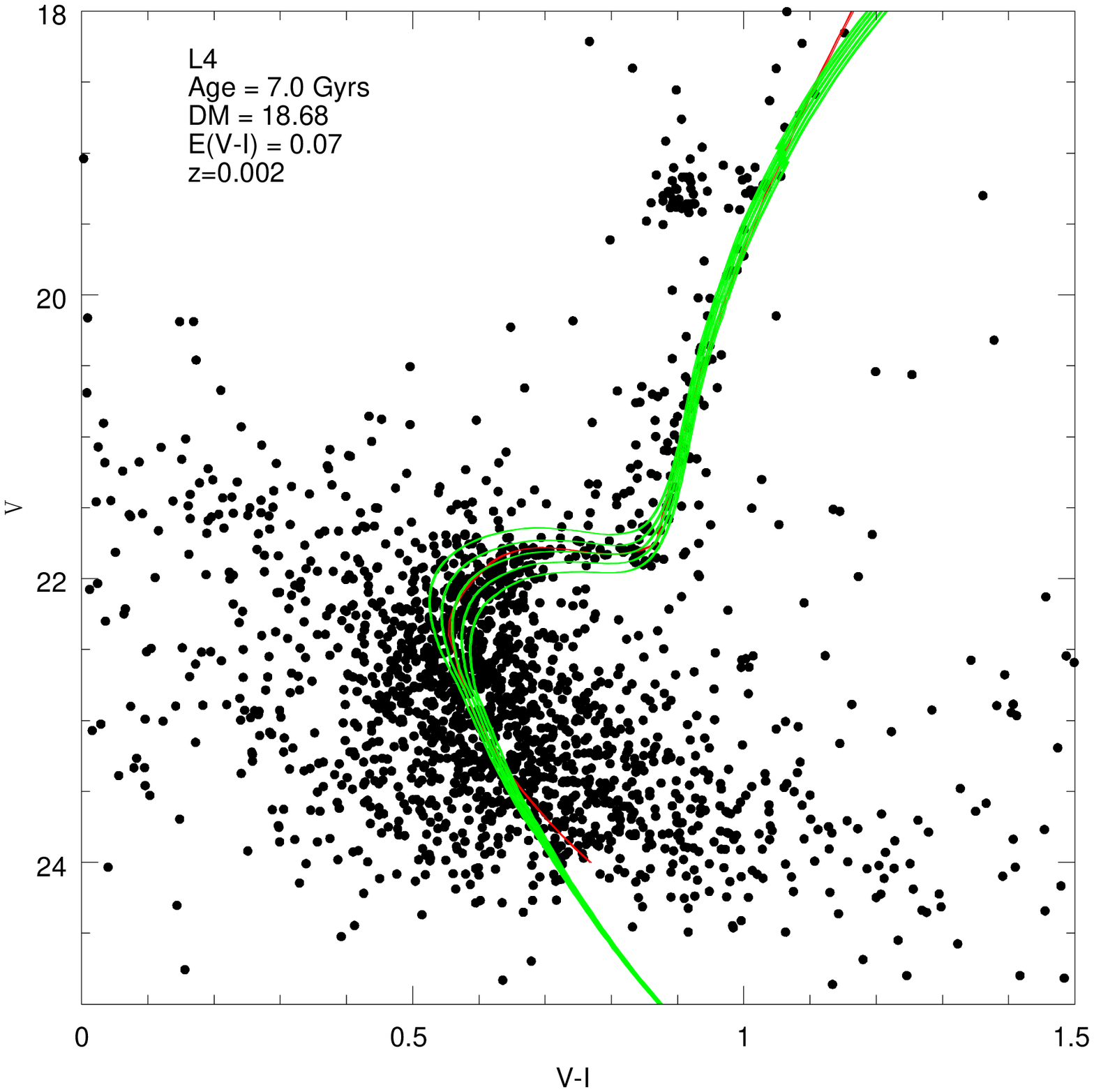}
\caption{\small{CMD of L\,4. {\it Upper panel:} we show Teramo isochrones as solid lines (colored green in the online version). 
They cover an age range from 5.5 to 7.7 Gyrs in steps of 0.5 Gyrs. The central 
isochrone (z=0.002), which best reproduces the fiducial ridgeline (red line in the online version), is the one that we have adopted. Fitting parameters are listed in the plot.
{\it Lower panel:} Same as upper panel but for Dartmouth isochrones in age steps of 6, 6.5, 7, 7.5 and 8 Gyrs.}}
\label{f:l4.iso}
\end{centering}
\end{figure}

\clearpage

\begin{figure}
\begin{centering}
\includegraphics[width=10.cm]{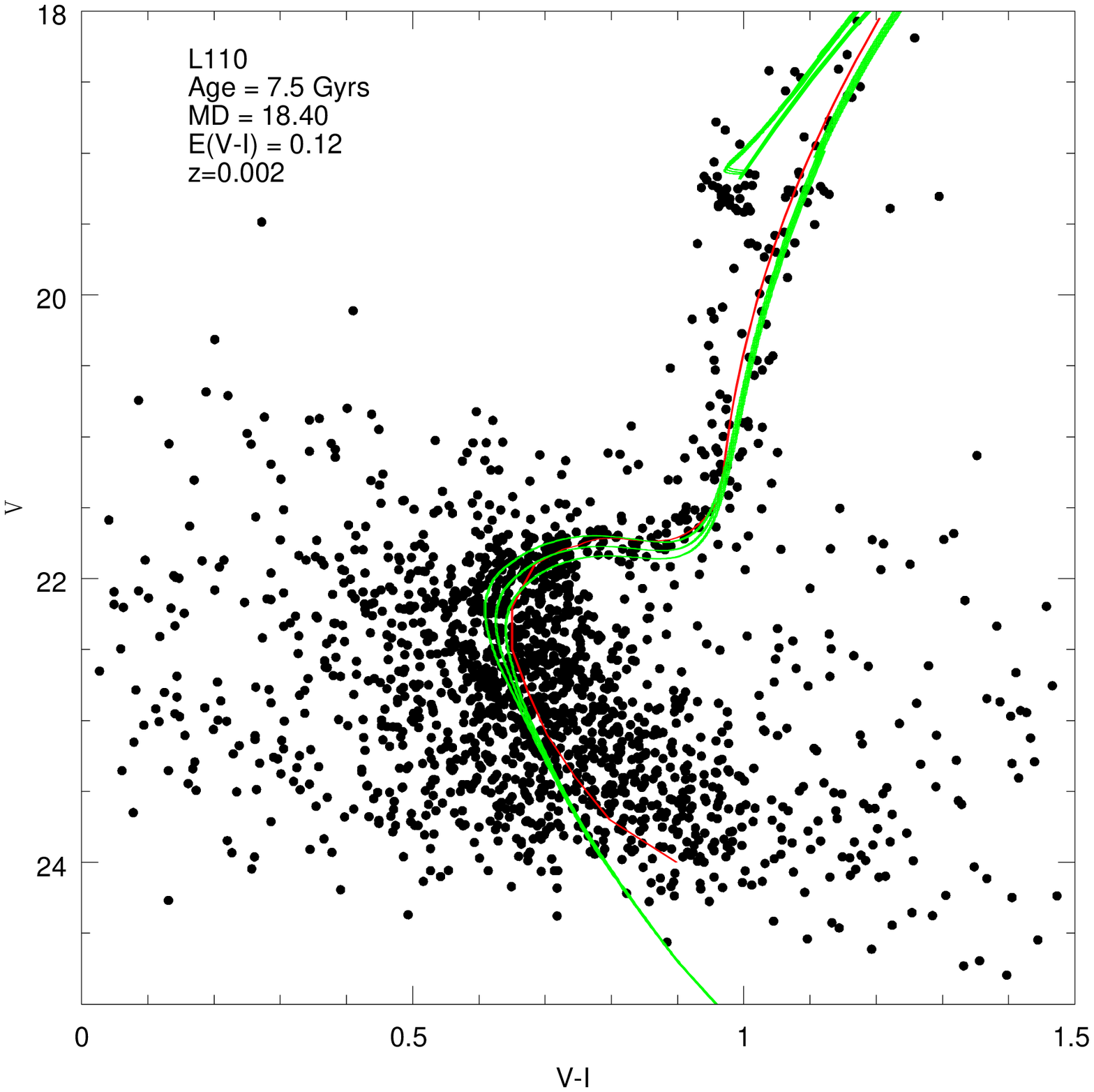}
\includegraphics[width=10.cm]{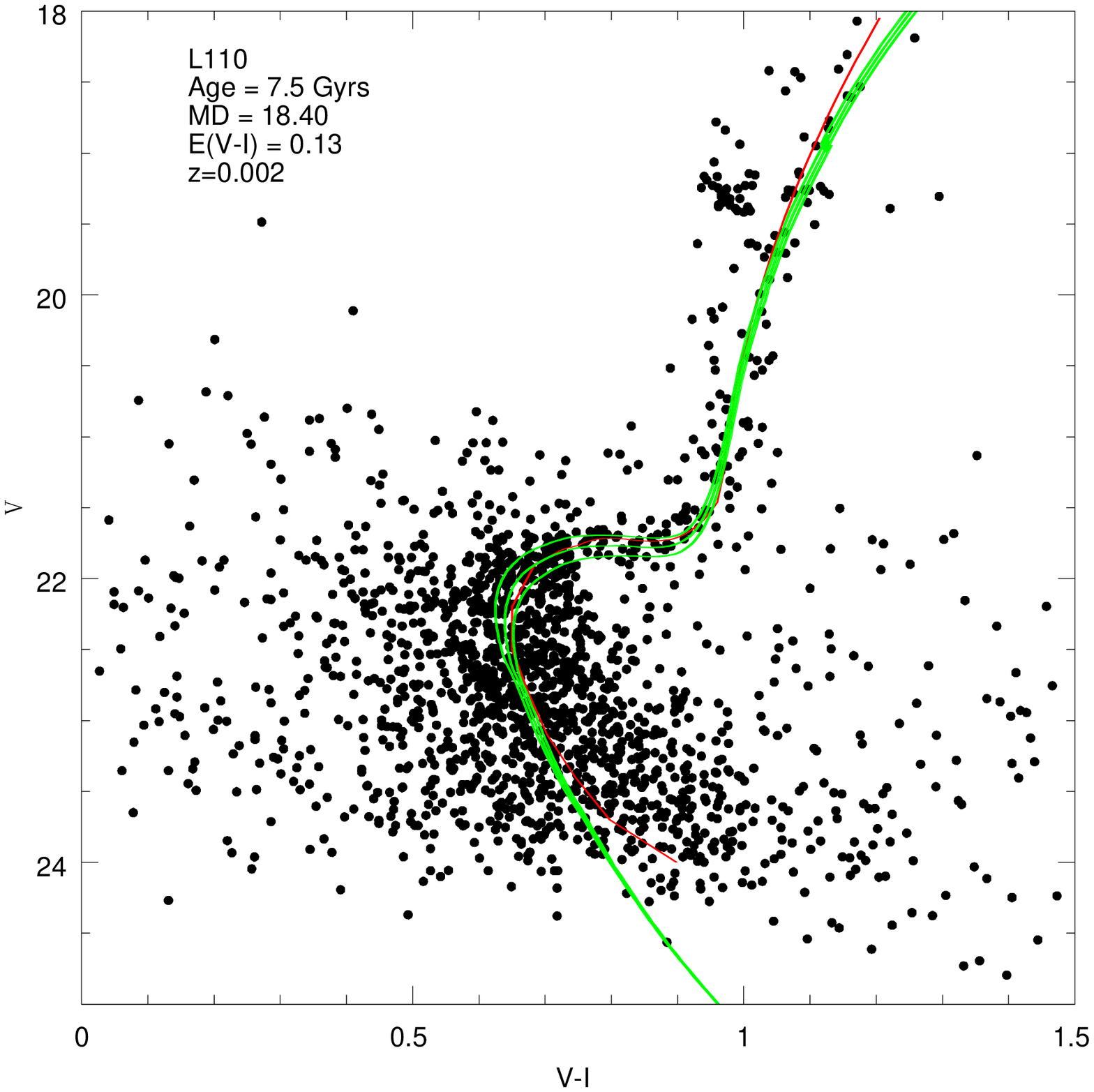}
\caption{\small{CMD of L\,110. {\it Upper panel:} we show Teramo isochrones as solid lines (colored green in the online version).
They cover an age range from 7 to 8 Gyrs in steps of 0.5 Gyrs. The central
isochrone (z=0.002), which best reproduces the fiducial ridgeline (red line in the online version), is the one that we have adopted. Fitting parameters are listed in the plot.
{\it Lower panel:} Same as upper panel but for Dartmouth isochrones in age steps of 7, 7.5 and 8 Gyrs.}}
\label{f:l110.iso}
\end{centering}
\end{figure}

\clearpage

\begin{figure}
\begin{centering}
\includegraphics[width=10.cm]{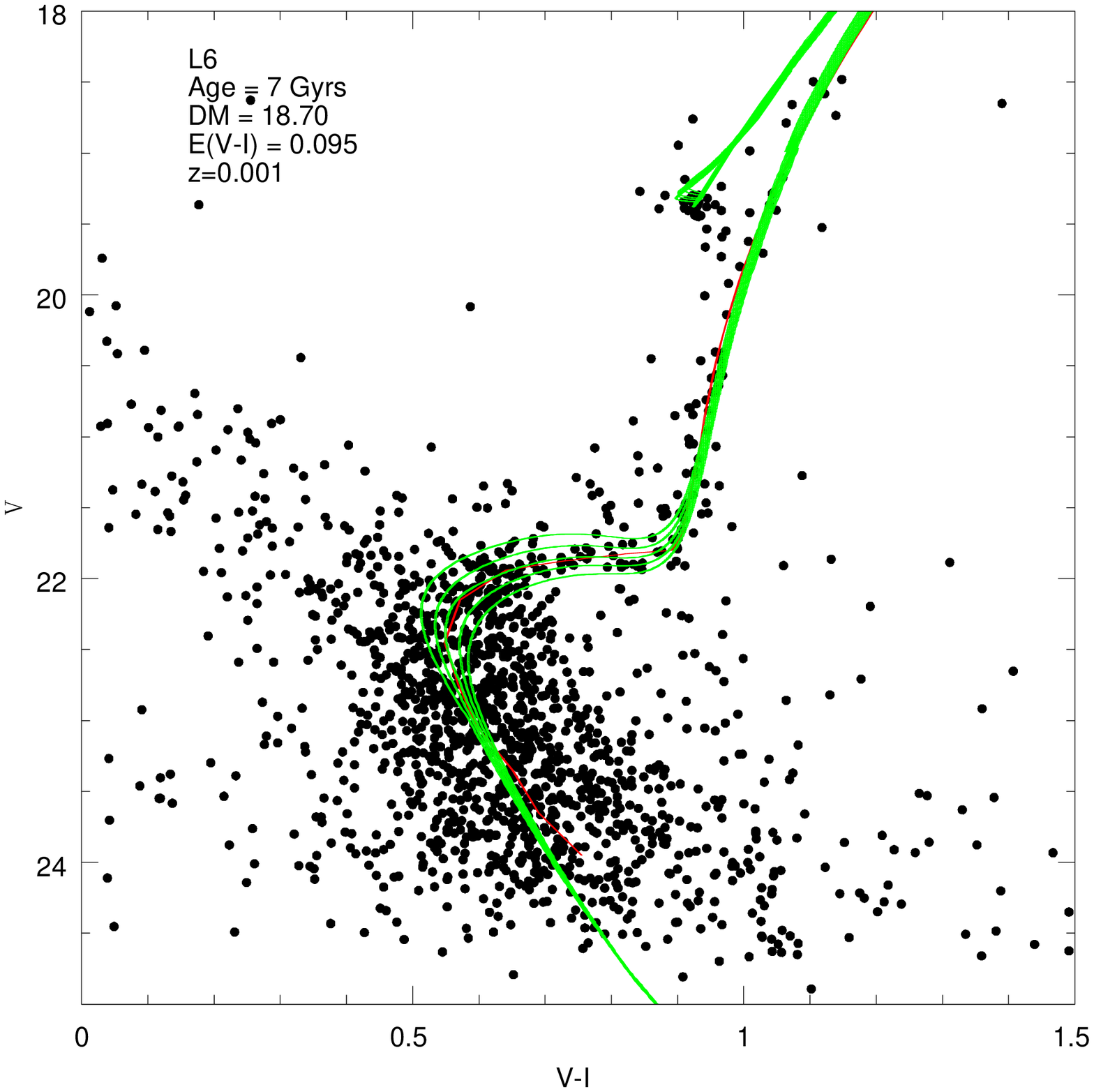}
\includegraphics[width=10.cm]{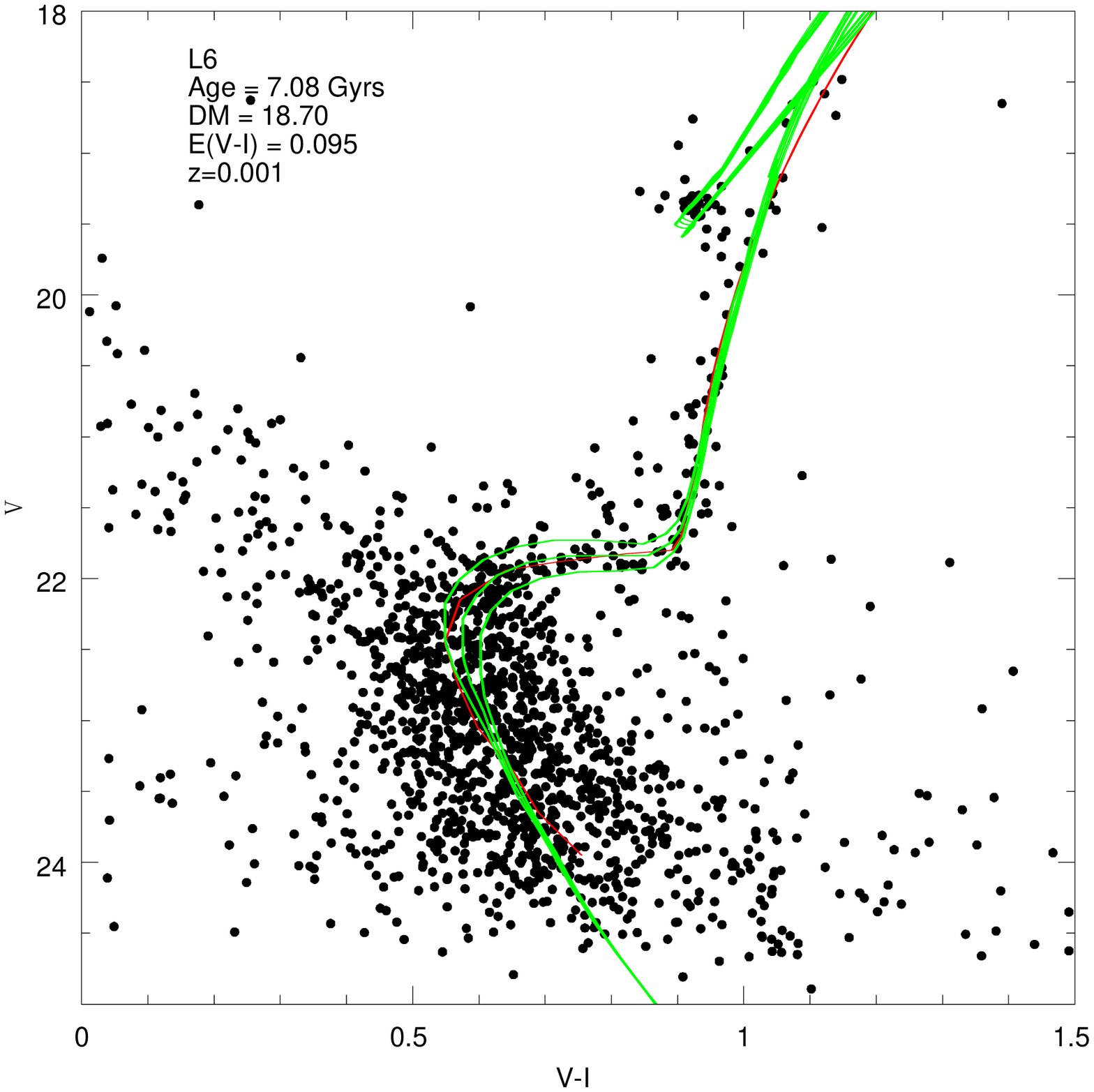}
\caption{\small{CMD of L\,6. {\it Upper panel:} we show Teramo isochrones as solid lines (colored green in the online version). 
They cover an age range from 6 to 8 Gyrs in steps of 0.5 Gyrs. The central
isochrone (z=0.001), which best reproduces the fiducial ridgeline (red line in the online version), is the one that we have adopted. Fitting parameters are listed in the plot.
{\it Lower panel:} Same as upper panel but for Padova isochrones in age steps of 6.3, 7.08 and 7.94 Gyrs.}}
\label{f:l6.iso}
\end{centering}
\end{figure}

\clearpage

\begin{figure}
\begin{centering}
\includegraphics[width=15.cm]{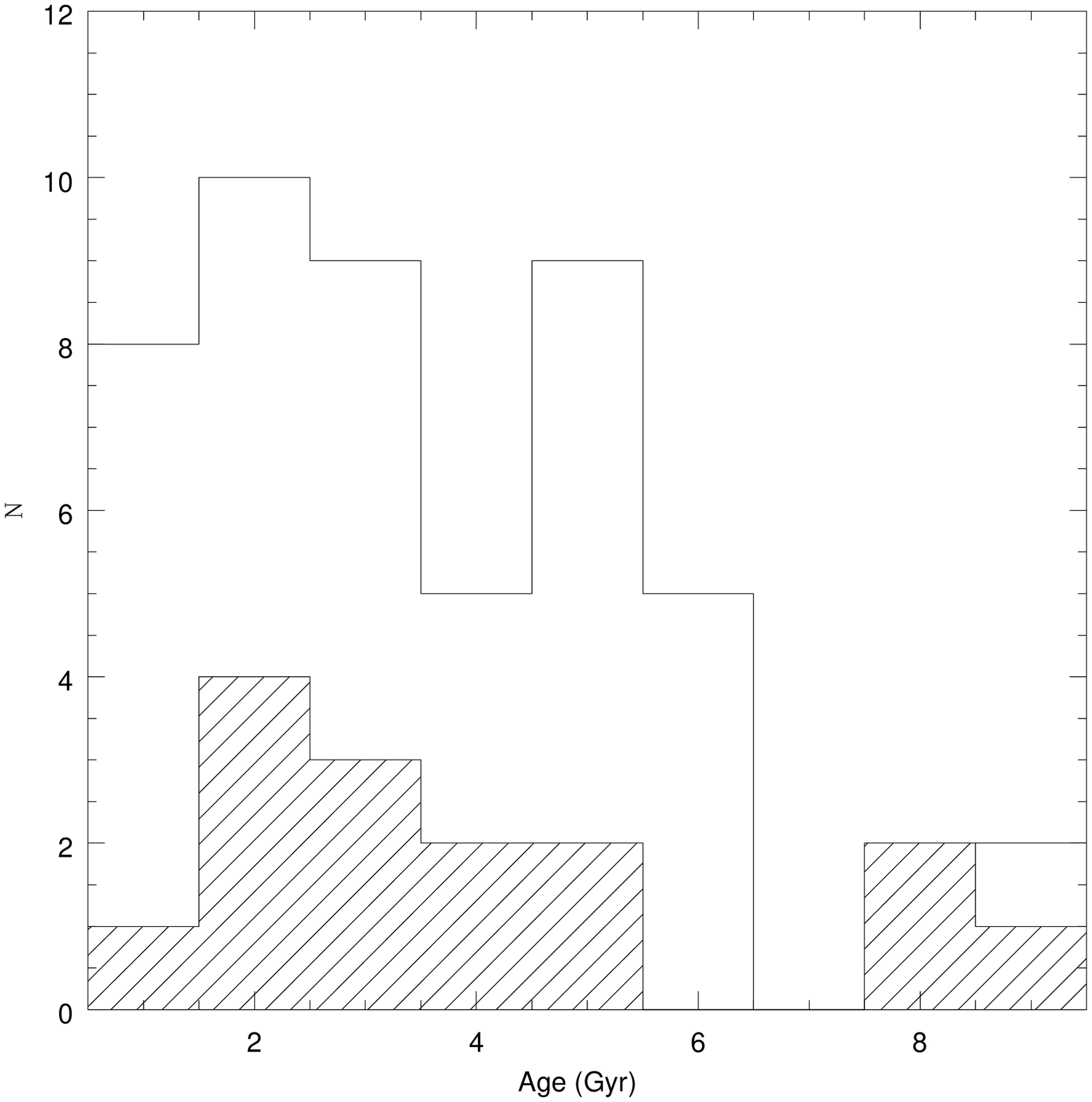}
\caption{\small{SMC Age Distribution. Lined histogram: 15 clusters from the present work. Empty histogram:
5 from \citet{pia01}, 2 from \citet{pia07b}, 9 from \citet{pia11a}, 7 from \citet{pia11b}, 11 from \citet{pia11c}, 
ESO 51-SC09 (This work and \citealt{pia12b}) and 15 from the present work.
}}
\label{f:AD}
\end{centering}
\end{figure}

\begin{figure}
\begin{centering}
\includegraphics[width=15.cm]{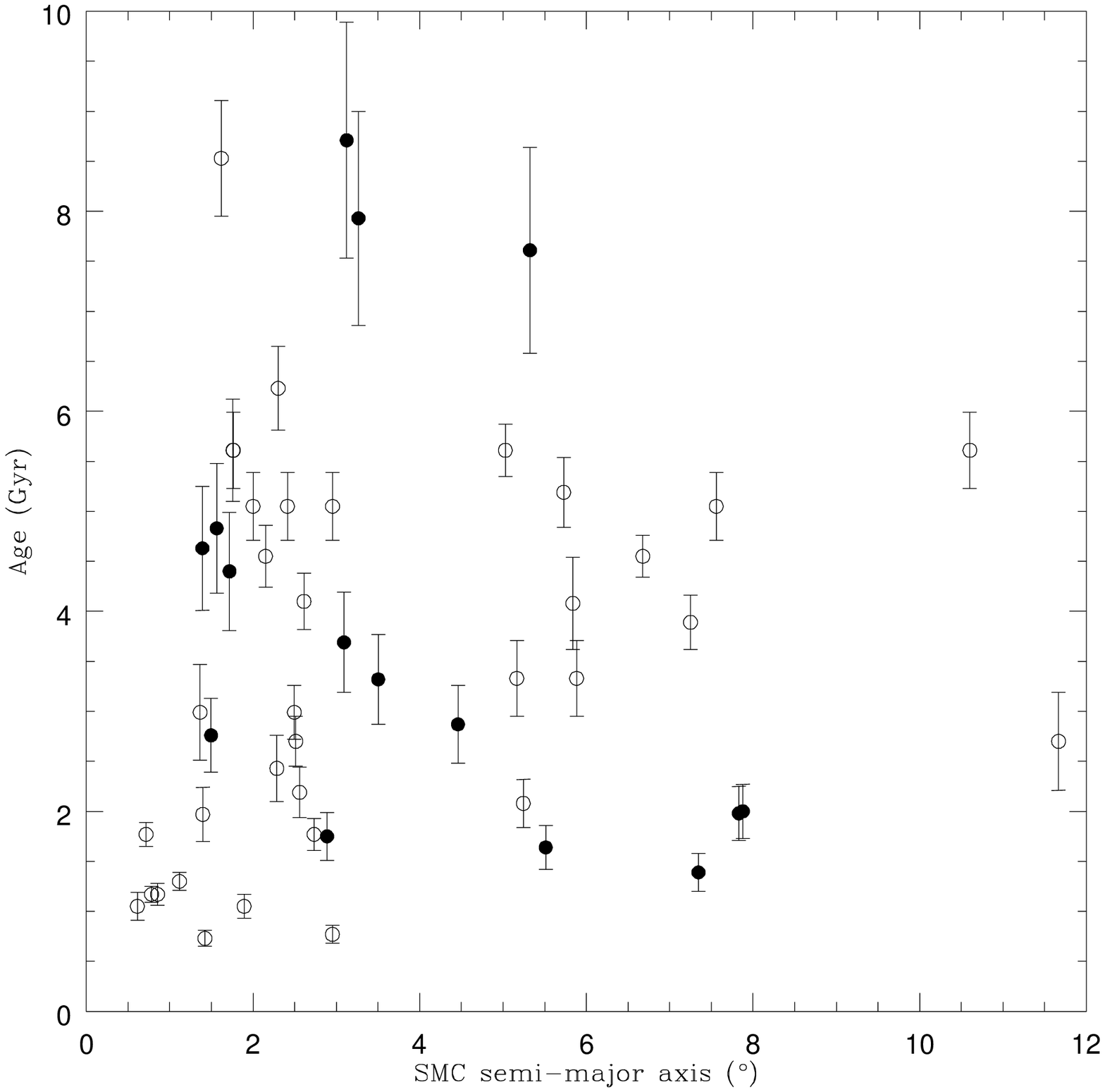}
\caption{\small{Age vs. semi-major axis $a$ for 15 clusters from the present work (filled circles) and 5 from \citet{pia01}, 2 from \citet{pia07b}, 
9 from \citet{pia11a}, 7 from \citet{pia11b}, 11 from \citet{pia11c}, ESO 51-SC09 (This work and \citealt{pia12b}) (open circles).
}}
\label{f:AG}
\end{centering}
\end{figure}

\begin{figure}
\begin{centering}
\includegraphics[width=15.cm]{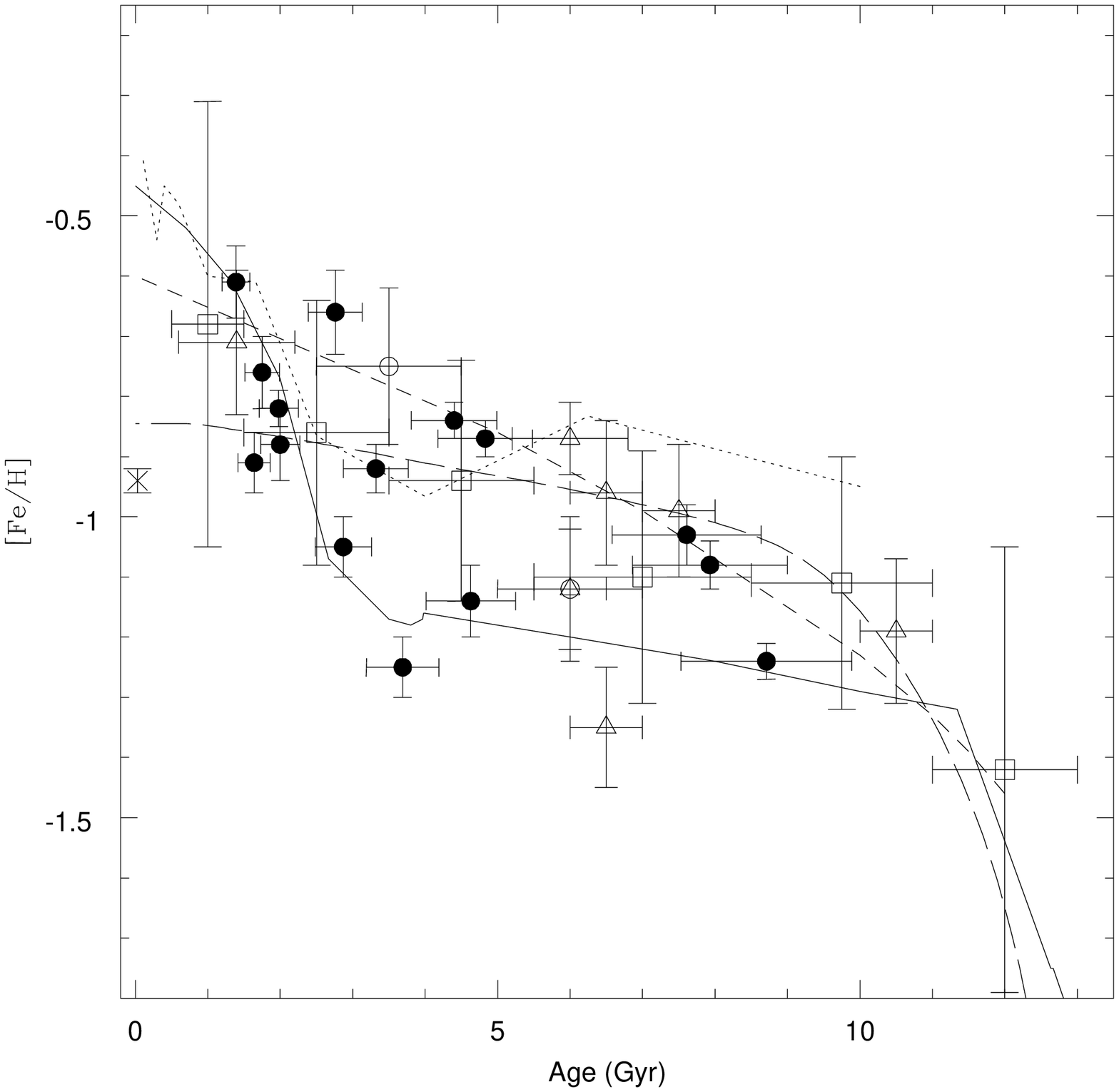}
\caption{\small{Filled circles stand for clusters of our sample. The cross marks cluster NGC\,330 \citep{gon99}. Squares show the mean 
metallicities in six age bins calculated by \citet{car08}. The short-dashed line represents the model of closed box continuous star 
formation computed by \citet{dch98}. The long-dashed line shows the best-fit model derived by \citet{car05}. The solid line represents 
the bursting model of \citet{pag98}. Finally, the dotted line stands for the AMR obtained by \citet{har04}.
}}
\label{f:age_met}
\end{centering}
\end{figure}

\clearpage

\begin{thebibliography}{}

\bibitem[Anthony-Twarog \& Twarog (1985)]{att85} Anthony-Twarog, B.J., \& Twarog, B.A. 1985, ApJ, 291, 595
\bibitem[Beasley et al. (2002)]{betal02} Beasly, M., Hoyle, F., \& Sharples, R.M. 2002, MNRAS, 336, 168
\bibitem[Bertelli et al. (1994)]{ber94} Bertelli, G., Bressan, A., Chiosi, C., Fagotto, F., \& Nasi, E. 1994, A\&AS, 106, 275	
\bibitem[Cannon (1970)]{can70} Cannon, R.D. 1970, MNRAS, 150, 111
\bibitem[Carraro \& Chiosi (1994)]{cc94} Carraro, G., \& Chiosi, C. 1994, A\&A, 287, 761 (CC94)
\bibitem[Carrera (2005)]{car05} Carrera, R. 2005, Ph. D. Thesis, Departamento de Astrof{\'\i}sica,
     Universidad de La Laguna, Espa\~na.
\bibitem[Carrera et al. (2008)]{car08} Carrera, R., Gallart, C., Aparicio, A., et al. 2008, AJ, 136, 1039
\bibitem[Carretta \& Gratton (1997)]{cg97} Carretta, E., \& Gratton, R.G. 1997, A\&AS, 121, 95
\bibitem[Cole et al. (2004)]{col04} Cole, A.A., Smecker-Hane, T.A., Tolstoy, E., Bosler, T.L., \& Gallagher,
    J.S. 2004, MNRAS, 347, 367
\bibitem[Crowl et al. (2001)]{cro01} Crowl, H.H., Sarajedinini, A., Piatti, A.E., et al. 2001, AJ, 122, 220  
\bibitem[Da Costa (1991)]{dac91} Da Costa, G.S. 1991, in IAU Symp. 148, The Magellanic Clouds, ed. R.
         Haynes \& D. Milne (Dordrecht: Kluwer), 183
\bibitem[Da Costa \& Hatzidimitriou (1998)]{dch98} Da Costa, G.S., \& Hatzidimitriou, D. 1998, AJ, 115, 1934 (DH98)
\bibitem[Dotter et al. (2007)]{dot07} Dotter, A., Chaboyer, B., Jevremovi\'c, D., et al. 2007, AJ, 134, 376
\bibitem[Dotter et al. (2010)]{dot10} Dotter, A., Sarajedini, A., Anderson, J., et al. 2010, ApJ, 708, 698
\bibitem[Dutra et al. (1999)]{dut99} Dutra, C.M., Bica, E., Clari\'a, J.J., \& Piatti, A.E., 1999, MNRAS, 305, 373
\bibitem[Ferraro et al. (1995)]{fer95} 	Ferraro, F.R., Fusi Pecci, F., Testa, V., et al. 1995, MNRAS, 272, 391
\bibitem[Geisler et al. (1997)]{gei97} Geisler, D., Bica, E., Dottori, H., et al. 1997, AJ 114, 1920
\bibitem[Girardi et al. (2000)]{gir00} Girardi, L., Bressan, A., Bertelli, G., \& Chiosi, C. 2000, A\&AS, 141,371
\bibitem[Girardi (2008)]{gir08} Girardi, L. Dalcanton, J., Williams, B., et al. 2008, PASP, 120, 583
\bibitem[Glatt et al. (2008a)]{gla08a} Glatt, K., Gallagher, J.S., III, Grebel, E.K., et al. 2008a, AJ, 135, 1106
\bibitem[Glatt et al. (2008b)]{gla08b} Glatt, K., Grebel, E.K., Sabbi, E. et al.  2008b, AJ, 136, 1703
\bibitem[Gonzalez \& Wallerstein (1999)]{gon99} Gonzalez, G., \&, Wallerstein, G. 1999, AJ, 117, 2286
\bibitem[Grocholski et al. (2006)]{gro06} Grocholski, A.J., Cole, A.A, Sarajedini, A., Geisler, D., \& Smith, V.
 2006, AJ, 132, 1630
\bibitem[Harris \& Zaritsky (2004)]{har04} Harris, J., \& Zaritsky, D. 2004, AJ, 127, 1531
\bibitem[Janes \& Phelps (1994)]{jp94} Janes, K.A., \& Phelps, R.L. 1994, AJ, 108, 1773 (JP94)
\bibitem[Landolt (1992)]{l92} Landolt, A. 1992. AJ, 104, 340
\bibitem[Marigo et al. (2008)]{mar08} Marigo, P., Girardi, L., Bressan, A., et al.  2008, A\&A, 482, 883
\bibitem[Mighell et al. (1998)]{mig98} 	Mighell, K.J., Sarajedini, A., \& French, R. S. 1998, AJ, 116, 2395
\bibitem[Olszewski et al. (1996)]{ols96} Olszewski, E.W., Suntzeff, N.B., \& Mateo, M. 1996, ARA\&A, 34, 511 
\bibitem[Pagel \& Tautvai\v{s}ien\.{e} (1998)]{pag98} Pagel, B.E.J., \& Tautvai\v{s}ien\.{e}, G.
      1998, MNRAS, 299, 535 (PT98)
\bibitem[Parisi et al. (2009)]{par09} Parisi, M.C., Grocholski A.J., Geisler, D., Sarajedini, A., \& Clari\'a, J.J.
2009, AJ, 138, 517 (Paper I)
\bibitem[Phelps et al. (1994)]{pjm94} Phelps, R.L., Janes, K.A., \& Montgomery, K.A. 1994, AJ, 107, 1079
\bibitem[Piatti (2011a)]{pia11a} Piatti, A.E. 2011a, MNRAS, 416, L89
\bibitem[Piatti (2011c)]{pia11c} Piatti, A.E. 2011c, MNRAS, 418, L69
\bibitem[Piatti (2012)]{pia12b} Piatti, A.E. 2012, ApJ, 756, L32 
\bibitem[Piatti et al. (2011b)] {pia11b} Piatti, A.E., Clari\'a, J.J., Bica, E., et al. 2011b, MNRAS, 417, 1559 
\bibitem[Piatti et al. (2001)]{pia01} Piatti, A.E., Santos, J.F.C., Clari\'a, J.J., et al. 2001, MNRAS, 325, 792
\bibitem[Piatti et al. (2007a)]{pia07a} Piatti, A.E., Sarajedini, A., Geisler, D., Clark, D., \& Seguel, J. 2007a,
       MNRAS, 377, 300
\bibitem[Piatti et al. (2007b)]{pia07b} Piatti, A.E., Sarajedini, A., Geisler, D., Gallart, C., \& Wischnjewsky,
       M. 2007b,  MNRAS, 381, L84
\bibitem[Piatti et al. (2007c)]{pia07c} Piatti, A.E., Sarajedini, A., Geisler, D., Gallart, C., \& Wischnjewsky, M.
       2007c, MNRAS, 382, 1203
\bibitem[Piatti et al. (2005)]{pia05a} Piatti, A.E., Sarajedini, A., Geisler, D., Seguel, J., \& Clark, D. 2005,
        MNRAS, 358, 1215
\bibitem[Pietrinferni et al. (2004)]{pie04} Pietrinferni, A., Cassisi, S., Salaris, M., \& Castelli, F. 2004, ApJ, 612, 168
\bibitem[Rafelski \& Zaritsky (2005)]{rz05} Rafelski, M., \& Zaritsky, D. 2005, AJ, 129, 2701
\bibitem[Rich et al. (2000)]{ri00} Rich, R.M., Shara, M., Fall, S.M., \& Zurek, D. 2000, AJ, 119, 197
\bibitem[Rosenberg et al. (1999)]{ros99} Rosenberg, A., Saviane, I., Piotto, G., \& Aparicio, A. 1999, AJ, 118, 2306
\bibitem[Salaris \& Weiss (1997)]{sw97} Salaris, M., \& Weiss, A. 1997, A\&A, 327, 107
\bibitem[Salaris et al. (2004)]{swp04} Salaris, M., Weiss, A., \& Percival, M. 2004, A\&A, 414, 163 (S04)
\bibitem[Sarajedini \& Demarque (1990)]{sd90} Sarajedini, A., \& Demarque, P. 1990, ApJ, 365, 219
\bibitem[Stetson (1987)]{ste87} Stetson, P.B. 1987, PASP, 99, 191
\bibitem[Storm et al. (2004)]{sto04} Storm, J., Carney, B.W., Gieren, W.P., et al.  2004,
          A\&A, 415, 531
\bibitem[Stryker et al. (1985)]{str85} Stryker, L.L., Da Costa, G.S., \& Mould, J.R. 1985, ApJ, 298, 544
\bibitem[van den Bergh (1991)]{vdb91} van den Bergh, S. 1991, ApJ, 369, 1
\bibitem[Vandenberg et al. (1990)]{van90} Vandenberg, D.A., Bolte, M., \& Stetson, P.B. 1990, AJ, 100, 445
\bibitem[Weisz et al. (2013)] {wei13} Weisz, D.R., Dolphin, A.E., Skillman, E.D., et al. 2013, 2013arXiv1301.7422

\end{thebibliography}
\end{document}